\documentclass[aps,twocolumn,pra,superscriptaddress,amsmath,showpacs,tightenlines]{revtex4-2}
\usepackage{epsfig,graphicx,times}
\usepackage{amstext}
\usepackage{amsmath}            
\usepackage{amssymb}            
\usepackage{graphicx}           
\usepackage{latexsym}
\usepackage{color}
\usepackage{bm}

\begin{document}

\title{Control the qubit-qubit coupling in the superconducting circuit with  double-resonator couplers}

\author{Hui Wang}
\email{wanghuiphy@126.com}
\affiliation{Shandong Inspur Intelligence Research Institute Co., Ltd., Jinan, China }
\affiliation{Shandong Yunhai Guochuang Innovative Technology Co., Ltd, Jinan, China}
\author{Yan-Jun Zhao}
\email{zhao_yanjun@bjut.edu.cn}
\affiliation{Key Laboratory of Opto-electronic Technology, Ministry of Education, Beijing University of
Technology, Beijing, China}

\author{Hui-Chen Sun}
\affiliation{Institute for Quantum Computing and Department of Physics and Astronomy, University of Waterloo,
Waterloo, Ontario N2L 3G1, Canada}

\author{Xun-Wei Xu}
\affiliation{Key Laboratory of Low-Dimensional Quantum Structures and Quantum Control of Ministry of Education,
Key Laboratory for Matter Microstructure and Function of Hunan Province,
Department of Physics and Synergetic Innovation Center for Quantum Effects and Applications,
Hunan Normal University, Changsha 410081, China}

\author{Yong Li}
\affiliation{Shandong Inspur Intelligence Research Institute Co., Ltd., Jinan, China }
\affiliation{Shandong Yunhai Guochuang Innovative Technology Co., Ltd}

\author{Yarui Zheng}
\email{ricezheng@qelement.xyz}
\affiliation{Quantum Element Technology (Shen Zhen) Co.,Ltd., Shenzhen 518048, China}

\author{Qiang Liu}
\affiliation{Shandong Inspur Intelligence Research Institute Co., Ltd., Jinan, China }
\affiliation{Shandong Yunhai Guochuang Innovative Technology Co., Ltd}

\author{Rengang Li}
\affiliation{Shandong Inspur Intelligence Research Institute Co., Ltd., Jinan, China }

\date{\today}

\begin{abstract}
We propose a theoretical scheme of using two-fixed frequency resonator couplers to tune the interaction  between two  Xmon qubits. The  indirect  interaction between two qubits induced by two resonators can cancel each other, so the direct  qubit-qubit coupling is not essential for the switching off. So we can suppress the static ZZ coupling with the weak direct qubit-qubit coupling and even eliminate the static ZZ coupling through the destructive interferences of the double-path couplers.
The cross-kerr resonance  can induce additional poles for the static ZZ coupling which should be kept away during the two-qubit gates.
The double-resonator couplers scheme could unfreeze some  restrictions during the design of superconducting quantum chips and mitigate the static ZZ coupling, which might supply a promising platform for future superconducting quantum chip.
  \end{abstract}

\maketitle
 \pagenumbering{arabic}

\section{Introduction}

In past several years, the superconducting quantum computing  develops quickly,  IBM announced 433 qubits superconducting quantum chip at the end of 2022, and  plan to launch  quantum chip with more than 1000 qubits in 2023.  The coherence time of superconducting qubits fabricated with new superconducting materials is  greatly enhanced\cite{Place,Xueg,Ren,Shen,Bal}, and the introduction of tunable coupler greatly enhances the fidelities of two-qubit  gates to above $99.5\%$ \cite{Chen,Yan,Sun,Tan1,Kandala,Zajac,Moskalenko}. The quantum supremacy of random circuit sampling and other multi-body quantum simulation experiments have been conducted on the superconducting quantum chip  with more than 50 qubits \cite{Martinis,Wu,Sirui,Shen}. But the fidelities of two-qubit gate  are still not high enough for the universal  quantum computer, and the state leakages and residual coupling are still need to be suppressed in superconducting qunatum chip.

The tunable coupler can switch off the interactions between adjacent  qubits, which can isolate  qubits from the surrounding environments for local quantum operations.  In the single-coupler circuit, the induced indirect qubit-qubit coupling (dispersive type interaction)   can  not be zero for finite frequency detuning between qubit and coupler, so the direct qubit-qubit interaction is required for  switching off.
 If the direct qubit-qubit coupling is very weak,  the switching off frequency should be very high, and this leaves narrow available frequency ranges for readout resonators (or qubits). For the case of strong direct qubit-qubit coupling, the state leakages and crosstalks  should be another perplex\cite{Chen,Yan,Sun}.
   So there are many limitations during the  design of single-coupler superconducting quantum chip, and the residual coupling  and state leakages  are still serious troubles\cite{Sung}.

In this article, we propose a theoretical scheme to dynamically tune the qubit-qubit coupling with
 the double-resonator  couplers in the superconducting quantum chip. As theoretically and experimentally demonstrated, the superconducting resonator  can function as a coupler\cite{wang,Wu2,Ming,IBM,McKay,Stehlik}.
 In particular, if the two resonator couplers take the respective  maximal and minimal frequencies, the induced indirect qubit-qubit coupling by two resonators are in opposite signs and can cancel each other. So the direct qubit-qubit couplings is not indispensable for  switching off in the double-resonator coupler circuit, which can hopefully unfreeze some restrictions on the superconducting quantum chip, such as the qubit-qubit coupling strengths, maximal frequencies of couplers, and so on.
The switching off positions can be  very close to   two-qubit gate regimes in the double-resonator couplers circuit, thus the maximal frequencies of couplers can be smaller.  So available frequency ranges for readout resonators or qubits can be wider in double-path coupler circuit, and this should relieve  the frequency crowding on the superconducting quantum chip.

We also study the effects of  superconducting artificial atom's high-excited states on qubits' energy levels and switching off positions.
The elimination of the static ZZ coupling through the destructive interferences of double-path couplers are also explored\cite{AHouck,Goto,Kandala,Sete}.   We find that the cross-kerr resonances through the virtual photon exchange could induce new poles of the static ZZ coupling,  which should be kept away from during the two-qubit gates.

The paper is organized as follows:  In Sec. II,  we first perform Numerical calculation of the circuit energy levels. In Sec. III, we then discuss the Switching off for  the qubit-qubit coupling. In Sec. IV,  we further study the suppression and cancellation of Static ZZ coupling. Finally, we summarize the results in Sec. V.

\begin{figure}
\includegraphics[bb=15 230 395 525, width=4.7 cm, clip]{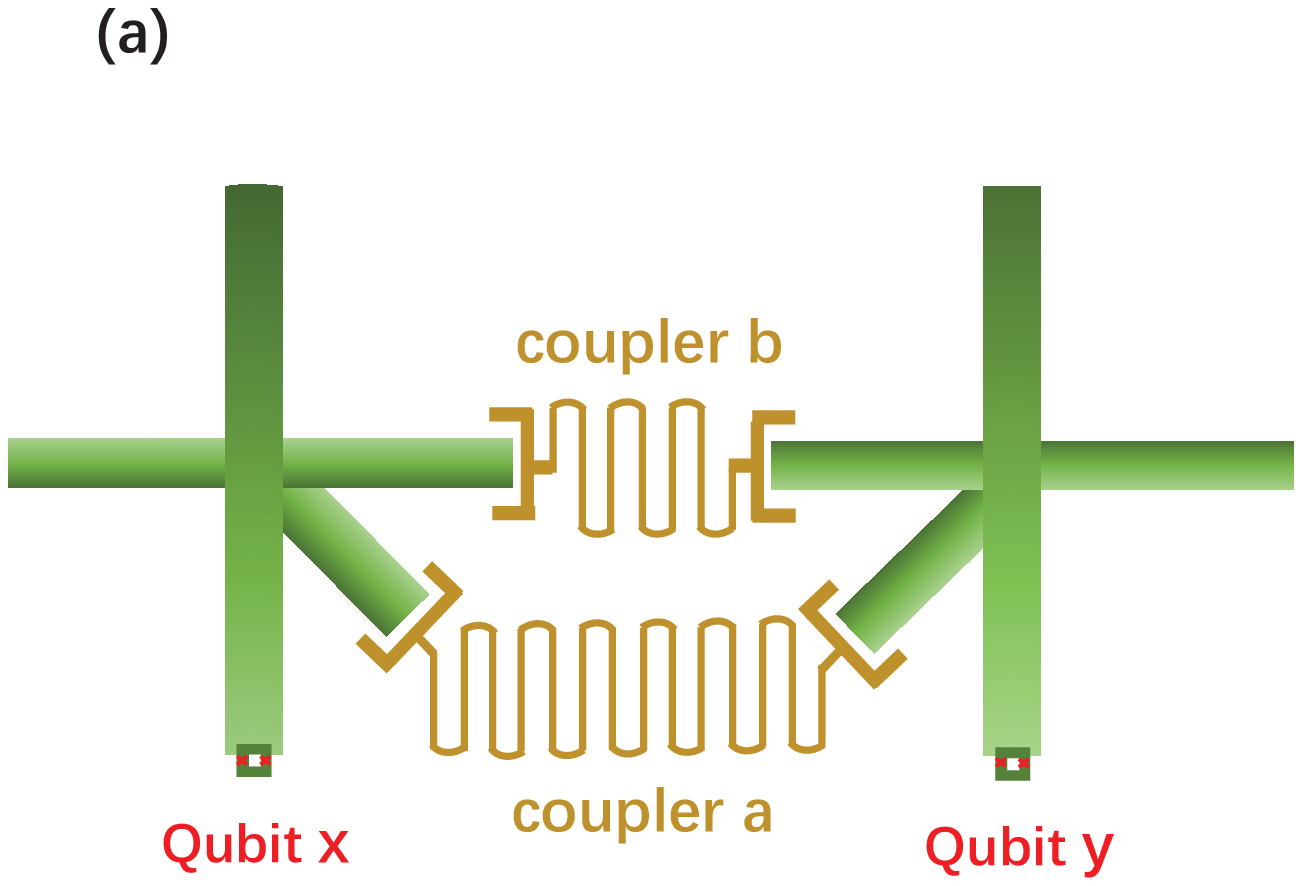}
\includegraphics[bb=0 20 593 530, width=3.85 cm, clip]{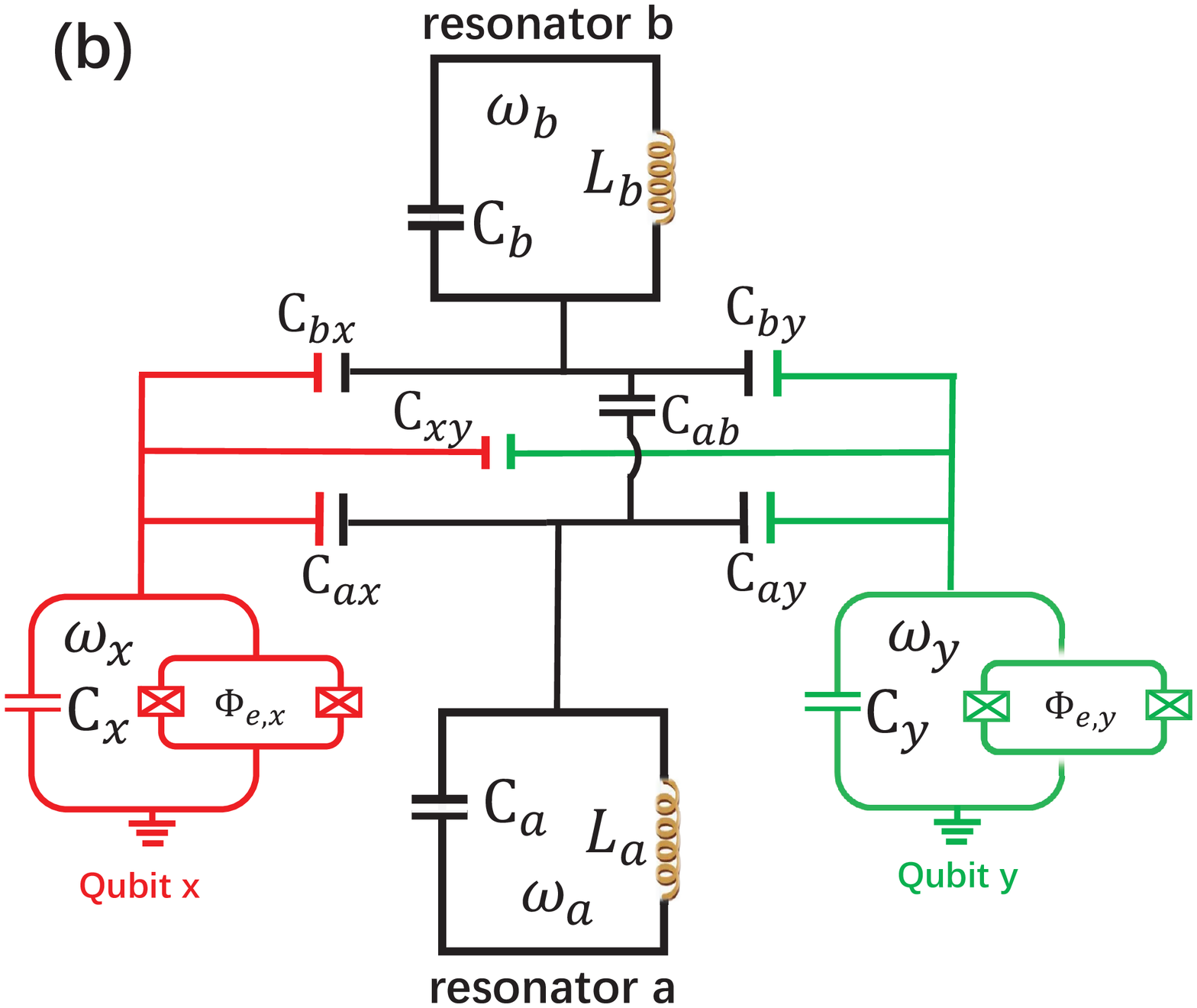}
\caption{(Color online) (a) Schematic diagram. The double-resonator couplers superconducting circuit consists of two Xmon qubits coupling to two common fixed-frequency resonators.  (b) The equivalent circuit. The $\omega_{\eta}$ is the frequency of a qubit or resonator,  with $\eta=a,b,x,y$.  The inductive type coupling is neglected, and there are only capacitive type interactions in the superconducting circuit.
  $C_{\eta}$ is the capacitances of a qubit  or resonator, and the two-body relative capacitance is $C_{\eta \eta^{\prime}}$, with $C_{\eta \eta^{\prime}}=C_{\eta^{\prime} \eta}$ and $\eta \neq \eta^{\prime}$.  $L_{a}$ and $L_{b}$ are the respective inductances of resonators \textbf{a} and \textbf{b}, while $\Phi_{e,x}$ and $\Phi_{e,y}$ are the external magnetic fluxes applying on the superconducting loops of qubits \textbf{x} and \textbf{y}, respectively.
}
\label{fig1}
\end{figure}

\section{Circuit Energy levels}

 In this section, we numerically calculate the energy levels of qubits for the superconducting  circuit in Fig.~\ref{fig1} with the QuTiP software \cite{Johansson1,Johansson2,SunH,Sung}.
  The superconducting circuit consists of two Xmon qubits coupling to two common resonator couplers. The two-body interactions are all assumed as capacitive type,  and the direct qubit-qubit and resonator-resonator interactions  are very weak. The two-body interactions in the superconducting circuit are all assumed as capacitive type.
Because of the small anharmonicities,  the high-excited states of superconducting artificial atoms  should also make contributions to the energy levels of qubits and couplers. Truncated to the second-excited states of  atoms  and third-excited states of  resonators,  the curved surfaces  of  qubits and resonators'  energy levels are plotted  in  Appendix \textbf{A}  (see Fig.~\ref{fig10}).

\begin{figure}
\includegraphics[bb=0 0 440 330, width=4.28 cm, clip]{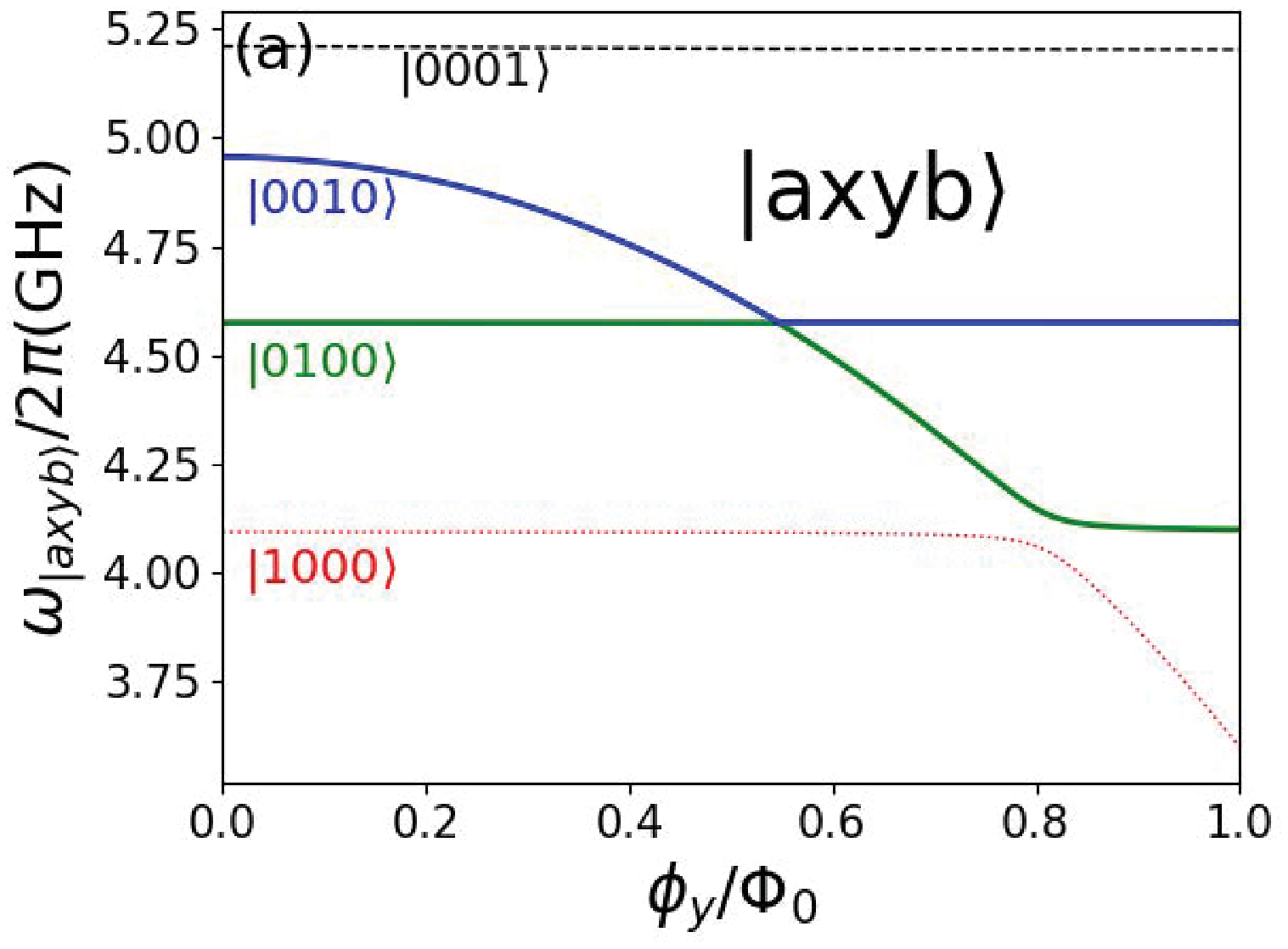}
\includegraphics[bb=0 0 440 330, width=4.28 cm, clip]{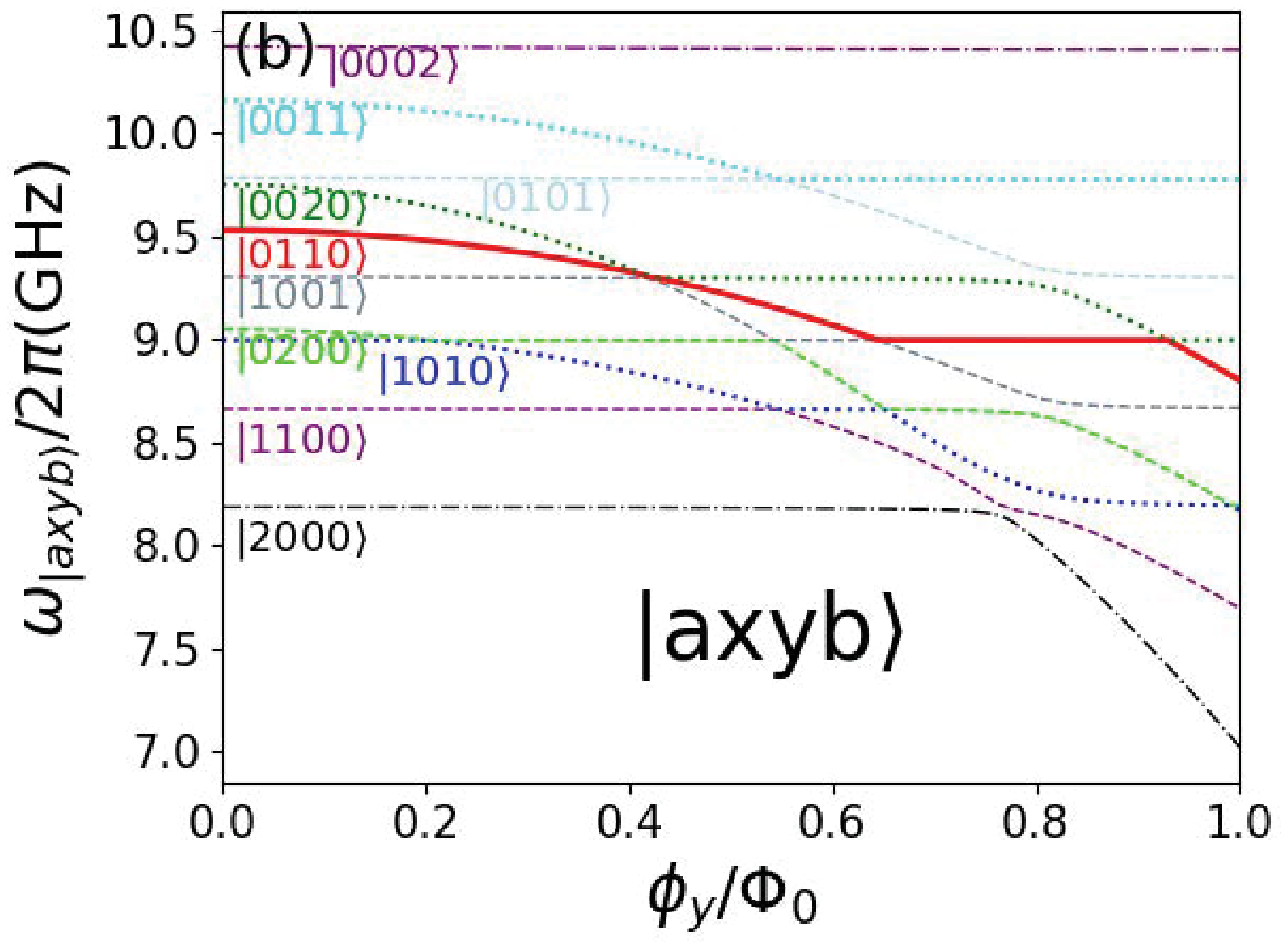}
\caption{(Color online)
The  energy level diagram. The energy level curves for (a) single and (b) double-excited states, here we set the transition frequency of qubit \textbf{x} as $\omega_x/2\pi=4.56$ GHz.
The maximal frequencies of qubits \textbf{x} and \textbf{y} are respective $\omega^{(max)}_x/2\pi=4.56$ GHz and $\omega^{(max)}_y/2\pi=5.12$ GHz, their corresponding anharmonicities are $\alpha_x/2\pi=-175$ MHz and $\alpha_y/2\pi=-195$ MHz. The resonant frequencies of resonators \textbf{a} and \textbf{b} are  $\omega_a/2\pi=4.10$ GHz and $\omega_b/2\pi=5.20$ GHz, respectively. The direct qubit-qubit and direct resonator-resonator coupling strengths are $g_{xy}/2\pi=1.0$ MHz  and the $g_{ab}/2\pi=0.1$ MHz,  respectively.
The coupling strengths of qubits  with resonator \textbf{a}  are $g_{ax}/2\pi=g_{ay}/2\pi=32$ MHz,
and  $g_{bx}/2\pi=g_{by}/2\pi=30$ MHz label the coupling strengths of qubits with  resonator  \textbf{b}.
}
\label{fig2}
\end{figure}

 For simplicity, we focus on the special case that the resonant frequency $\omega_a$ of resonator \textbf{a},  resonant frequency  $\omega_b$ of resonator \textbf{b} , and the  transition frequency $\omega_x$ of qubit \textbf{x}  are fixed, and only the transition frequency  $\omega_y$ for qubit \textbf{y}  is tuned by the external magnetic flux $\Phi_{e,y}$. By setting $\omega_x/(2\pi)=4.56$ GHz,  the energy level curves of qubits and resonators' single and double-excited states  are plotted in Figs.\ref{fig2}(a) and \ref{fig2}(b), respectively. During the  numerical calculations with QuTiP software , the transition frequencies of qubit \textbf{x} and  qubit \textbf{y} are respectively chosen as $\omega_x/(2\pi)=4.56$ GHz and $\omega_y/(2\pi)=5.12$ GHz, the resonant frequencies of resonator \textbf{a} and resonator \textbf{b} are $\omega_a/(2\pi)=4.10$ GHz and $\omega_b/(2\pi)=5.20$ GHz, respectively. The coupling strengths between resonator \textbf{a} (or \textbf{b}) and two qubits are $g_{ax}/(2\pi)=g_{ay}/(2\pi)=32$ MHz (or $g_{bx}/(2\pi)=g_{by}/(2\pi)=30$ MHz), so the qubits and couplers are in the dispersive coupling regimes.

  The ket vector of four-body quantum state is defined as  $|m_a m_x m_y m_b\rangle$, and the values of $m_a, m_x, m_y, m_b$   respectively describe quantum numbers of resonator \textbf{a}, qubit \textbf{x}, qubit \textbf{y}, and resonator \textbf{b}. Because of the avoided crossing effect, each curve in  Fig.~\ref{fig2}  can not describe the whole energy level of a certain quantum state,  here we mark  each curve with the corresponding state at the zero magnetic flux.
 In Fig.\ref{fig2}(a), the transition frequency of qubit \textbf{y} decreases under magnetic field,  and it becomes anticrossing  with qubit \textbf{x} at the frequency regimes  close to 4.56 GHz and  with resonator \textbf{a} at the regimes close to 4.10 GHz.  For the double-excited states, the energy levels and avoided crossing gaps  can be seen in Fig.~\ref{fig2}(b). The energy level structure in Fig.\ref{fig2} is important for analyzing the switching off and the static ZZ coupling on the superconducting quantum chip as shown in the follow sections.

\section{switching off}

\subsection{Circuit Quantization}

Figure~\ref{fig1}(a) describes the superconducting circuit consisting of two Xmon qubits coupling to two common fixed frequency resonators, the frequencies of qubits are tunable. Besides the direct  interactions, the   qubit-resonator couplings can also induce  indirect qubit-qubit and resonator-resonator interactions. The   resonant frequencies of resonators \textbf{a} and \textbf{b} are respective $\omega_a$ and  $\omega_b$,  while $\omega_x$ and  $\omega_y$ label the transition frequencies of qubits \textbf{x} and \textbf{y}, respectively. In the case of zero  magnetic fluxes, the frequencies of qubits and resonators satisfy $\omega_{a}<\omega_{x}<\omega_{y}<\omega_{b}$. We expect a large distance between two resonators and neglect their  inductive coupling, then the interactions  are all regarded as  capacitive type  as shown in Fig.~\ref{fig1}(b).  The capacitances of resonators should be distributed type and proportional to their lengths, but in the article we simply label the total capacitances of resonators \textbf{a} and \textbf{b} as $C_a $ and $C_b$, respectively.

Inspired by  previous work\cite{wang,Wu2},  the kinetic energy of superconducting circuit with
double-resonator couplers can be written as $T=\sum_{\eta=a,b,x,y}C_\eta\dot{\phi}^2_\eta/2+\sum_{\eta,\eta^{\prime}=a,b,x,y \atop\eta\neq\eta^{\prime}} C_{\eta\eta^{\prime}}\left( \dot{\phi}_\eta-\dot{\phi}_{\eta^{\prime}}\right)^2/4$,
where $C_\eta$ is the capacitance of qubit or resonator, and $C_{\eta\eta^{\prime}}$ ($C_{\eta\eta^{\prime}}=C_{\eta^{\prime}\eta}$) is relative capacitance between arbitrary two devices among qubits and resonators, with $\eta,\eta^{\prime}=a,b,x,y$ and $\eta\neq\eta^{\prime}$.
The $\phi_a$ and $\phi_b$ are the respective magnetic fluxes of circuit nodes for resonators \textbf{a} and \textbf{b}, while $\phi_x$ and $\phi_y$ are the respective   node fluxes of  qubits \textbf{x} and \textbf{y}, and they can be tuned  by the  external magnetic fluxes $\Phi_{e,x}$ and $\Phi_{e,y}$\cite{Filipp,Devore,Girvin}.
 If we label the  $L_a$ and $L_b$ as the respective inductances of resonators \textbf{a} and \textbf{b}, thus the  potential energy of superconducting circuit can be written as $U=\sum_{\lambda=a,b}\phi^2_\lambda/(2L_\lambda)+\sum_{\beta=x,y}E_{J_{\beta}}\left[1-\cos\left(2\pi\phi_\beta/\Phi_0\right)\right]$, the subscript $\lambda=a,b$ label the respective variables of resonators \textbf{a} and \textbf{b}, while  $\beta=x,y$ describe the variables of qubit \textbf{x} and qubit \textbf{y}, respectively.
The $E_{J_{\beta}}=I_{c\beta}\Phi_{0}/(2\pi)$ is the Josephson energy of  qubit $\beta$, where the $I_{c\beta}$ is the corresponding critical current, and  $\Phi_0=h/2e$ is the flux quantum  with the planck constant \textbf{h}  and an electron charge \textbf{e}.

With the  kinetic energy $T$ and potential energy $U$, the Lagrangian of the superconducting circuit in Fig.~\ref{fig1} can be formally written as  $L = T-U$. If we define the generalized momentum operators as $q_{\eta}= \partial L/\partial \dot{\phi}_{\eta}=C \dot{\phi}_{\eta}$ (with $\eta=a,b,x,y$), under the conditions $C_{ab}\ll C_{xy}\ll C_{ax}, C_{ay}, C_{bx}, C_{by}\ll C_{x}, C_{y} \ll C_{a},C_{b}$,  we  obtain the expression of Hamiltonian  (see Appendix \textbf{B})
  \begin{eqnarray}\label{eq:1}
 H&=&4\sum_{\lambda=a,b} \left[E_{C_\lambda}(n_\lambda)^2+\frac{\phi^2_\lambda}{8L_\lambda}\right]\nonumber\\
 &+&\sum_{\beta=x,y}\left[E_{C_\beta}(n_\beta)^2-E_{j\beta}\cos\left(\frac{2\pi}{\Phi_0}\phi_\beta\right)\right] \nonumber\\
 &+ &8\sum_{\lambda=a,b \atop \beta=x,y}\frac{C_{\lambda\beta}}{\sqrt{C_{\lambda}C_{\beta}}}\sqrt{E_{C_\lambda}E_{C_\beta}}(n_\lambda n_\beta)\\
 &+ &8\left(1+\frac{C_{ax}C_{bx}}{C_{x}C_{ab}}+\frac{C_{ay}C_{by}}{C_{y}C_{ab}}\right)\frac{C_{ab}}{\sqrt{C_{a}C_{b}}}\sqrt{E_{C_a}E_{C_b}}(n_a n_b)\nonumber\\
 &+ &8\left(1+\frac{C_{ax}C_{ay}}{C_{a}C_{xy}}+\frac{C_{bx}C_{by}}{C_{b}C_{xy}}\right)\frac{C_{xy}}{\sqrt{C_{x}C_{y}}}\sqrt{E_{C_x}E_{C_y}}(n_x n_y),\nonumber
 \end{eqnarray}
where $n_{\eta}=q_{\eta}/2e$  is the Cooper-pair number operator of a qubit or resonator,
 and the corresponding charging energy is  $E_{C_\eta}=e^2/2C_{\eta}$.  The  transition frequencies of resonators and qubits are respectively defined as $\omega_{\lambda}=1/\sqrt{C_{\lambda} L_{\lambda}}$ and $\omega_{\beta}=(\sqrt{8 E_{J_{\beta}}E_{C_{\beta}}}-E_{C_{\beta}})/\hbar$, while  $\alpha_{\beta}=-E_{C_{\beta}}/\hbar$ labels the anharmonicity of qubit $\beta$.
As shown in Appendix \textbf{B}, the two-body coupling strengths among qubits and resonators can be defined as
\begin{eqnarray}\label{eq:2}
g_{\lambda\beta}&=&\frac{1}{2}\frac{C_{\lambda\beta}}{\sqrt{C_{\lambda}C_{\beta}}}\sqrt{\omega_{\lambda}\omega_{\beta}},\\
g_{ab}&=&\frac{1}{2}\left(1+\frac{C_{ax}C_{bx}}{C_{x}C_{ab}}+\frac{C_{ay}C_{by}}{C_{y}C_{ab}}\right)\frac{C_{ab}}{\sqrt{C_{a}C_{b}}}\sqrt{\omega_a \omega_b},\quad\\
g_{xy}&=&\frac{1}{2}\left(1+\frac{C_{ax}C_{ay}}{C_{a}C_{xy}}+\frac{C_{bx}C_{by}}{C_{b}C_{xy}}\right)\frac{C_{xy}}{\sqrt{C_{x}C_{y}}}\sqrt{\omega_x \omega_y}.
\end{eqnarray}
The two-body interactions  are mainly decided by their relative capacitances  $C_{\eta \eta^{\prime}}$ with
$\eta, \eta^{\prime}=a, b, x, y$ and $\eta\neq \eta^{\prime}$.
The qubit-resonator coupling strength $g_{\lambda\beta}$ in Eq.(2) could induce  indirect  interaction between two qubits, so the Eq.~(4) can not describe the complete interaction between two qubits. In the  single-coupler superconducting quantum chip, there are many restrictions on the capacitances and frequencies of qubits (or couplers), but these limitations might be unfrozen in the double-coupler circuit as will be discussed in the follow sections.

\begin{figure}
\includegraphics[bb=0 -15 420 305, width=4.48 cm, clip]{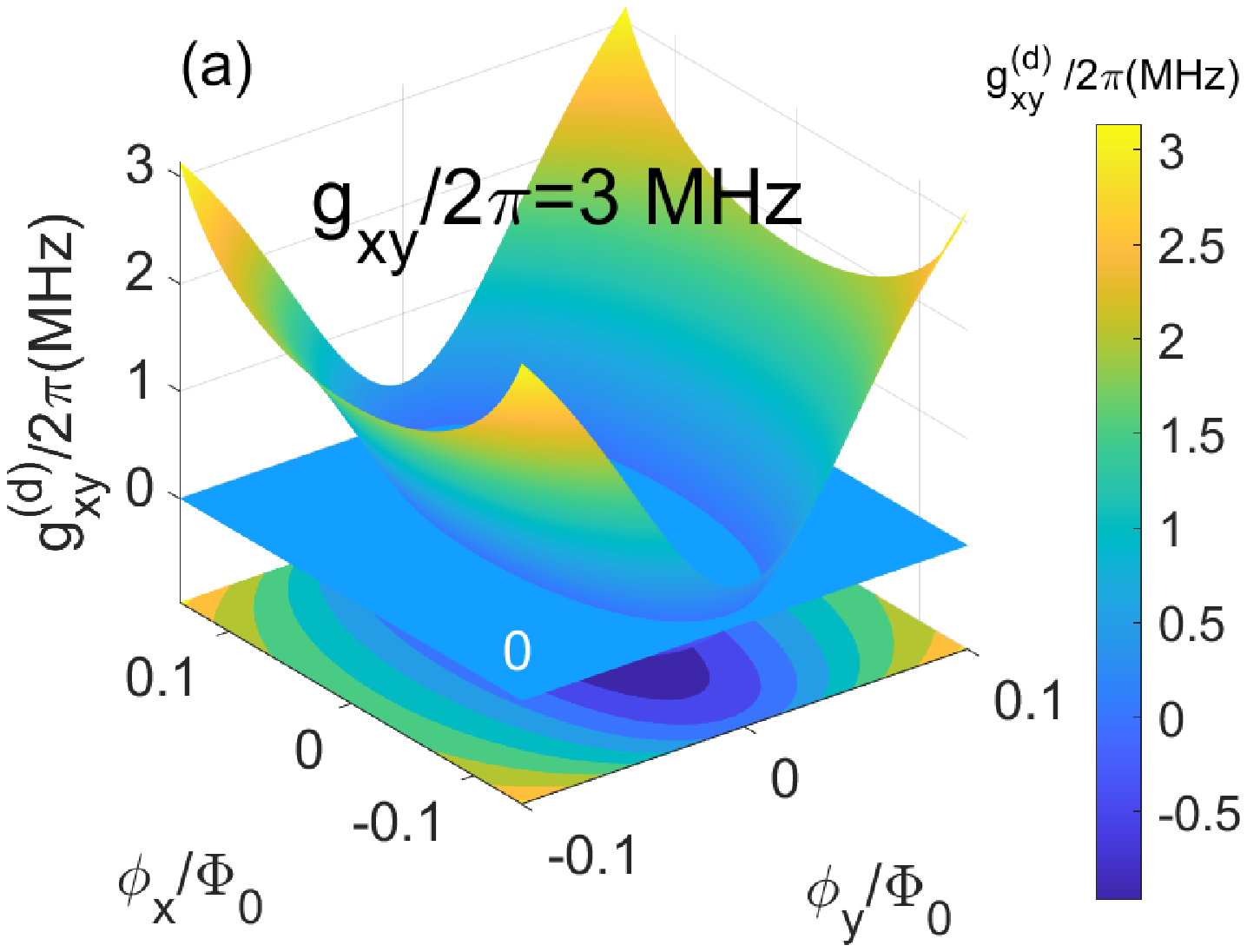}
\includegraphics[bb=-5 0 465 430, width=3.98 cm, clip]{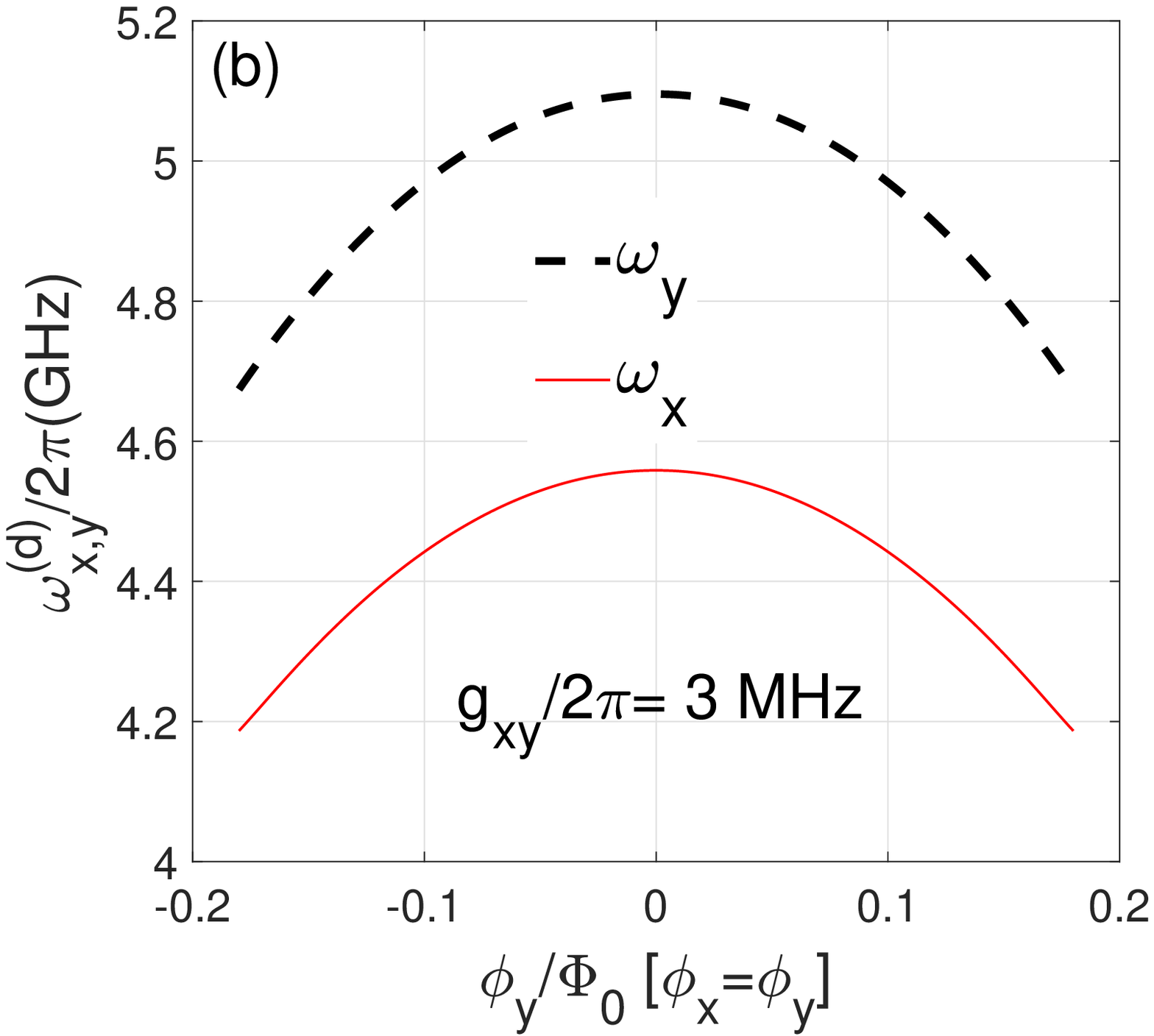}\\
\includegraphics[bb=0 -15 400 300, width=4.43 cm, clip]{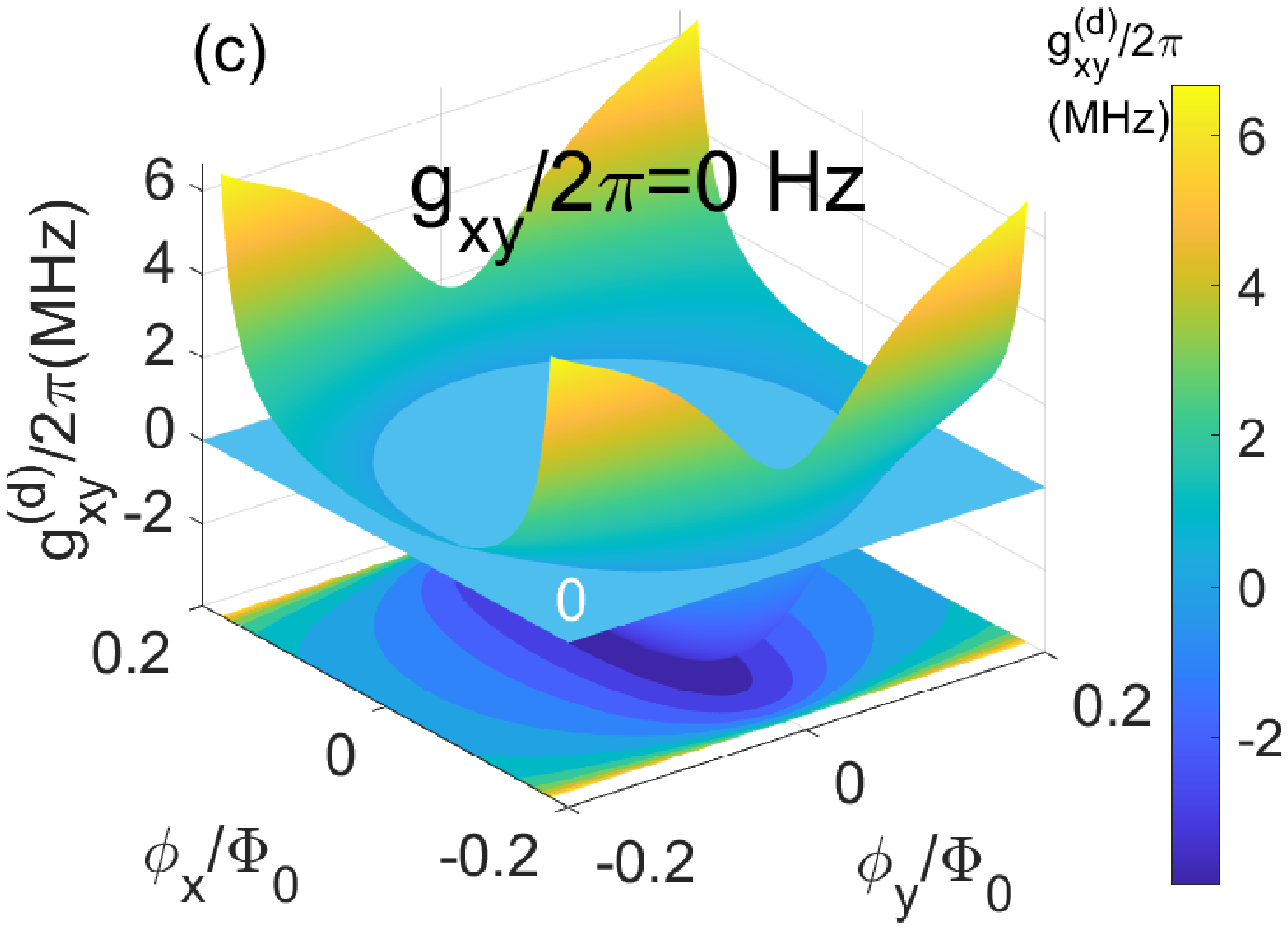}
\includegraphics[bb=-5 0 470 430, width=4.02 cm, clip]{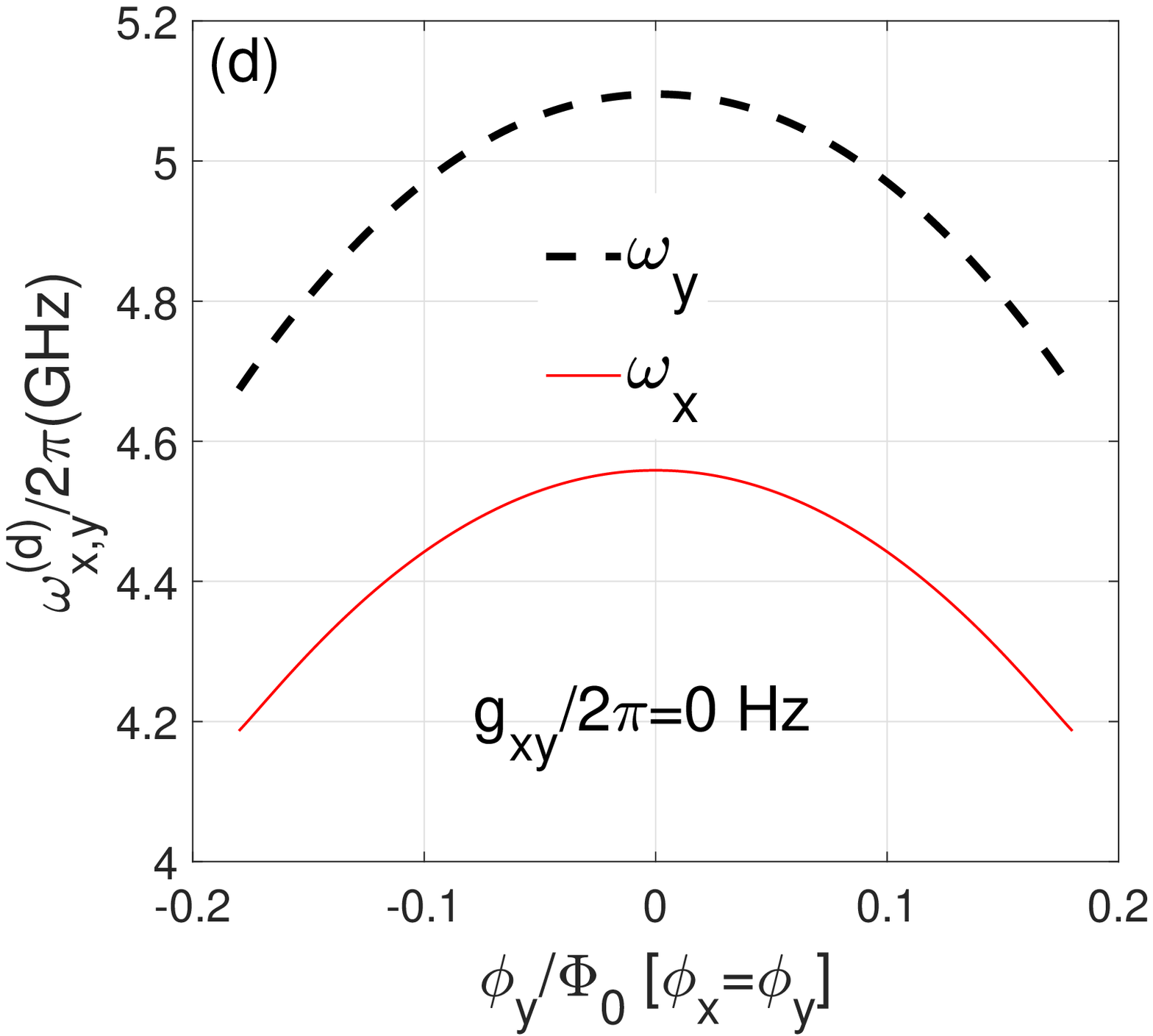}
\caption{(Color online) The decoupled qubit-qubit coupling strengths.
The curved surfaces of  $g^{(d)}_{xy}$ are plotted in (a) $g_{xy}/2\pi=3$ MHz and (c) $g_{xy}/2\pi=0$ Hz,
while (b) and (d) describe the corresponding decoupled frequencies of qubits along the diagonal line direction in the $\phi_x$-$\phi_y$ plane.
The other parameters are the same as in Fig.2.
}
\label{fig3}
\end{figure}

\subsection{Effective coupling}

To get the effective qubit-qubit coupling, we try to decouple the  qubit-resonator interactions in this section.
For the Xmon qubit, the Josephson energy is much larger than its charging energy, $E_{J_{\beta}}/E_{C_{\beta}}\gg 1$, and then the $\phi_\beta$ should be very small and we can  use the approximate equation: $\cos(\phi_\beta)=1-\phi^2_\beta/2+\phi^4_\beta/24-...$.
If we introduce the creation and  annihilation operators by the definitions
   $\phi_\beta=\sqrt[4]{2E_C/E_{J_{\beta}}}(a^{\dagger}_\beta+a_\beta)$,
  and $n_\beta=(i/2)\sqrt[4]{2E_C/E_{J_{\beta}}}(a^{\dagger}_\beta-a_\beta)$,
 the second-quantization  Hamiltonian  can be obtained as $H=\sum_{\lambda=a,b}{H_{\lambda}}+\sum_{\beta=x,y}{H_{\beta}}+\sum_{\lambda=a,b \atop \beta=x,y}H_{\lambda\beta}+H_{ab}+H_{xy}$, with
 \begin{eqnarray}\label{eq:3}
 H_{\lambda}/\hbar&=&\frac{\omega_{\lambda}}{2} c^{\dagger}_{\lambda} c_{\lambda},\\
 H_{\beta}/\hbar
&=& \frac{\omega_{\beta}}{2} a^{\dagger}_{\beta} a_{\beta}+\frac{\alpha_{\beta}}{2} a^{\dagger}_{\beta} a^{\dagger}_{\beta} a_{\beta} a_{\beta},\\
H_{\lambda\beta}/\hbar&=& g_{\lambda\beta}(c^{\dagger}_{\lambda}a_{\beta}+c_{\lambda}a^{\dagger}_{\beta}-c^{\dagger}_{\lambda}a^{\dagger}_{\beta}-c_{\lambda}a_{\beta}),\\
H_{ab}/\hbar&=& g_{ab}(c^{\dagger}_{a}c_{b}+c_{a}c^{\dagger}_{b}-c^{\dagger}_{a}c^{\dagger}_{b}-c_{a}c_{b}),\\
H_{xy}/\hbar&=& g_{xy}(a^{\dagger}_{x}a_{y}+a_{x}a^{\dagger}_{y}-a^{\dagger}_{x}a^{\dagger}_{y}-a_{x}a_{y}).
 \end{eqnarray}
We define $\alpha_{\beta}=-E_{C_{\beta}}/\hbar$ to describe anharmonicity of qubit $\beta$, and the nonlinear term $(\alpha_{\beta}/2) a^{\dagger}_{\beta} a^{\dagger}_{\beta} a_{\beta} a_{\beta}$ reflects the effects of high-excited states of superconducting artificial atom.  We define the $\Delta_{\lambda\beta}=\omega_{\beta}-\omega_{\lambda}$ to describe the frequency detuning between the qubit $\beta$ and resonator $\lambda$, while $\Sigma_{\lambda\beta}=\omega_{\beta}+\omega_{\lambda}$ is the frequency summation of qubit $\beta$ and resonator $\lambda$.   $\Delta_{xy}=\omega_{y}-\omega_{x}$ describes the frequency detuning between two qubits, and  $\Delta_{ab}=\omega_{b}-\omega_{a}$ labels the frequency detuning between two resonators.

Separating the Hamiltonian as $H=H_0+H_{int}$,  the free term is defined as $H_0=\sum_{\lambda=a,b}H_{\lambda}+\sum_{\beta=x,y}H_{\beta}$, while the interaction term is $H_{int}=H_{ab}+H_{xy}+\sum_{\lambda=a,b \atop \beta=x,y}H_{\lambda\beta}$. In the  qubit-resonator  dispersive coupling regimes $g_{\lambda\beta}/|\Delta_{\lambda\beta}|\ll1$
and $g_{\lambda \beta}/\Sigma_{\lambda \beta} \ll 1$, we define $S=\sum_{\lambda=a,b \atop \beta=x,y}[(g_{\lambda\beta}/\Delta_{\lambda\beta})(c^{\dagger}_{\lambda} a_{\beta}-c_{\lambda} a^{\dagger}_{\beta})-(g_{\lambda\beta}/\Sigma_{\lambda\beta}) (c^{\dagger}_{\lambda} a^{\dagger}_{\beta}-c_{\lambda} a_{\beta})]$.
Under the Schrieffer–Wolff transformation, if we choose $H^{(d)}=\exp{(S)}H\exp{(-S)}$ and  $H_{int}+[S,H_0]=0$, thus
the  decoupled Hamiltonian  becomes $H^{(d)}=H_0-(1/2)[H_{int},S]+O(H_{int}^3)$ (see Appendix \textbf{C}), that is
\begin{eqnarray}\label{eq:4}
\frac{H^{(d)}}{\hbar}&=&\sum_{\lambda=a,b}\omega^{(d)}_{\lambda}c^{\dagger}_{\lambda}c_{\lambda}+ \sum_{\beta=x,y}\left(\omega^{(d)}_{\beta}a^{\dagger}_{\beta}a_{\beta}+\frac{\tilde{\alpha}_\beta}{2} a^{\dagger}_{\beta}a^{\dagger}_{\beta}a_{\beta}a_{\beta}\right)\nonumber\\
&+&g^{(d)}_{xy}(a^{\dagger}_{x}a_{y}+a^{\dagger}_{y}a_{x})+ g^{(d)}_{ab}(c^{\dagger}_{a}c_{b}+c^{\dagger}_{b}c_{a}).
\end{eqnarray}
  Since $g_{a b}, g_{xy}\ll g_{\lambda\beta}$, the contributions of $ H_{ab}$ and $H_{xy}$  have been neglected. Following the method of  previous work\cite{Yan}, we  assumed  $\tilde{\alpha}_{\beta} \approx \alpha_{\beta}$ during the  derivations of Eq.(10), thus the contributions of superconducting artificial atoms' high-excited states are neglected.

 The decoupled frequencies of qubits and resonators  can be respectively  defined as $\omega^{(d)}_{\beta}=\omega_{\beta}+\sum_{\lambda=a,b}(g^2_{\lambda \beta}/\Delta_{\lambda \beta}-g^2_{\lambda \beta}/\Sigma_{\lambda \beta})$ and $\omega^{(d)}_{\lambda}=\omega_{\lambda}-\sum_{\beta=x,y}(g^2_{\lambda \beta}/\Delta_{\lambda \beta}-g^2_{\lambda x}/\Sigma_{\lambda \beta})$ (as shown in  Appendix \textbf{C},).
And  the decoupled qubit-qubit coupling strength can be obtained as
\begin{eqnarray}\label{eq:5}
g^{(d)}_{xy}&=&\frac{1}{2}\sum_{\lambda=a,b \atop \beta=x,y}\left(\frac{g_{\lambda x}g_{\lambda y}}{\Delta_{\lambda \beta}}-\frac{g_{\lambda x}g_{\lambda y}}{\Sigma_{\lambda \beta}}\right)+g_{xy}.
\end{eqnarray}
Since  $\Delta_{\lambda \beta}$ and $\Sigma_{\lambda \beta}$ depend the frequency of qubit $\beta$,
so the induced qubit-qubit coupling $g^{(in)}_{\lambda,xy}=(1/2)\sum_{\atop \beta=x,y}(g_{\lambda x}g_{\lambda y}/\Delta_{\lambda \beta}-g_{\lambda x}g_{\lambda y}/\Sigma_{\lambda \beta})$ can be tuned  by the external magnetic fluxes $\Phi_{e,x}$ and $\Phi_{e,y}$. To switch off the qubit-qubit coupling ($g^{(d)}_{xy}/(2\pi)=0$ Hz), we should find parameters to satisfy $-g^{(in)}_{xy}=g_{xy}$.

From the expression of $g^{(in)}_{\lambda,xy}$, both two resonators  make contributions to the effective qubit-qubit coupling, and their contributions might cancel each other ($g^{(in)}_{a,xy}+g^{(in)}_{b,xy}=0$) if the qubit frequency of qubits satisfy certain conditions.
Thus the direct qubit-qubit coupling might be not necessary for the switching off in the double-resonator couplers  circuit.
 The qubits could also induce indirect interactions between the two resonators, and  the decoupled resonator-resonator coupling strength can be defined as $g^{(d)}_{ab}=(1/2)\sum_{\lambda=a,b \atop \beta=x,y}\left( g_{a \beta}g_{b \beta}/\Delta_{\lambda \beta}-g_{a \beta}g_{b \beta}/\Sigma_{\lambda \beta}\right)+g_{ab}$ (see Appendix \textbf{C}). Because of the large frequency detuning between two resonators ($|\Delta_{ab}|\gg g^{(d)}_{ab}$), the effective resonator-resonator interaction makes little effect on the energy levels of qubits and  resonators.

With the parameters in Fig.~\ref{fig2}, we can get $g_{by}/\min{(|\Delta_{by}|)}\sim 1/3$ and $g_{ax}/\min{(|\Delta_{ax}|)}\sim 1/15$  in the  idling states of qubits (without external magnetic field). This means that the qubits and resonators are in the dispersive or weak-dispersive coupling regimes, thus the perturbation method can be used to calculate the effective qubit-qubit coupling .
If we choose $\omega_a<\omega_x<\omega_y<\omega_b$,  the signs of $\Delta_{a\beta}$ and $\Delta_{b\beta}$ are opposite. As indicated by  $g^{(in)}_{\lambda,xy}$,   the resonator  \textbf{b}  will induce negative indirect qubit-qubit coupling, and the  contributions of resonators \textbf{a} is positive.
With  Eq.~(11), the  curved surfaces of  $g^{(d)}_{xy}$ are plotted in Figs.~\ref{fig3}(a) and \ref{fig3}(c), and the corresponding decoupled frequencies of qubits  are shown in  Figs.~\ref{fig3}(b) and \ref{fig3}(d).
 The curved surface of  $g^{(d)}_{xy}$ has many crossing points  with the zero value plane
($g^{(d)}_{xy}/(2\pi)=0 $ Hz)  in Fig.~\ref{fig3}(a) ($g_{xy}/(2\pi)=3$ MHz), which correspond to switch off positions for the qubit-qubit coupling. The multiple switching off points can be used to optimize the quantum operation parameters and reduce the effects of adjoint qubits.
For the case of nonzero direct qubit-qubit  coupling ($g_{xy}\neq 0$),   the distribution of crossing points  forms an approximately elliptical curve in $\phi_x$-$\phi_y$ plane as shown in Fig.~\ref{fig3}(a), and this means that the contributions of resonator \textbf{b} to  induced qubit-qubit coupling (in amplitudes) is  larger than the contributions of resonator  \textbf{a}.

In the case of $g_{xy}/(2\pi)=0 $ Hz, the effective qubit-qubit can also be zero if the indirect qubit-qubit couplings induced by resonators \textbf{a} and  \textbf{b} are the same  in amplitudes but opposite in signs ($g^{(in)}_{a,xy}=-g^{(in)}_{b,xy}$).
 As shown in Fig.~\ref{fig3}(c),  the curved surface of $g^{(d)}_{xy}$ can also cross with the zero values plane ($g^{(d)}_{xy}/(2\pi)=0$ Hz) in the case of $g_{xy}/(2\pi)=0 $ Hz, and this means that the switching off can be realized  without the direct qubit-qubit interaction in the double-resonator couplers circuit. The distribution of switching off  points  approximately forms a circle in $\phi_x$-$\phi_y$ plane in Fig.~\ref{fig3}(c), which indicates the approximate equal contributions (in amplitudes) of two resonators  to the effective  qubit-qubit couplings.  The decoupled frequencies of qubits are plotted in   Fig.~\ref{fig3}(b) ($g_{xy}/(2\pi)=3$ MHz) and   Fig.~\ref{fig3}(d)($g_{xy}/(2\pi)=0 $ Hz), and the effects of direct qubit-qubit coupling to the transition frequencies of qubits seems not very large.

  The switching off positions are not totally decided by the direct qubit-qubit coupling in the double-resonator coupler circuit, thus we can take arbitrary small or even zero direct qubit-qubit coupling strength in principally,  which might be helpful to suppress the state leakages and  crosstalks on the superconducting quantum chips.
  And the restrictions on the direct qubit-qubit coupling strength and coupler's frequency   can be  unfreezed on the double-resonator couplers superconducting quantum chip). If we choose the  switching off positions close to  two-qubit gate regimes, thus the maximal frequencies of  couplers can be smaller, and this might create wider available  frequency ranges for the readout resonators and relieve the frequency crowding on the superconducting quantum chip.

\subsection{High-excited states corrections}

In current theoretical model for tunable coupler circuit, the nonlinear term $H_{nl,\beta}=(\alpha_{\beta}/2) a^{\dagger}_{\beta} a^{\dagger}_{\beta} a_{\beta} a_{\beta}$ term  is regarded as invariant during dynamical decoupling processes for qubit-resonator interactions\cite{Yan,Sung}.
 This approximation in fact neglects the effects of superconducting artificial atoms' high-excited states, so the  $g^{(d)}_{xy}$ in  Eq.~(11)  does not contain the information of anharmonicity $\alpha_{\beta}$. Because of the small anharmonicity for Xmon qubit (between 200 MHz and 400 MHz),  the interactions between the resonators and
high-excited states of superconducting artificial atoms should make corrections to the qubits'  energy levels and effective qubit-qubit coupling.

 \begin{figure}
\includegraphics[bb=0 0 460 390, width=4.25 cm, clip]{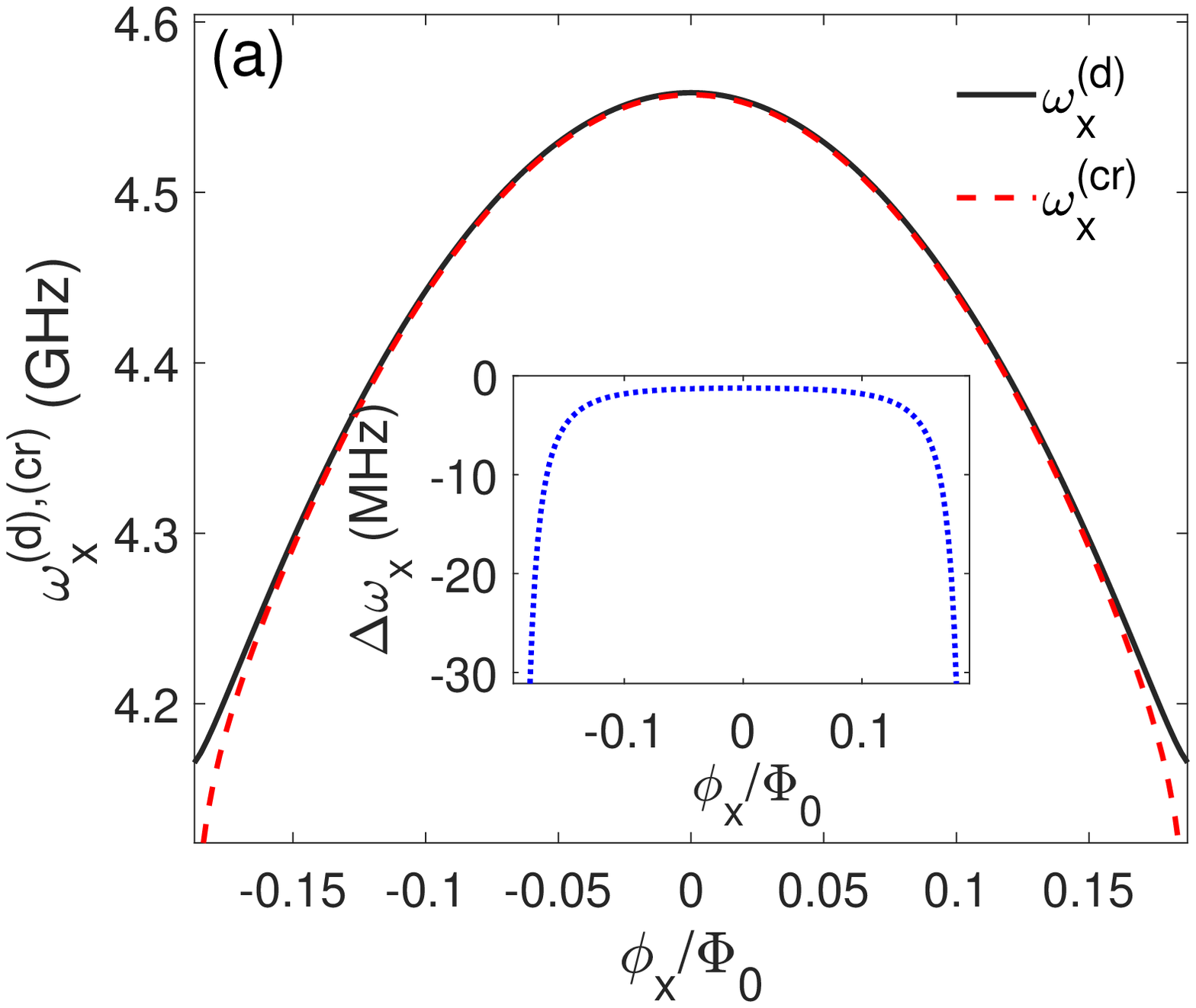}
\includegraphics[bb=0 0 460 385, width=4.27 cm, clip]{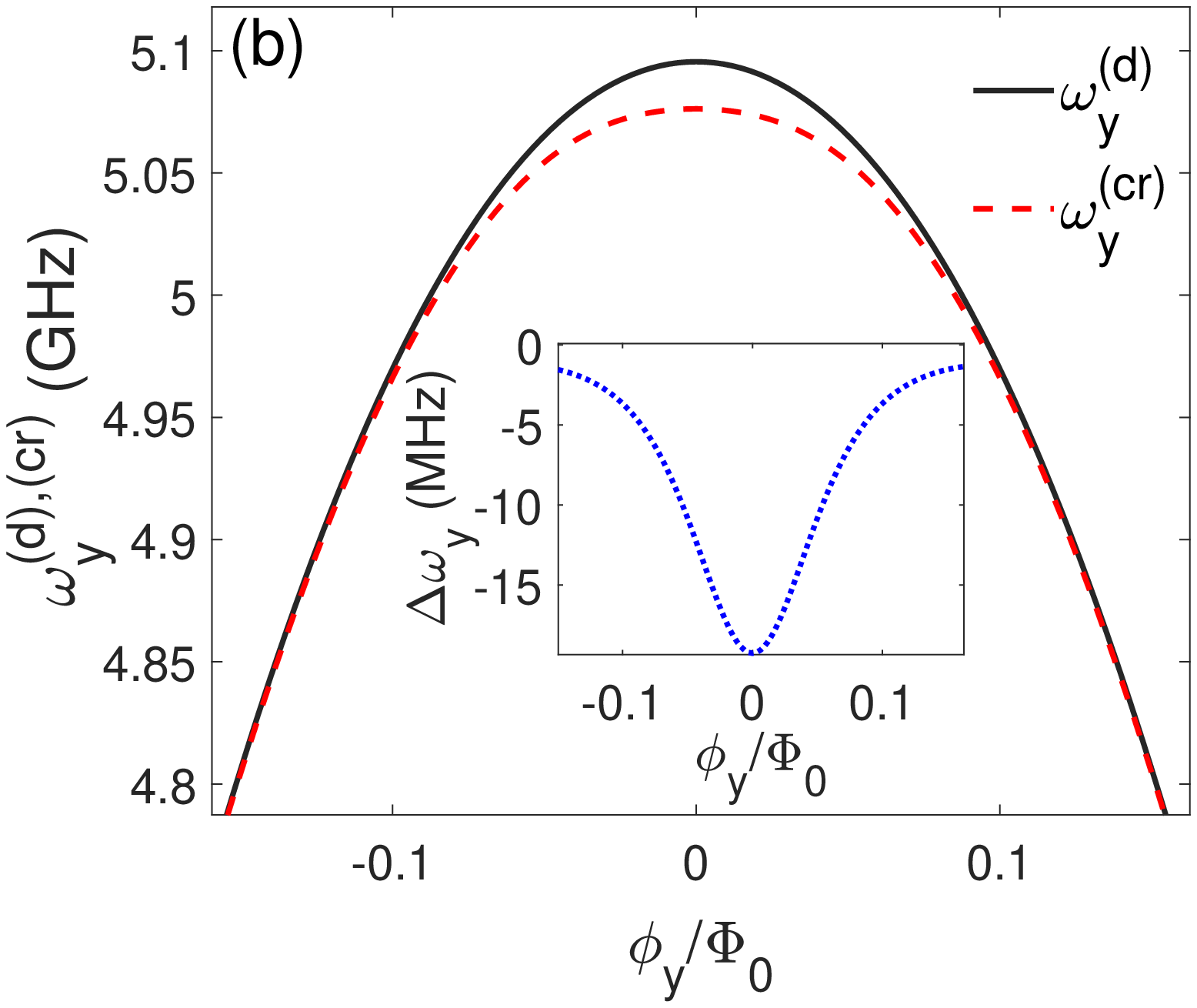}\\
\includegraphics[bb=0 0 385 300, width=4.22 cm, clip]{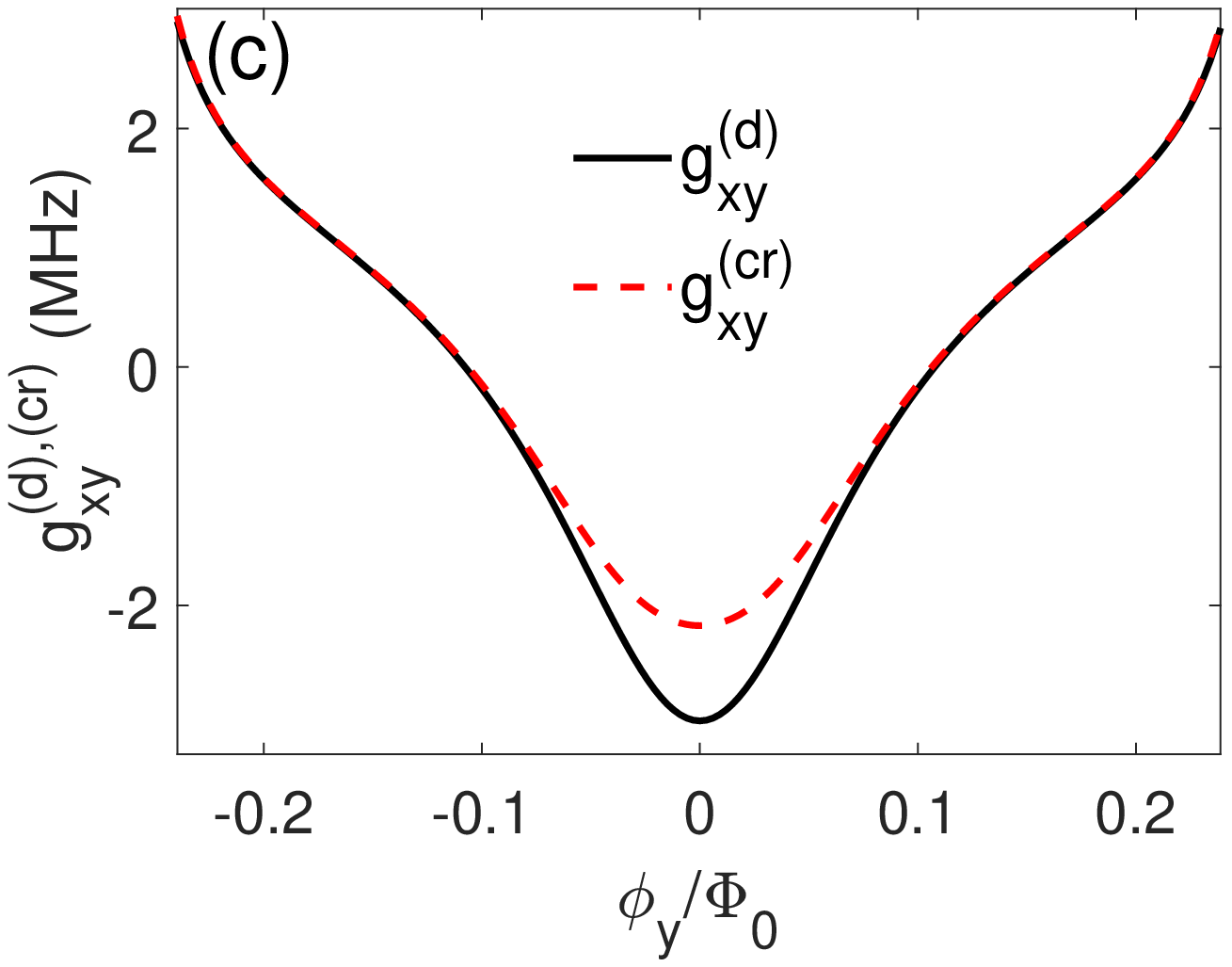}
\includegraphics[bb=0 0 390 300, width=4.2 cm, clip]{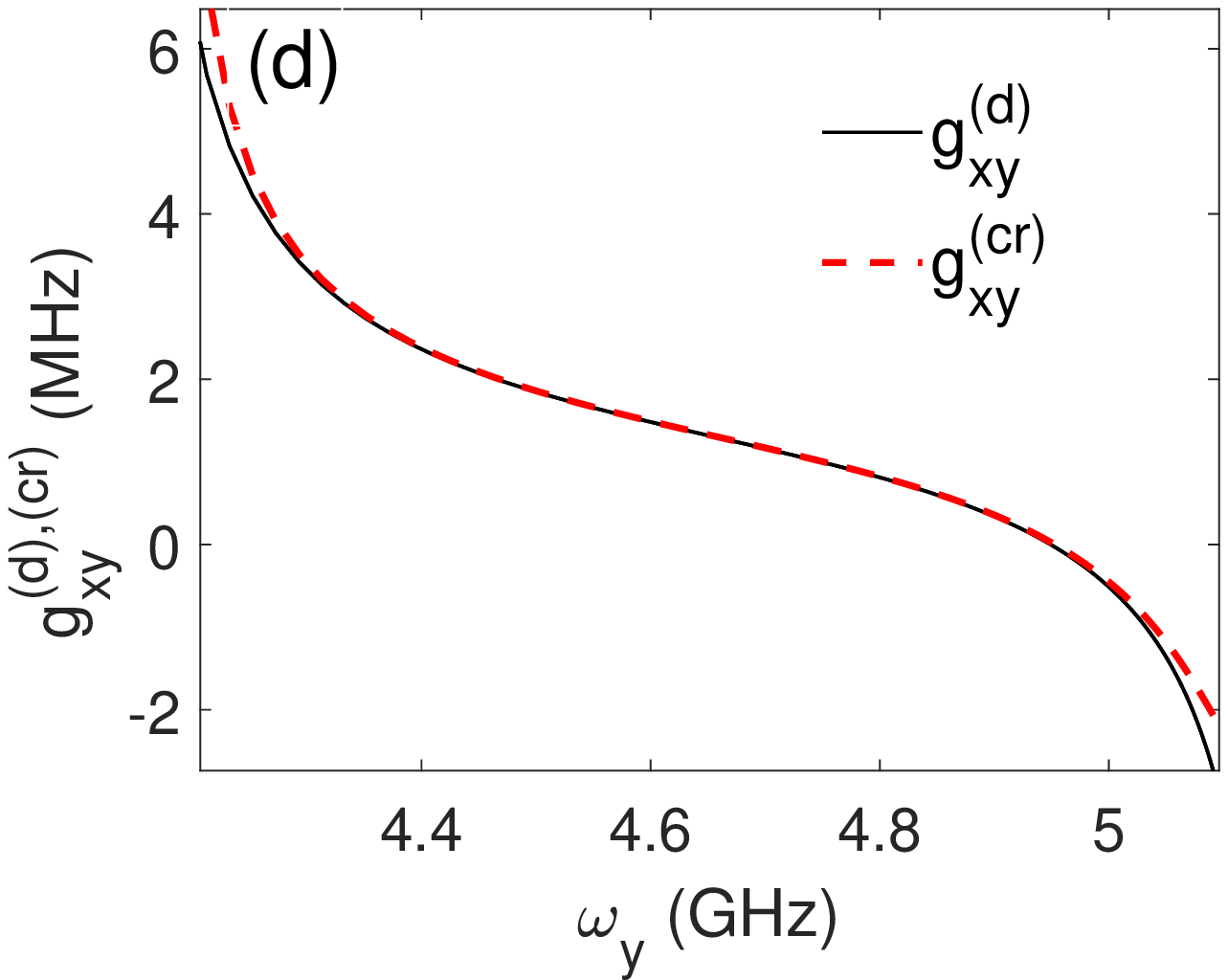}
\caption{(Color online) Corrections by the high-excited states. The  black-solid curves in (a) qubit \textbf{x} and (b) qubit \textbf{y} describe the decoupled  frequencies $\omega^{(d)}_{\beta}$, while the red-dashed curves label the corresponding corrected frequencies $\omega^{(cr)}_{\beta}(=\omega^{(d)}_{\beta}+\Delta\omega_{\beta})$,
 and the  insert figures show the frequency shifts  $\Delta\omega_{\beta} $.
 The effective qubit-qubit coupling strengths $g^{(d)}_{xy}$ (black-solid curve) and $g^{(cr)}_{xy}$ (red-dashed curve)  are plotted
 via  (c) node phase $\phi_{y}$ and (d) transition frequency $\omega_{y}$.
 Here $\omega_x/2\pi=4.56$ GHz and $g_{xy}/2\pi=1$ MHz,
 the other parameters are the same as in Fig.2.
}
\label{fig4}
\end{figure}

The  Bogoliubov transformation has been used to analyze the variations of nonlinear term $H_{nl,\beta}$ during the  decoupling processes\cite{Blais,Ferguson}, and the derived  self-kerr and cross-kerr  resonant terms under the unitary transformation reflect the contributions of superconducting artificial atoms' high-excited states.
 To maintain consistency, in this section we continue to use the  Schrieffer–Wolff transformation to calculate the effects of the nonlinear term $H_{nl,\beta}$ (see appendix \textbf{D}).
In the qubit-resonator dispersive coupling regimes, $g_{\lambda \beta}/|\Delta_{\lambda \beta}|\ll 1$ and $g_{\lambda \beta}/|\Sigma_{\lambda \beta}| \ll 1$,  we define $S_{\lambda\beta}=(g_{\lambda\beta}/\Delta_{\lambda\beta}) (c^{\dagger}_{\lambda} a_{\beta}-c_{\lambda} a^{\dagger}_{\beta})-(g_{\lambda\beta}/\Sigma_{\lambda\beta}) (c^{\dagger}_{\lambda} a^{\dagger}_{\beta}-c_{\lambda} a_{\beta})$ with  $S=\sum_{\lambda=a,b \atop \beta=x,y} S_{\lambda\beta}$.
Since $H_{nl,\beta}$ is a  small quantity, we will separately conduct the Unitary transform $H^{\prime}_{nl,\beta}=\exp(S)H_{nl,\beta}\exp(-S)$ to study its contributions to the high-order effects, such as cross-kerr resonance, self-kerr resonance, and so on\cite{Blais}.
With tedious calculations (see Appendix \textbf{D}), up to the second-order perturbation expansion terms, we get
\begin{eqnarray}\label{eq:6}
H^{\prime}_{nl,\beta} &\approx &\sum_{\lambda=a,b}\left(\frac{g^2_{\lambda \beta}\alpha_{\beta}}{\Sigma^2_{\lambda \beta}}
-\frac{g^2_{\lambda \beta}\alpha_{\beta}}{\Delta^2_{\lambda \beta}}\right) a^{\dagger}_{\beta}a^{\dagger}_{\beta}a_{\beta} a_{\beta}\\
&+&\sum_{\lambda=a,b}\left[\frac{2 g^2_{\lambda x}\alpha_{\beta}}{\Delta^2_{\lambda \beta}} c^{\dagger}_{\lambda}c_{\lambda}a^{\dagger}_{\beta}a_{\beta}+\frac{2 g^2_{\lambda \beta}\alpha_{\beta}}{\Sigma^2_{\lambda \beta}} c_{\lambda}c^{\dagger}_{\lambda} a^{\dagger}_{\beta}a_{\beta} \right].\nonumber
\end{eqnarray}
 Since $g_{xy}, g_{ab}\ll g_{\lambda\beta}$, the indirect interaction induced by the weakly direct qubit-qubit and resonator-resonator interactions have been neglected.
 The first and second lines in the right side of Eq.(12) respectively describe the self-kerr  and cross-kerr resonance terms,  and the complete calculation results can be seen in Appendix \textbf{D}.
There are no external pump fields for resonator couplers, so the cavity photon numbers should be very small ($n_{\lambda}=a^{\dagger}_{\lambda}a_{\lambda}\ll 1$). Thus we can get the approximate frequency shift for qubit ${\beta}$  induced by the nonlinear terms,
\begin{eqnarray}\label{eq:7}
\Delta\omega_{\beta}=\sum_{\lambda=a,b}\left(\frac{g^2_{\lambda \beta}}{\Delta^2_{\lambda \beta}}+\frac{g^2_{\lambda \beta}}{\Sigma^2_{\lambda \beta}}\right)\alpha_{\beta}.
\end{eqnarray}
We can see that the frequency  shift $\Delta\omega_{\beta}$ for qubit $\beta$ is proportional to the qubit's anharmonicity $\alpha_{\beta}$, which reflects the effects of the second-excited state of superconducting artificial atoms. Adding the frequency shift induced by the nonlinear term $H_{nl,\beta}$,
we can approximately get the corrected  frequency $\omega^{(cr)}_{\beta}$ of qubit $\beta$ in decoupled coordinate frame
\begin{eqnarray}\label{eq:8}
 \omega^{(cr)}_{\beta}&=&\omega_{\beta}+\sum_{\lambda=a,b}\left(\frac{g^2_{\lambda \beta}}{\Delta_{\lambda \beta}}-\frac{g^2_{\lambda \beta}}{\Sigma_{\lambda \beta}}\right)\nonumber\\
 & &+\sum_{\lambda=a,b}\left(\frac{g^2_{\lambda \beta}}{\Delta^2_{\lambda \beta}}+\frac{g^2_{\lambda \beta}}{\Sigma^2_{\lambda \beta}}\right)\alpha_{\beta}.\nonumber\\
\end{eqnarray}

The transition frequency of qubit \textbf{x} in decoupled coordinate is plotted in Fig.~\ref{fig4}(a), the deviation between $\omega^{(cr)}_{x}$ and $\omega^{(d)}_{x}$
 is larger at the regimes far from the zero magnetic flux points, which coincides with the curve of $\Delta\omega_{x}$ in insert figure.
  On the contrary,  the  maximal deviation between  $\omega^{(cr)}_{y}$ and $\omega^{(d)}_{y}$  appears at the regime close to the zero magnetic flux ($\Phi_{e,y} \rightarrow 0$ or $\phi_y \rightarrow 0$) in Fig.~\ref{fig4}(b), this also coincides with the curve of $\Delta\omega_{y}$ in the insert figure.

If we replace the $\omega^{(d)}_{\beta}$ with $\omega^{(cr)}_{\beta}$ in Eq.(11), we can get the corrected  effective qubit-qubit coupling $g^{(cr)}_{xy}$.
The resonators' resonant frequencies $\omega_a$ and $\omega_b$ are  fixed in this article, so the effective qubit-qubit coupling  are mainly tuned by the qubits' transition frequencies $\omega_x$ and $\omega_y$. By setting $\omega_x/(2\pi)=4.56$ GHz,
we plot the curves of effective qubit-qubit coupling $g^{(cr)}_{xy}$ and $g^{(d)}_{xy}$  on the respective $\phi_{y}$ and  $\omega_{y}$ in Figs.~\ref{fig4}(c) and ~\ref{fig4}(d), the points satisfying $g^{(cr)}_{xy}/(2\pi)=0$ Hz or $g^{(d)}_{xy}/(2\pi)=0$ Hz  corresponds to the switching off position for the qubit-qubit interaction.  The zero value points of  $g^{(cr)}_{xy}$ and $g^{(cr)}_{xy}$ are different, which reflects the effects of the nonlinear term $H_{nl}$ ( also the second-excited state of superconducting artificial atom) on the switching off positions.  The calculation results of the corrections to qubits' frequencies and effective qubit-qubit coupling strength can help to accurately design the superconducting quantum chip.

\subsection{Switching off the qubit-qubit coupling }

\begin{figure}
\includegraphics[bb=10 0 550 450, width=4.3 cm, clip]{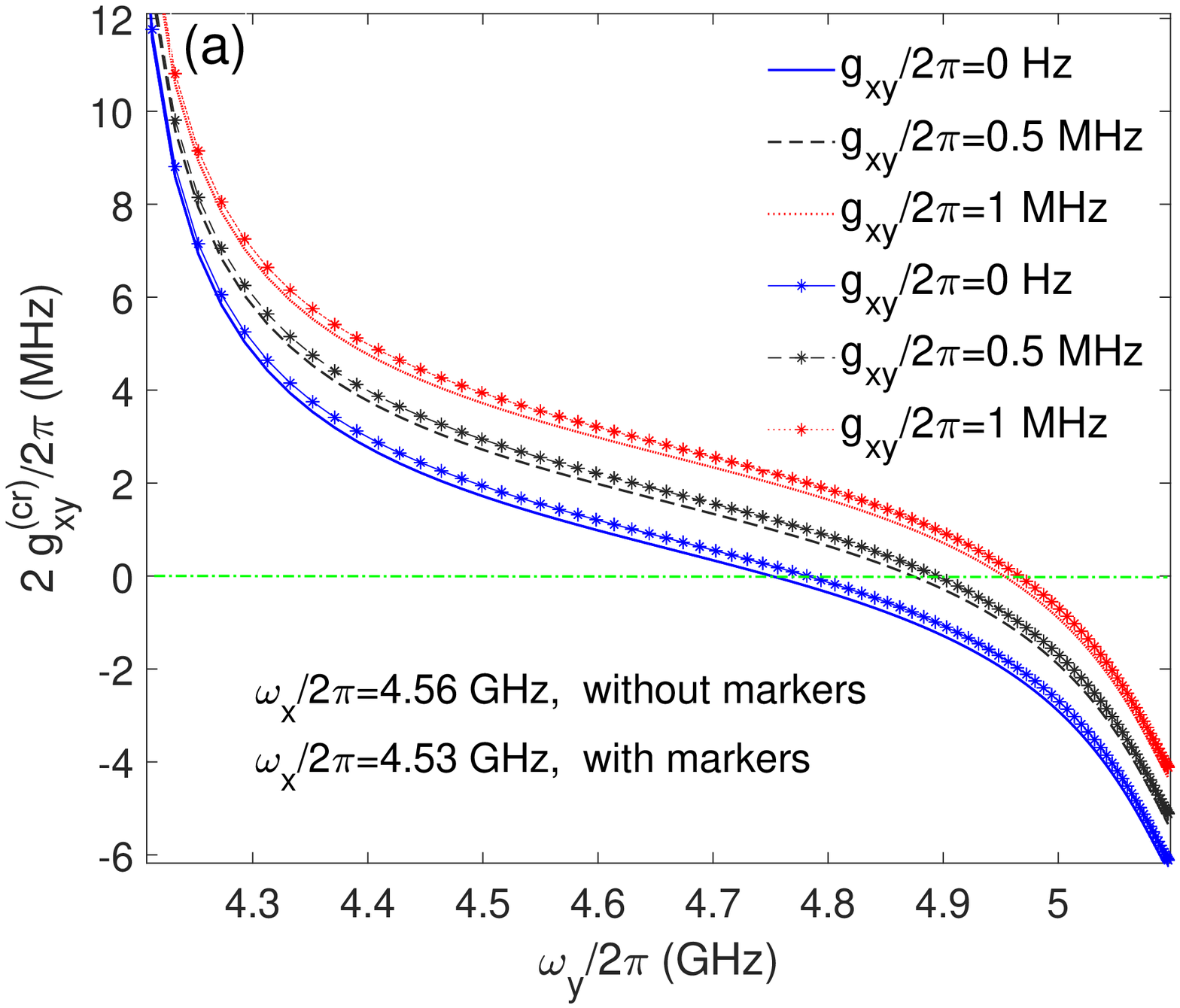}
\includegraphics[bb=10 0 505 425, width=4.25 cm, clip]{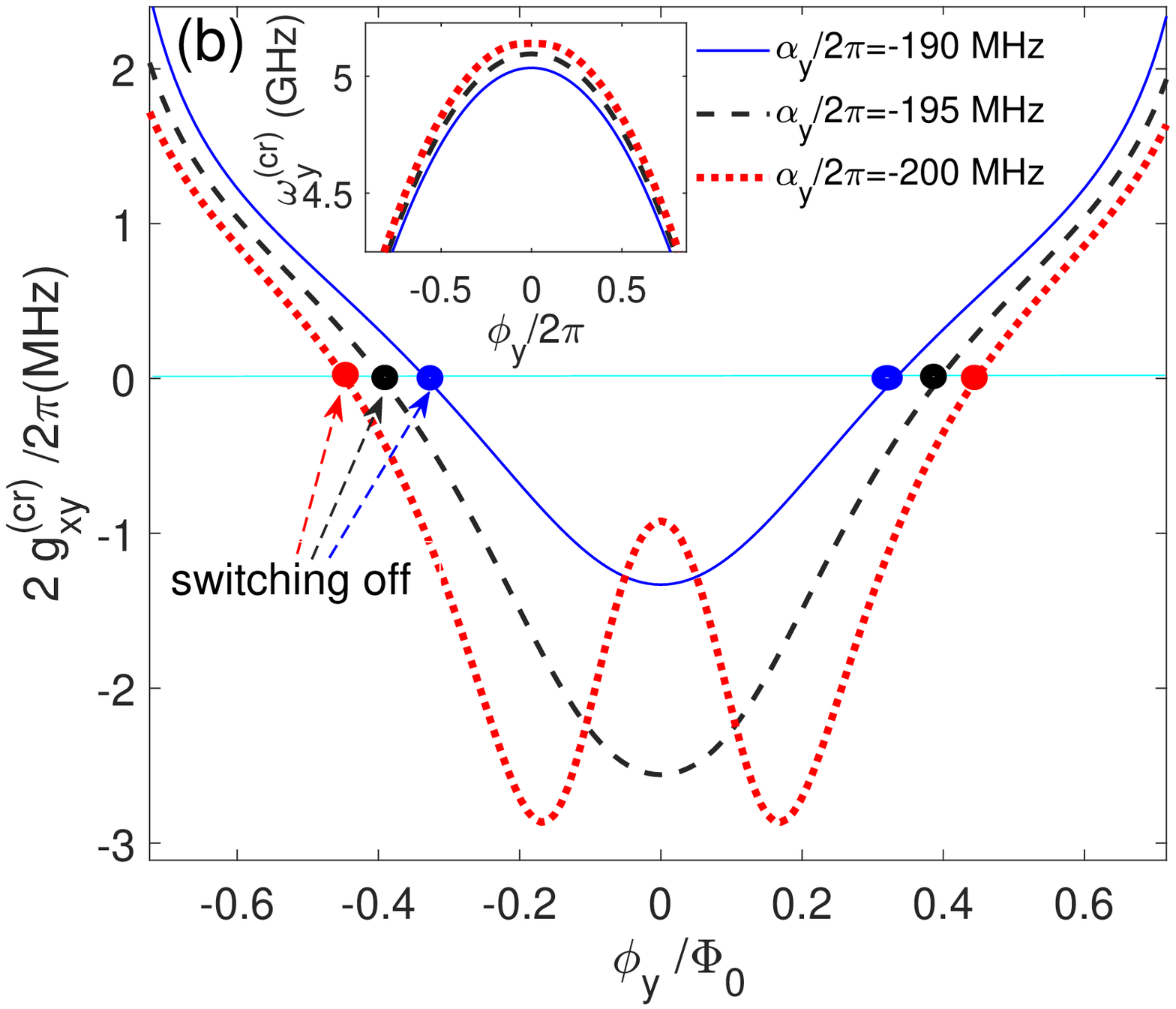}
\caption{(Color online) Switching off the interaction between two qubits.
The effects of  (a) direct qubit-qubit coupling strengths and (b) qubits' anharmoncities on
   the switching off positions. (a) The three types of line styles correspond to different direct qubit-qubit coupling strengths: (1)  $g_{xy}/2\pi= 0$ Hz for blue-solid curves;
  (2) $g_{xy}/2\pi= 0.5$ MHz for black-dashed curves;
 (3) $g_{xy}/2\pi= 1$ MHz red-dotted curves.  The $\omega_x/2\pi=4.56$ GHz for the  curves without markers,
    while $\omega_x/2\pi=4.53$ GHz  for the curves with star markers.
(b) The three curves correspond to different anharmonicities:
  (1)  $\alpha_{y}/2\pi=-190$ MHz for the blue-solid curve; (2) $\alpha_{y}/2\pi=-195$ MHz for the black-dashed curve;
(3) $\alpha_{y}/2\pi=-200$ MHz for the red-dotted curve.  The variation of $\omega_y$  on the node phase $\phi_y$ is shown in the insert figure.
 We choose $\omega_x/2\pi=4.56$ GHz and $g_{xy}/2\pi= 0.5$ MHz in (b). The other parameters of (a) and (b) are the same as in Fig.2.}
\label{fig5}
\end{figure}

 In this section, we study the effects of direct qubit-qubit coupling $g_{xy}$ and  qubit's anharmonicitiy $\alpha_{\beta}$ on the effective qubit-qubit coupling $g^{(cr)}_{xy}$ and the switching off position. If we take the parameters of Fig.~\ref{fig2}, we can get $g_{by}/\min{(|\Delta_{by}|)}\sim 1/3$ and $g_{ax}/\min{(|\Delta_{ax}|)}\sim 1/15$  in the  idling states of qubits (without external magnetic field), so the qubits and resonators are in the dispersive or weak-dispersive coupling regimes.  To see clearer the  working mechanism of switching-off processes in the double-resonator couplers circuit, we plot the one-dimensional curves of effective qubit-qubit  coupling $g^{(cr)}_{xy}$ with the variation of qubit transition frequency $\omega_y$ in Fig.~\ref{fig5}(a). By fixing $\omega_x/(2\pi)=4.56$ GHz, the three curves without markers in  Fig.~\ref{fig5}(a)  correspond to different  direct qubit-qubit coupling strengths: $g_{xy}/(2\pi)=0$ Hz in the blue-solid curve, $g_{xy}/(2\pi)=0.5$ MHz in black-dashed curve,  and $g_{xy}/(2\pi)=1$ MHz in the red-dotted curve. The crossing points of  three curves with zero value line ( $g^{(cr)}_{xy}/(2\pi)=0$ Hz)  are  different, this means that the direct qubit-qubit coupling could affect the switching off positions. If we  set $\omega_x/(2\pi)=4.53$ GHz, the switching off points in the three marked curves show  considerable  shifts  relative to the corresponding  same color curves without markers.

In Fig.~\ref{fig5}(b), we plot the curves of effective qubit-qubit coupling $g^{(cr)}_{xy}$ on the node phase $\phi_y$ with $\omega_x/(2\pi)=4.56$ GHz. For the same   ranges of node phase $\phi_{y}$, the qubit's maximal frequencies $\omega^{(max)}_y$  are not the same for  different  anharmonicities  $\alpha_{y}$ as shown in the insert figure. For  different anhamonicities $\alpha_y$, the shifts of switching off positions on  three curves  reflect the effects of  superconducting artificial atom's second-excited states.

The frequency of qubit \textbf{y} should be  tuned close to  $\omega_y\approx \omega_x $ for the iSWAP  gate and $\omega_y+\alpha_y \approx \omega_x $ for the Controlled-Z gate.
For the single-path coupler circuit,  the switching off point is usually close to the idling coupler frequency (usually about 6.0 GHz) which is  far from the two-qubit gate regimes (usually below 5.0 GHz). In the double-resonator couplers circuit,  the switching off positions can be  very close to the two-qubit gate regimes as shown in Fig.~\ref{fig5}(a). So  the maximal frequencies of  couplers can be smaller in the double-resonator couplers superconducting circuit, this leaves wider available ranges for readout resonators (or qubits) and might relieve  the frequency crowding on superconducting quantum chip.

\section{Static ZZ coupling}

The tunable coupler  could isolate the qubits from surrounding environments for local quantum operations and reduce the accumulated phases for the quantum state preparations, and  this can greatly enhance the fidelity of two-qubit gate\cite{Tan1,Kandala,Zajac,Moskalenko,Martinis,Wu,Sirui,Shen}.
Because of the small anharmonicity of the Xmon qubit and  the high-order quantum state exchanges (originating from the  qubit-qubit and qubit-coupler interactions), the quantum state leakages and the Parasitic crosstalks are still important obstructions for the further enhancement the fidelity of two-qubit gate\cite{Sung}.  Suppressing the residual coupling and the Parasitic crosstalks among neighbour qubits are the leading tasks for enhancing the quality of  superconducting quantum chip\cite{wang,Hazard,Xue,Santos,Xin}.

The residual ZZ coupling  consists the  static type ZZ coupling and the dynamic type ZZ coupling, but the  dynamic ZZ coupling is usually suppressed by optimizing  the microwave pulse shapes and not the interest of this article\cite{Sung,Santos}. In the section, we mainly focus on the  Static ZZ coupling which can be mitigated by the designing structures and working parameters of qubits and tunable couplers\cite{AHouck,wang,Sete,Goto,Hazard,Xue,Kandala,Santos,Xin,Jaseung,Marxer,Leroux}.  In the double-coupler superconducting quantum circuit, the direct qubit-qubit coupling can be arbitrarily small in principally, and this should be helpful for suppressing the static ZZ coupling. And the destructively interferences between double-path couplers might eliminate the static ZZ coupling \cite{AHouck,Goto,Kandala,Sete}.

\subsection{Analytic calculations}

In Figs.~\ref{fig2}(a)-\ref{fig2}(b), we have numerically calculated the  energy level curves of states $|0100\rangle$, $|0010\rangle$, and $|0110\rangle$,  in principally  the static ZZ coupling can be easily calculated through the definition $\xi_{ZZ}=\omega_{|0110\rangle}-\omega_{|0100\rangle}-\omega_{|0010\rangle}+\omega_{|0000\rangle}$.
In practically, it is difficult  to  accurately fit the energy level curves of qubits because the avoided crossing gaps are affected by  multi-body interactions.
By setting $\omega_x/(2\pi)=4.56$ GHz, if we tune the frequency of qubit \textbf{y} to be near resonant with qubit \textbf{x}  ($\omega_y\approx\omega_x$), thus we can get
 $g_{by}/|\Delta_{by}|\approx g_{bx}/|\Delta_{bx}|\sim 1/20$ and $g_{ax}/|\Delta_{ax}|\approx g_{ay}/|\Delta_{ay}|\sim 1/15$. Thus the qubit-resonator are dispersive coupling regimes, and the perturbation method can be used to analyze  the static ZZ coupling close to the two-qubit gate regimes.

For convenience and  consistency, we still use  $\omega_{\beta}$ to describe the energy levels of  first-excited state of qubit $\beta$ in this section,  and  the energy level for second-excited state is $2\omega_{\beta}+\alpha_{\beta}$, where the $\alpha_{\beta}$ is the qubit's anharmoncity .
If we temporarily disregard the weak direct qubit-qubit and direct resonator-resonator interactions,  up to the fourth-order perturbation theory the effective Hamiltonian on the qubits' eigenstates space can be obtained as  \cite{Blais,Ferguson}

 \begin{eqnarray}\label{eq:9}
& &H^{\prime}_{m}/\hbar
= \sum_{\lambda=a,b}\omega_{\lambda}c^{\dagger}_{\lambda} c_{\lambda}\nonumber\\
& &+ \sum_{\beta=x,y}\sum_{\lambda=a,b \atop j_{\beta}=0,1,2,...}(\omega_{j_{\beta}}+\kappa_{\lambda,j_{\beta}}+\chi_{\lambda,j_{\beta}}c^{\dagger}_{\lambda}c_{\lambda})|j_{\beta}\rangle\langle j_{\beta}|\nonumber\\
& &+ \sum_{\beta=x,y }\sum_{ j_{\beta}=0,1,2,...}\bigg[\sum_{\lambda=a,b}\mu_{\lambda,j_{\beta}} (c^{\dagger}_{\lambda}c_{\lambda})^2\nonumber\\
& &+\nu_{ab,j_{\beta}} c^{\dagger}_{a}c_{a}c^{\dagger}_{b}c_{b}+ \nu_{ba,j_{\beta}} c^{\dagger}_{b}c_{b}c^{\dagger}_{a}c_{a}\bigg]|j_{\beta}\rangle\langle j_{\beta}|,
\end{eqnarray}

 The ket vector $|j_{\beta}\rangle$ describes the $j_{\beta}$-th excited state of qubit $\beta$, with $j_{\beta},j^{\prime}_{\beta}=0,1,2,...$. We define $g^{j_{\beta}j^{\prime}_{\beta}}_{\lambda}$ as the coupling strength between resonator $\lambda$ and the transition $|j_{\beta}\rangle\leftrightarrow |j^{\prime}_{\beta}\rangle$ of qubit $\beta$. Considering the selection rule, the resonator $\lambda$ can only interact with the neighbour quantum states of qubit $\beta$: $g^{j_{\beta}j^{\prime}_{\beta}}_{\lambda}=0$ for $j^{\prime}_{\beta}\neq j_{\beta}\pm 1$.
In this section, we neglect the small differences for the coupling strengths  between resonator $\lambda$  and different neighbour state transitions of qubit $\beta$ , then   $g^{j_{\beta}-1,j_{\beta}}_{\lambda}=g^{j_{\beta},j_{\beta}+1}_{\lambda}=g_{\lambda\beta}$ ( $j_{\beta}=1,2,...$).
 Defines $\chi^{j_{\beta}-1,j_{\beta}}_{\lambda}=j_{\beta}g_{\lambda\beta}/[\Delta_{\lambda\beta}+(j_{\beta}-1)\alpha_{\beta}]$,
 the $\kappa_{\lambda,j_{\beta}}= \chi^{j_{\beta}-1,j_{\beta}}_{\lambda}$ describes the level shifts of Lamb type for the quantum state $|j_{\beta}\rangle$  which is induced by the interaction between resonator $\lambda$ and qubit $\beta$ ( $|j_{\beta}-1\rangle\leftrightarrow |j_{\beta}\rangle$ and  $|j_{\beta}\rangle\leftrightarrow |j_{\beta}+1\rangle$), while $\chi_{\lambda,j_{\beta}}=\chi^{j_{\beta}-1,j_{\beta}}_{\lambda}-\chi^{j_{\beta},j_{\beta}+1}_{\lambda}$  describes the corresponding ac-stark type dispersive shifts for the quantum state $|j_{\beta}\rangle$ ($\chi_{\lambda,0_{\beta}}=-\chi^{0_{\beta},0_{\beta}+1}=-g^{2}_{\lambda\beta}/(2\Delta_{\lambda\beta})$)\cite{Ferguson,Blais}. If we  add the contributions of  second-excited states of superconducting artificial atoms, besides  the self-kerr resonant term  $\mu_{\lambda,j_{\beta}} (c^{\dagger}_{\lambda}c_{\lambda})^2$, the cross-kerr resonant terms $\nu_{ab,j_{\beta}} c^{\dagger}_{a}c_{a}c^{\dagger}_{b}c_{b}$  and $\nu_{ba,j_{\beta}} c^{\dagger}_{b}c_{b}c^{\dagger}_{a}c_{a}$ should also make contributions to the static ZZ coupling(as will be discussed in  follow sections)\cite{Ferguson,Blais,Pople}.

If we temporarily disregard the cross-kerr resonant terms, after adding the contributions of  weak direct qubit-qubit coupling, and then the static ZZ coupling in qubit-resonator dispersive coupling regimes can be obtained as \cite{Sun,wang,Sung,Sete},
\begin{eqnarray}\label{eq:10}
\xi^{(2)}_{ZZ}&=&\frac{2 (g_{xy})^2(\alpha_x+\alpha_y)}{(\Delta_{xy}+\alpha_y)(\Delta_{xy}-\alpha_x)},\\
\xi^{(3)}_{ZZ,\lambda}&=&2 g_{xy}g_{\lambda x}g_{\lambda y}\bigg[\frac{1}{\Delta_{\lambda y}}\left(\frac{1}{\Delta_{xy}}-\frac{2}{\Delta_{xy}+\alpha_y}\right)\nonumber\\
&- &\frac{1}{\Delta_{\lambda y}}\left(\frac{1}{\Delta_{xy}}-\frac{2}{\Delta_{xy}-\alpha_x}\right)\bigg],\\
\xi^{(4s)}_{ZZ,\lambda}&=&\frac{2(g_{\lambda y})^2(g_{\lambda x})^2}{\Delta_{\lambda y}+\Delta_{\lambda x}-\alpha_{\lambda}}\left(\frac{1}{\Delta_{\lambda y}}+\frac{1}{\Delta_{\lambda x}}\right)^2\nonumber\\
      &- &\frac{(g_{\lambda y})^2 (g_{\lambda x})^2}{\Delta^2_{\lambda y}} \left(\frac{1}{\Delta_{xy}}+\frac{1}{\Delta_{\lambda x}}-\frac{2}{\Delta_{xy}-\alpha_x}\right)\nonumber\\
      &- &\frac{(g_{\lambda y})^2(g_{\lambda x})^2}{\Delta^2_{\lambda x}} \left(\frac{2}{\Delta_{xy} +\alpha_y}-\frac{1}{\Delta_{xy}}+\frac{1}{\Delta_{\lambda y}}\right).\qquad
\end{eqnarray}
 The  $\xi^{(2)}_{ZZ}$ is the second-order static ZZ coupling between two qubits, and $\xi^{(3)}_{ZZ,\lambda}$ describes the third-order static ZZ coupling between two qubits intermediated  by the resonator $\lambda$.
We use $\xi^{(4s)}_{ZZ,\lambda}$ to label fourth-order  static ZZ coupling contributed by the self-kerr resonance  intermediated  by the resonator $\lambda$, and the $\xi^{(4c)}_{ZZ,\lambda}$ describe the static ZZ coupling induced by the cross-kerr resonance.

Even being listed together, the second-order  $\xi^{(2)}_{ZZ}$, third-order $\xi^{(3)}_{ZZ}=\sum_{\lambda=a,b}\xi^{(3)}_{ZZ,\lambda}$, and fourth-order (self-kerr) $\xi^{(4s)}_{ZZ}=\sum_{\lambda=a,b}\xi^{(4s)}_{ZZ,\lambda}$ static ZZ coupling terms come from different sources.  As shown in  Eqs.(16) and (17), the second-order term $\xi^{(2)}_{ZZ}$ originates from the  direct qubit-qubit coupling \cite{Chen,Yan,Sun,wang},
and the fourth-order term $\xi^{(4s)}_{ZZ}$ results from the perturbation expansion of qubit-resonator dispersive  coupling\cite{Ferguson,Pople}, while the third-order term $\xi^{(3)}_{ZZ}$ is joint effects of  direct qubit-qubit coupling and qubit-resonator interaction.
Here $\alpha_{\lambda}=\sum_{\beta=x,y} \alpha_{\beta}(g_{\lambda \beta}/\Delta_{\lambda \beta})^4$ describes the  nonlinearity of resonator $\lambda$ induced by the qubit-resonator  dispersive coupling\cite{Blais}.
 Since $g_{xy}, g_{ab}\ll g_{\lambda \beta}$, the  contributions of direct qubit-qubit and resonator-resonator couplings to fourth-order static ZZ coupling are neglected.

\begin{figure}
\includegraphics[bb=0 0 460 430, width=4.25cm, clip]{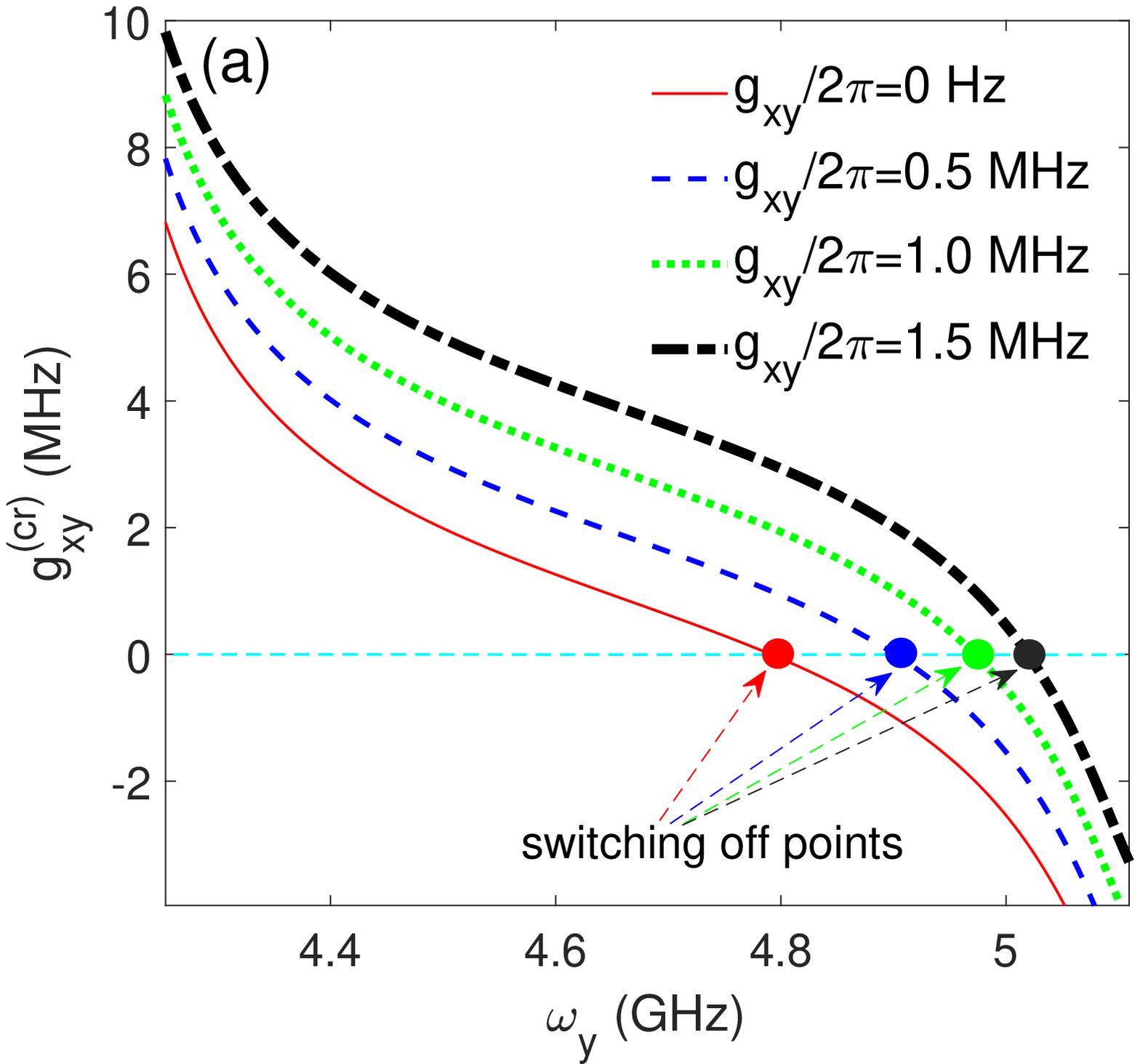}
\includegraphics[bb=-5 0 465 430, width=4.25cm, clip]{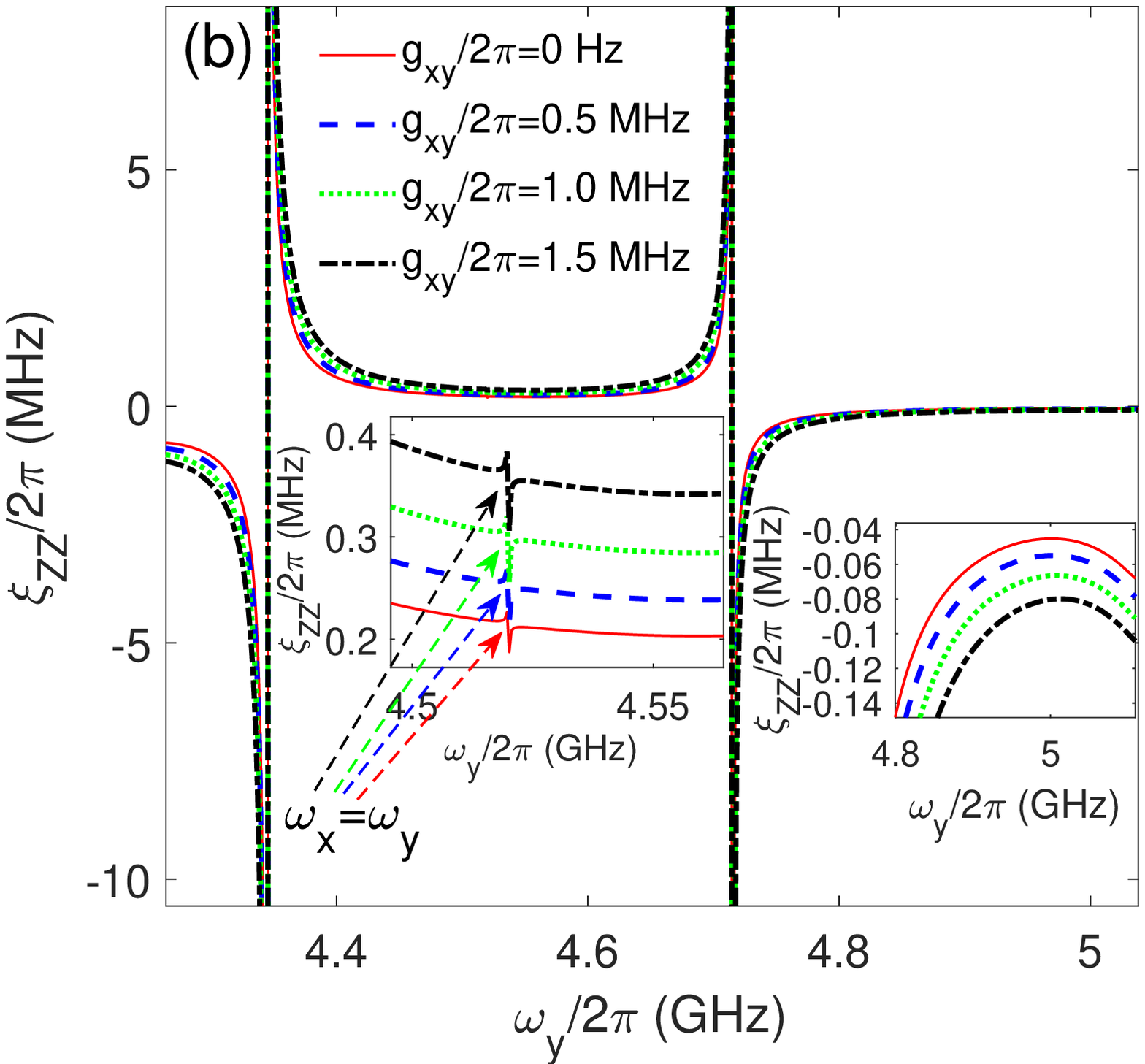}\\
\includegraphics[bb=0 0 465 430, width=4.25cm, clip]{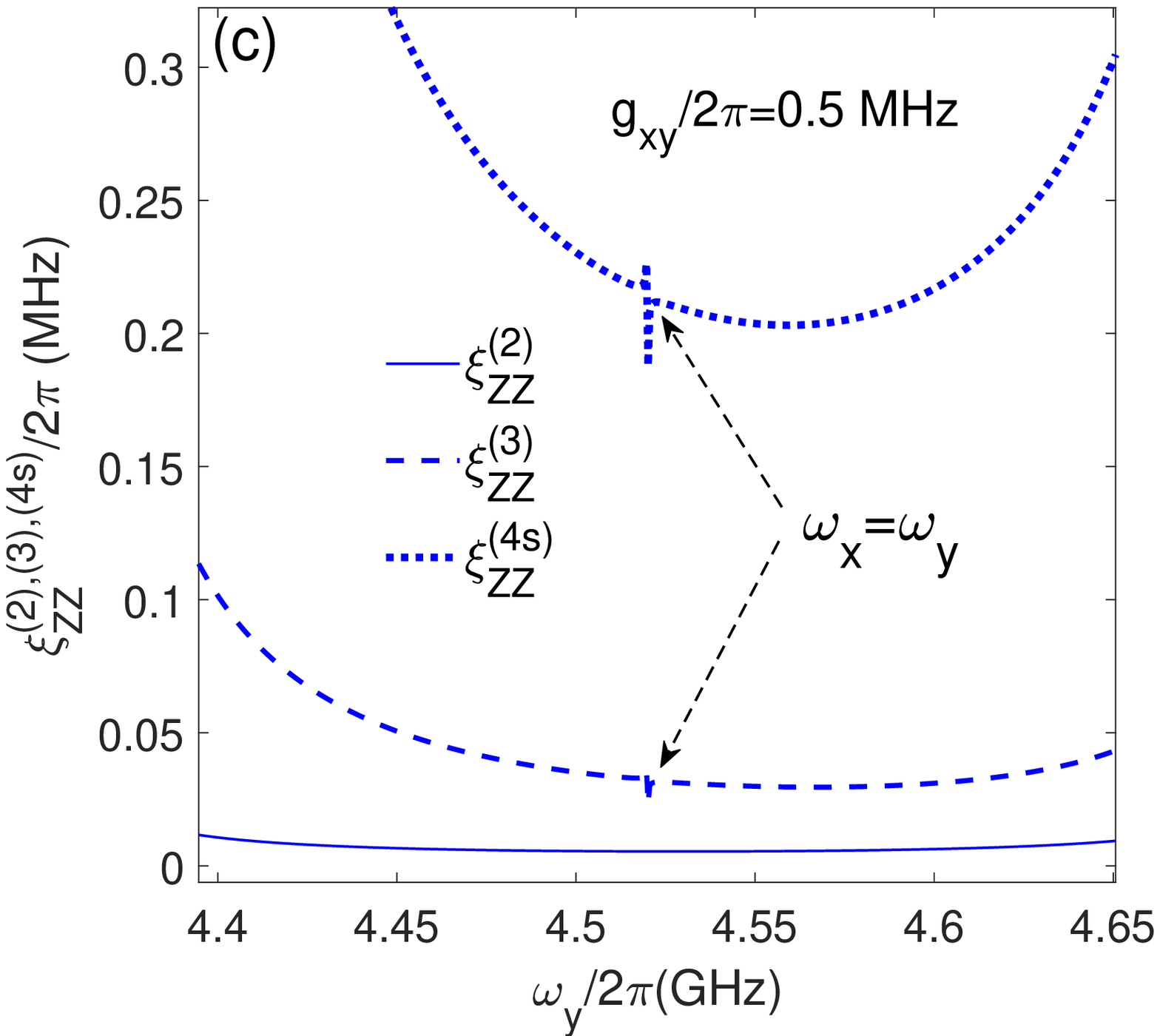}
\includegraphics[bb=0 0 460 430, width=4.25cm, clip]{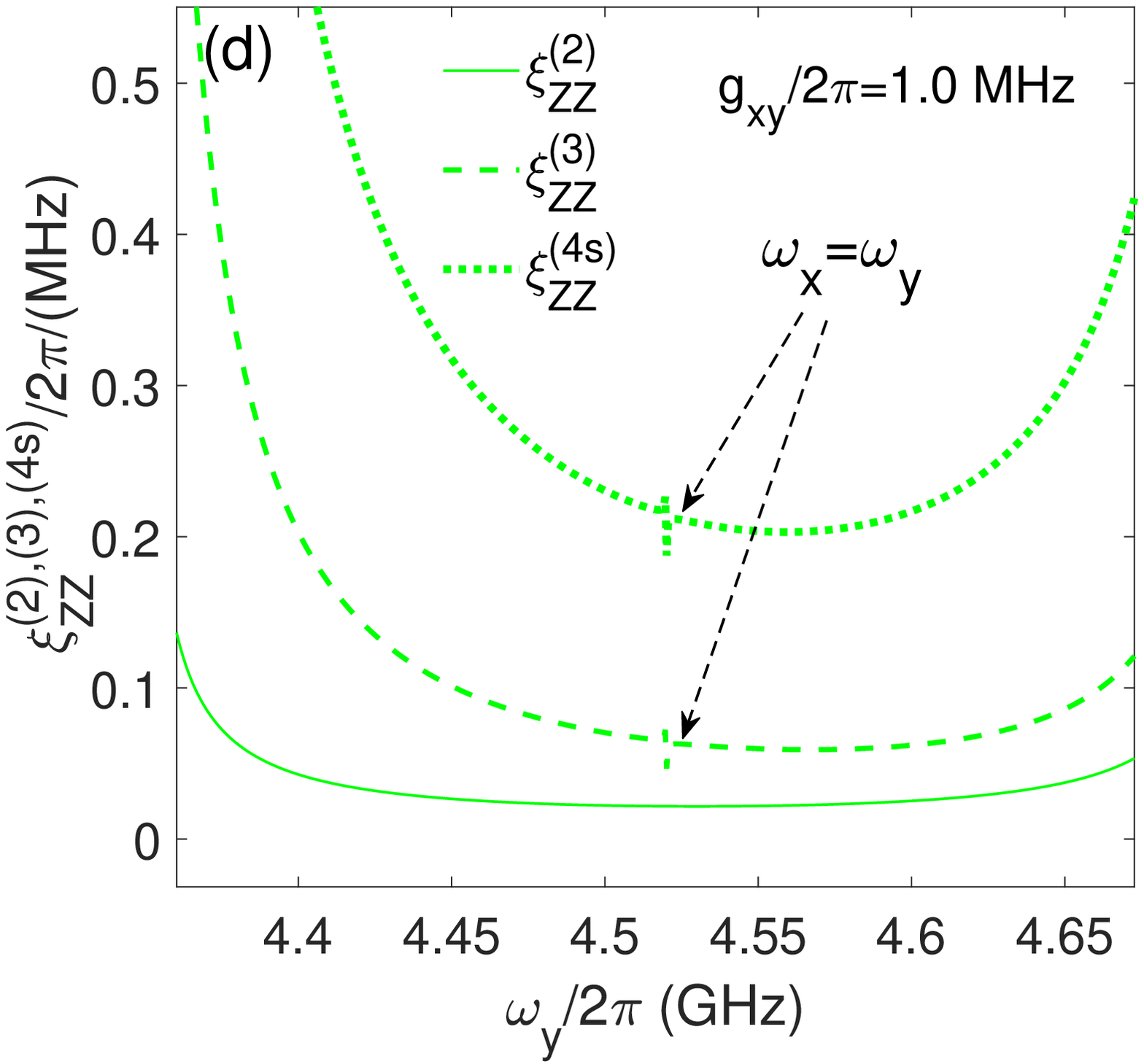}
\caption{(Color online) Suppression of the static ZZ coupling.
The  (a) effective qubit-qubit coupling  and (b) static ZZ
 coupling are plotted for different direct qubit-qubit coupling strengths:
 (1) $g_{xy}/2\pi=0 $ Hz (red-solid curve); (2) $g_{xy}/2\pi=0.5 $
  MHz (blue-dashed curve); (3) $g_{xy}/2\pi=1$ MHz (green-dotted curve);
(4) $g_{xy}/2\pi=1.5$ MHz (black-dashed-dot curve). The  insert figures are the partial enlarged drawing.
The  $\xi^{(2)}_{ZZ}$, $\xi^{(3)}_{ZZ}(=\sum_{\lambda=a,b}\xi^{(3)}_{ZZ,\lambda})$
  and $\xi^{(4s)}_{ZZ}(=\sum_{\lambda=a,b}\xi^{(4s)}_{ZZ,\lambda})$ are  plotted in (c) and (d),
they respectively take the parameters of blue-dashed and black dashed-dot curves in (b). The frequency of qubit \textbf{x} is fixed at  $\omega_x/2\pi=4.52$ GHz,
and the other parameters of (a)-(d) are the same as in Fig.2.
}
\label{fig6}
\end{figure}

\subsection{Suppression of the Static ZZ coupling}

\begin{figure}
\includegraphics[bb=-20 0 570 460, width=4.3 cm, clip]{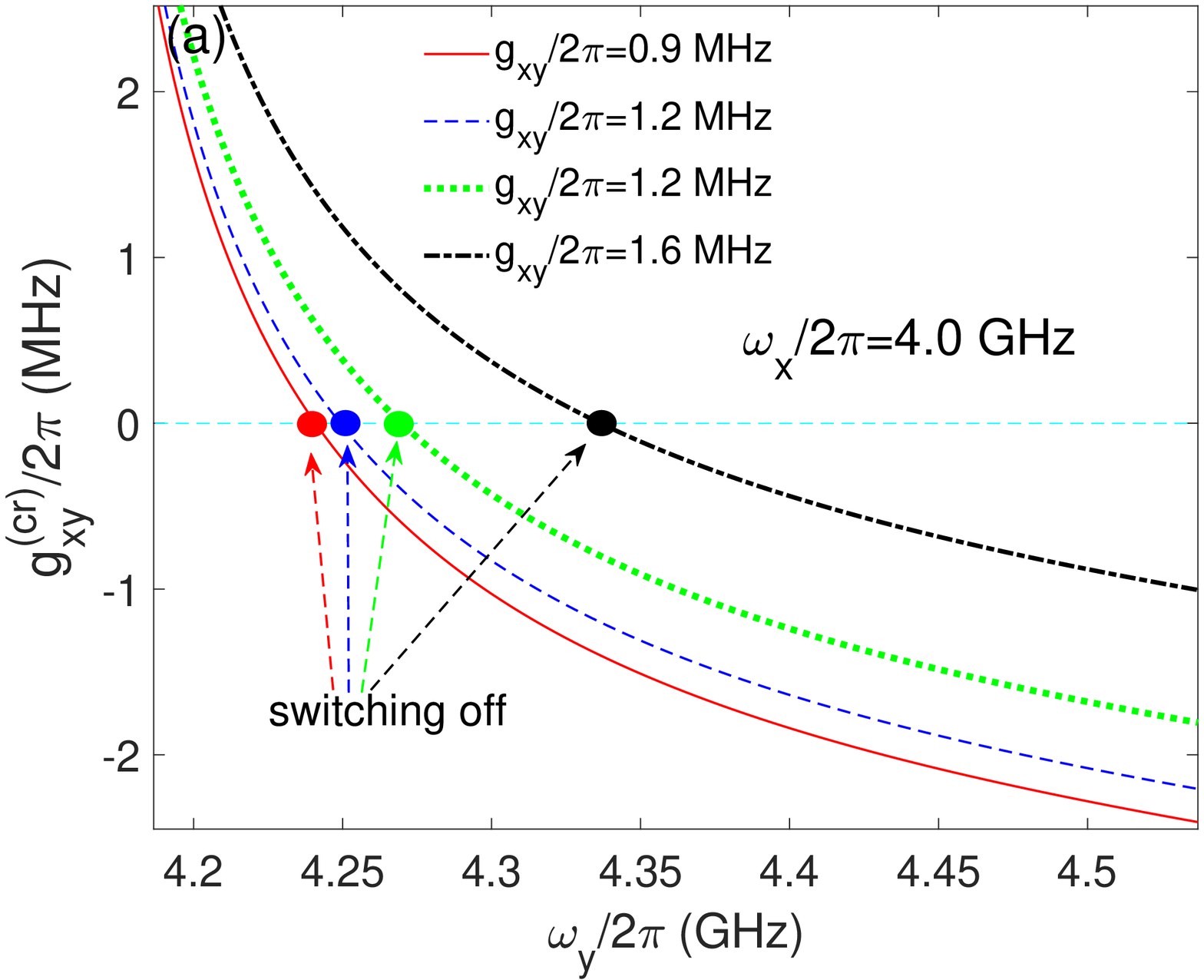}
\includegraphics[bb=-40 0 485 435, width=4.2 cm, clip]{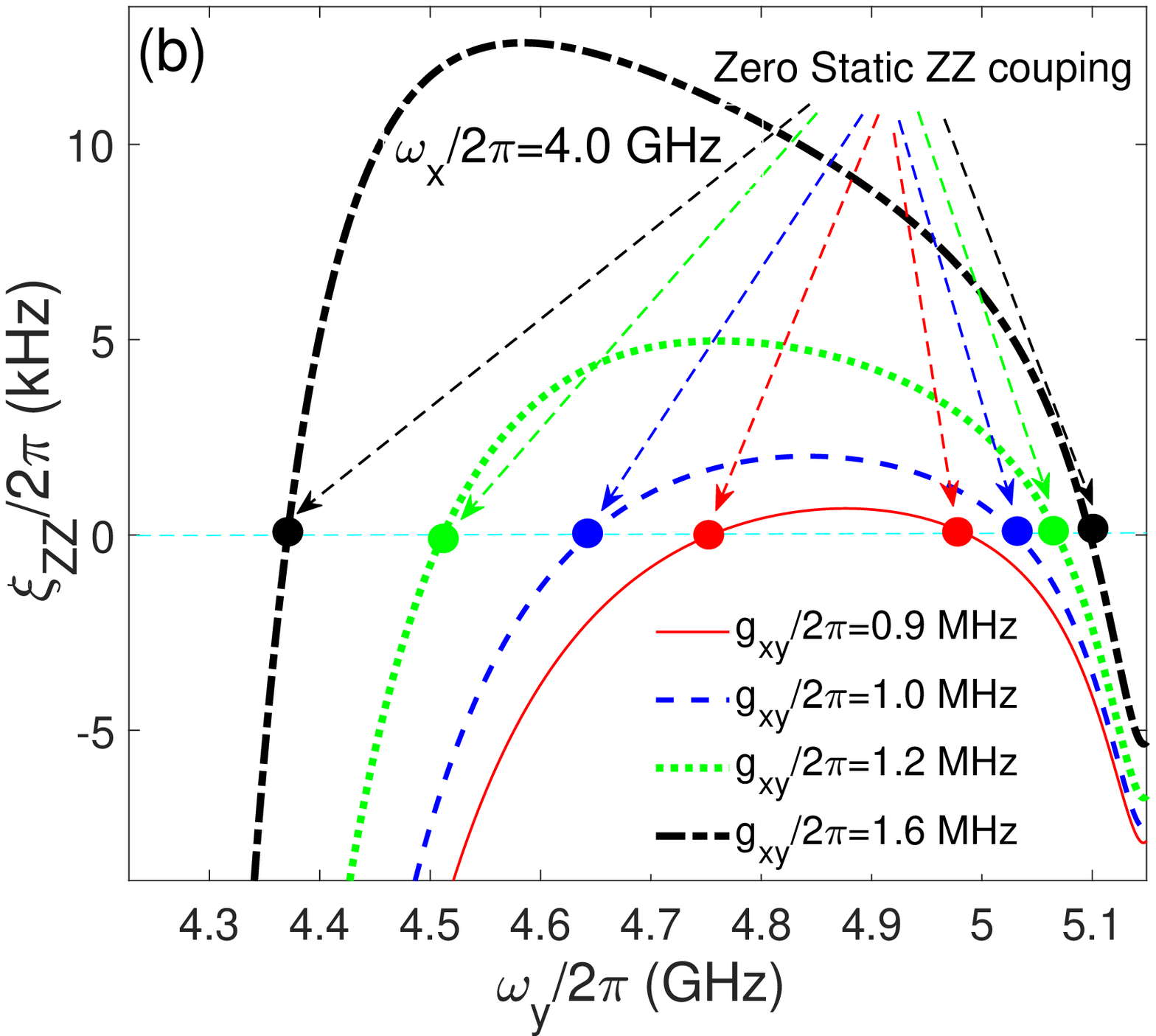}\\
\includegraphics[bb=-15 0 465 385, width=4.25 cm, clip]{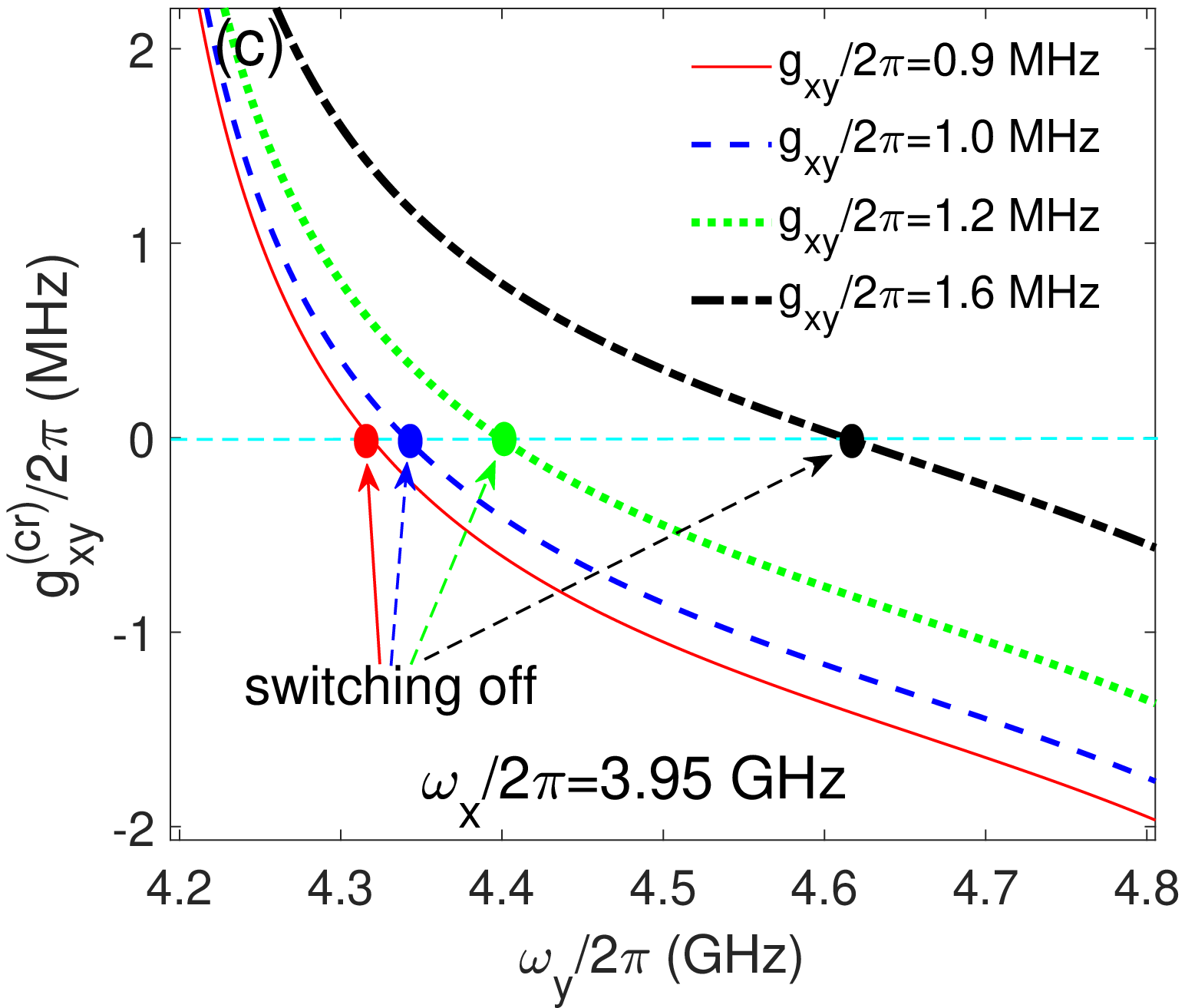}
\includegraphics[bb=-5 0 460 380, width=4.25 cm, clip]{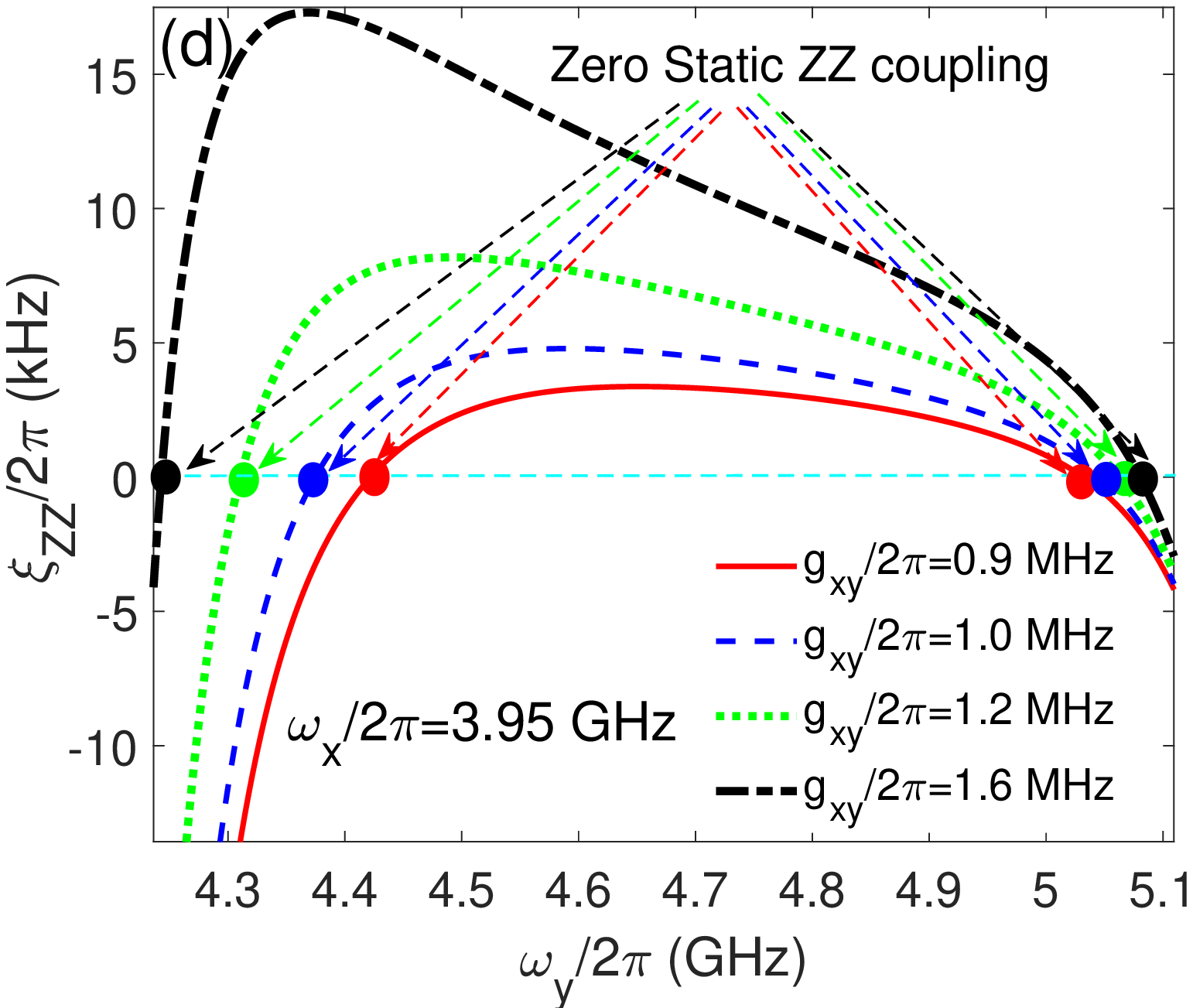}\\
\includegraphics[bb=-15 0 465 430, width=4.0 cm, clip]{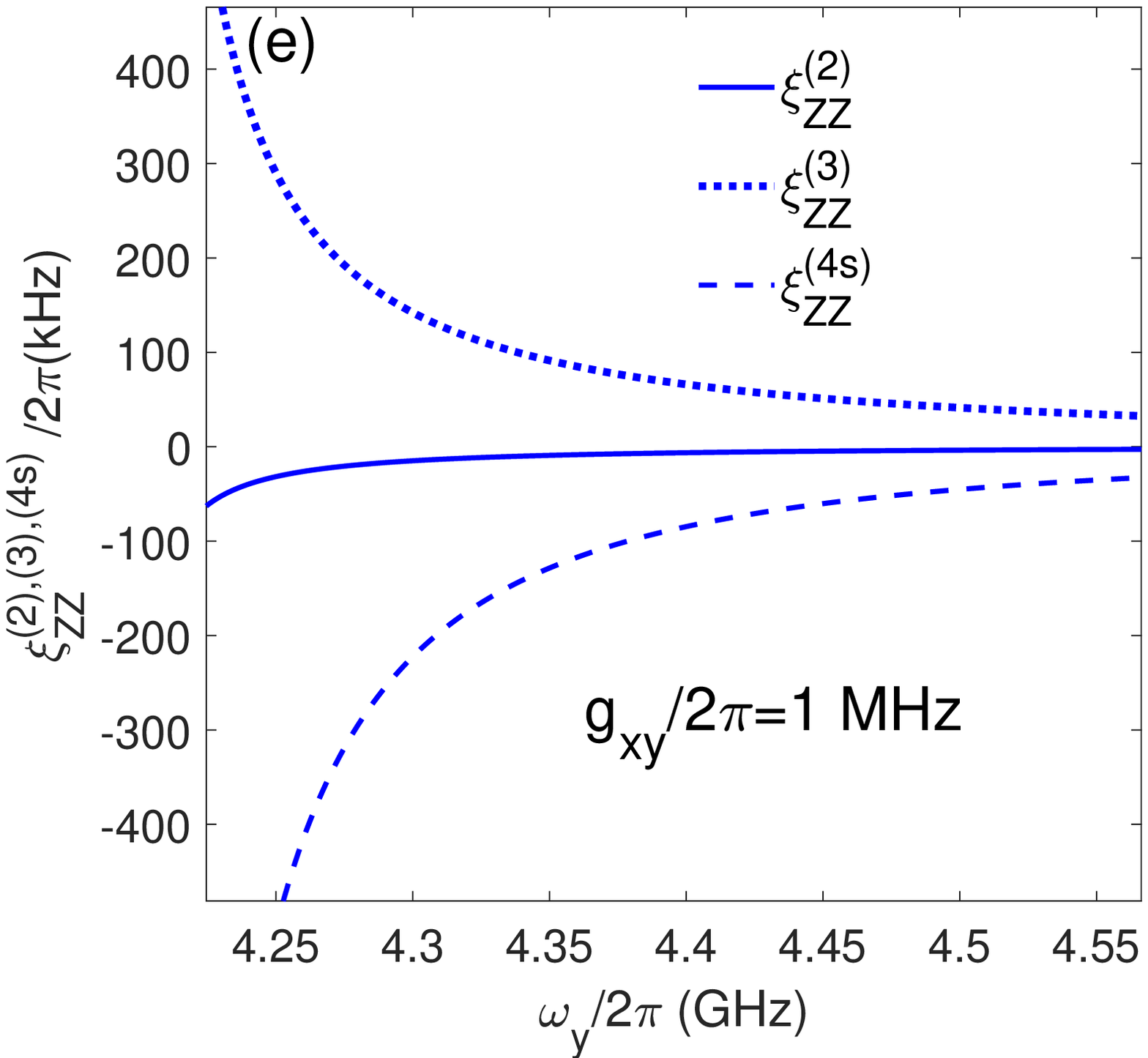}
\includegraphics[bb=-10 0 460 420, width=4.2 cm, clip]{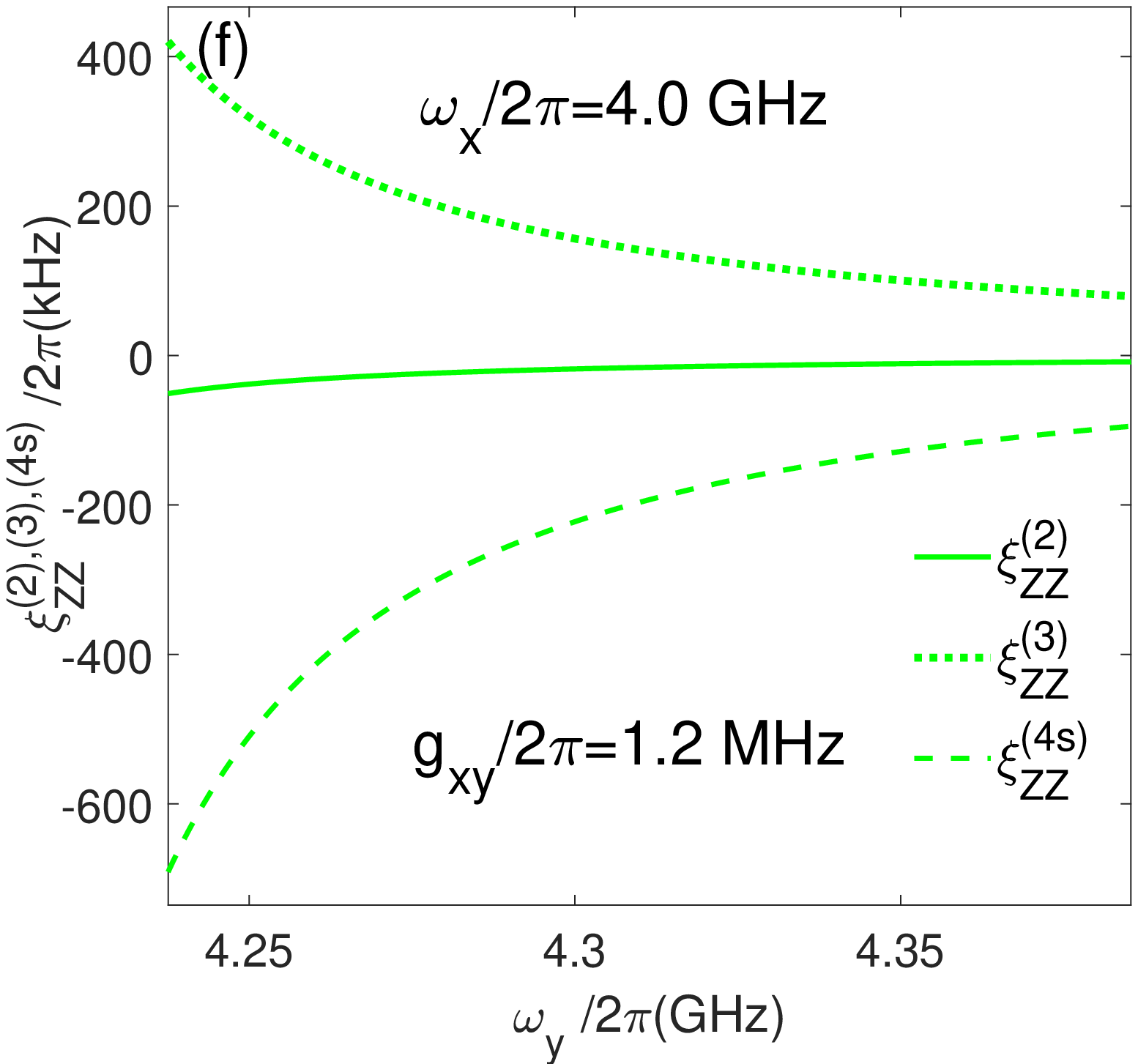}\\
\caption{(Color online) Cancellation of the Static ZZ coupling.
The   effective qubit-qubit coupling  and the Static ZZ coupling are respectively plotted in (a) and (b) in the case of $\omega_x/2\pi=4.0$ GHz,
the corresponding result for  $\omega_x/2\pi=3.95$ GHz are shown in (c) and (d).
The four curves in each figure of (a)-(d) correspond to different direct qubit-qubit coupling strengths:
 (1) $g_{xy}/2\pi=0.9 $ MHz (red-solid curves); (2) $g_{xy}/2\pi=1.0 $ MHz (blue-dashed curves);
 (3) $g_{xy}/2\pi=1.2 $ MHz (green-dotted curves); (4) $g_{xy}/2\pi=1.6 $ MHz (black dash-dotted curves).
  The $\xi^{(2)}_{ZZ}$, $\xi^{(3)}_{ZZ}$
  and $\xi^{(4s)}_{ZZ}$ are plotted in (e) and (f),
  they respectively take the parameters of  blue-dashed and green-dotted curves of (b).
The other parameters of (a)-(f) are the same as in Fig.2.
}
\label{fig7}
\end{figure}

In this section, we try to suppress the Static ZZ coupling with the  direct qubit-qubit coupling which can be arbitrary small in the double-coupler superconducting circuit.
By setting $\omega_x/(2\pi)=4.52$ GHz, the curves of static ZZ coupling $\xi_{ZZ}=\xi^{(2)}_{ZZ}+\xi^{(3)}_{ZZ}+\xi^{(4s)}_{ZZ}$ are plotted in  Fig.~\ref{fig6}(b) according to Eqs.(16)-(18).  The four curves correspond to different direct qubit-qubit coupling strengths. In the regimes suitable for the two-qubit gates, the values of static ZZ coupling  are apparently suppressed by the weaker direct qubit-qubit coupling as shown in the insert figure. And the values of $\xi^{(3)}_{ZZ}$ and $\xi^{(4s)}_{ZZ}$  are obviously suppressed by weaker direct qubit-qubit coupling in Fig.~\ref{fig6}(c)  ($g_{xy}/(2\pi)=0.5$ MHz) compared with the results in Fig.~\ref{fig6}(d) ($g_{xy}/(2\pi)=1.0$ MHz). As shown in  Fig.~\ref{fig6}(b)-\ref{fig6}(f), the static ZZ coupling  can be suppressed below Sub-MHz in double-resonator coupler circuit, which is the similar level with transmon-based coupler circuit\cite{Yan,Sung}.

For each static ZZ coupling curve in Fig.~\ref{fig6}(b), two  poles appear at  $\Delta_{xy}=\alpha_{x}$ and $\Delta_{xy}=-\alpha_{y}$  which originate from resonant state exchanges between the state $|0200\rangle\leftrightarrow|0110\rangle$ and $|0020\rangle\leftrightarrow|0110\rangle$, respectively.
There is also a pole locating at $\omega_y=\omega_x$ in each curve of static ZZ coupling in  Fig.~\ref{fig6}(b), and the poles also appears  in  the dashed curves (third-order static ZZ coupling) and  dotted curve  (fourth-order self-kerr resonance static ZZ coupling) in Figs.~\ref{fig6}(c) and ~\ref{fig6}(d), so it
 should originate the qubit-qubit resonance state exchanges as indicated by the  term $1/\Delta_{xy}$ containing in Eqs.(17)-(18). As shown in Fig.~\ref{fig6}(a), the switching off positions in double-resonator couplers circuit  can be below 5 GHz which is very close to the regimes of two-qubit gate.  If the frequency of qubit \textbf{y} is tuned away from the switching off positions,  the effective  qubit-qubit coupling quickly increase  to above 5 MHz  for the two-qubit gates.

\subsection{Cancellation of the Static ZZ coupling}

The nonzero residual coupling leads to unnecessary always-on quantum gates and additional accumulated phases, which are the dominant obstructions for the further enhancement of  two-qubit gate fidelities.  Recently, some  work announce to eliminate the static ZZ coupling in the superconducting quantum chip\cite{Sete,Xin,Jaseung}, and it might also be removed  in our proposed scheme  through the destructive interferences of double-path couplers.

 When we tune frequencies of  qubit \textbf{x}  to satisfy $\omega_x<\omega_a$,  and then the static ZZ  coupling $\xi_{ZZ}$ can be zeroes at some points  as shown in Figs.~\ref{fig7}(b) and~\ref{fig7}(d), thus the static ZZ coupling are eliminated in the double-resonator coupler circuit. The signs of second-order  $\xi^{(2)}_{ZZ}$, third-order  $\xi^{(3)}_{ZZ}$ and fourth-order  $\xi^{(4s)}_{ZZ}$ (self-kerr resonance) static ZZ coupling are different in  Figs.~\ref{fig7}(e) and ~\ref{fig7}(f), and they cancel  each other and  eliminate the static ZZ coupling at certain points.  It should be mentioned that the pole at $\omega_y=\omega_x$ in Figs.~\ref{fig6} do not appear in  Figs.~\ref{fig7}(b) just because they are outside  the scope of drawing.

As shown in Fig.~\ref{fig7}(a) (or Fig.~\ref{fig7}(c)) and Fig.\ref{fig7}(b) (or Fig.~\ref{fig7}(d)),   the static ZZ coupling are not switched off  together with the effective qubit-qubit coupling. This result is not difficult to understand for the perturbation calculation methods\cite{Sete,Sung}, because the effective qubit-qubit coupling is calculated only up to the second-order dispersive interaction (ac-Stark/Lamb Shifts), but static ZZ coupling  contains
 some the high-order effects, such as self-kerr resonance, cross-kerr resonance, the high-excited states corrections, and so on.
  The intervals between zero value positions  of  static ZZ coupling and effective qubit-qubit coupling change for different  direct qubit-qubit coupling strengths, and the  black-dash-dotted curves   ($g_{xy}/(2\pi)=1.6$  MHz)  get the smallest interval in  Figs.~\ref{fig7}(a) and \ref{fig7}(b). When we set $\omega_x/(2\pi)=3.95$ GHz,  the blue-dashed curves ($g_{xy}/(2\pi)=1.0$  MHz) get the smallest interval in Figs.~\ref{fig7}(c) and \ref{fig7}(d). So the interval between the switching off and  zero static ZZ coupling points can be tuned by the direct qubit-qubit coupling and the frequencies of qubits, and it is possible to conduct the switching off and  two-qubit gates both at the  Zero static ZZ coupling regimes  in the double-coupler superconducting circuit.

\subsection{Corrections to Static ZZ coupling}

\begin{figure}
\includegraphics[bb=0 80 565 510, width=6.8 cm, clip]{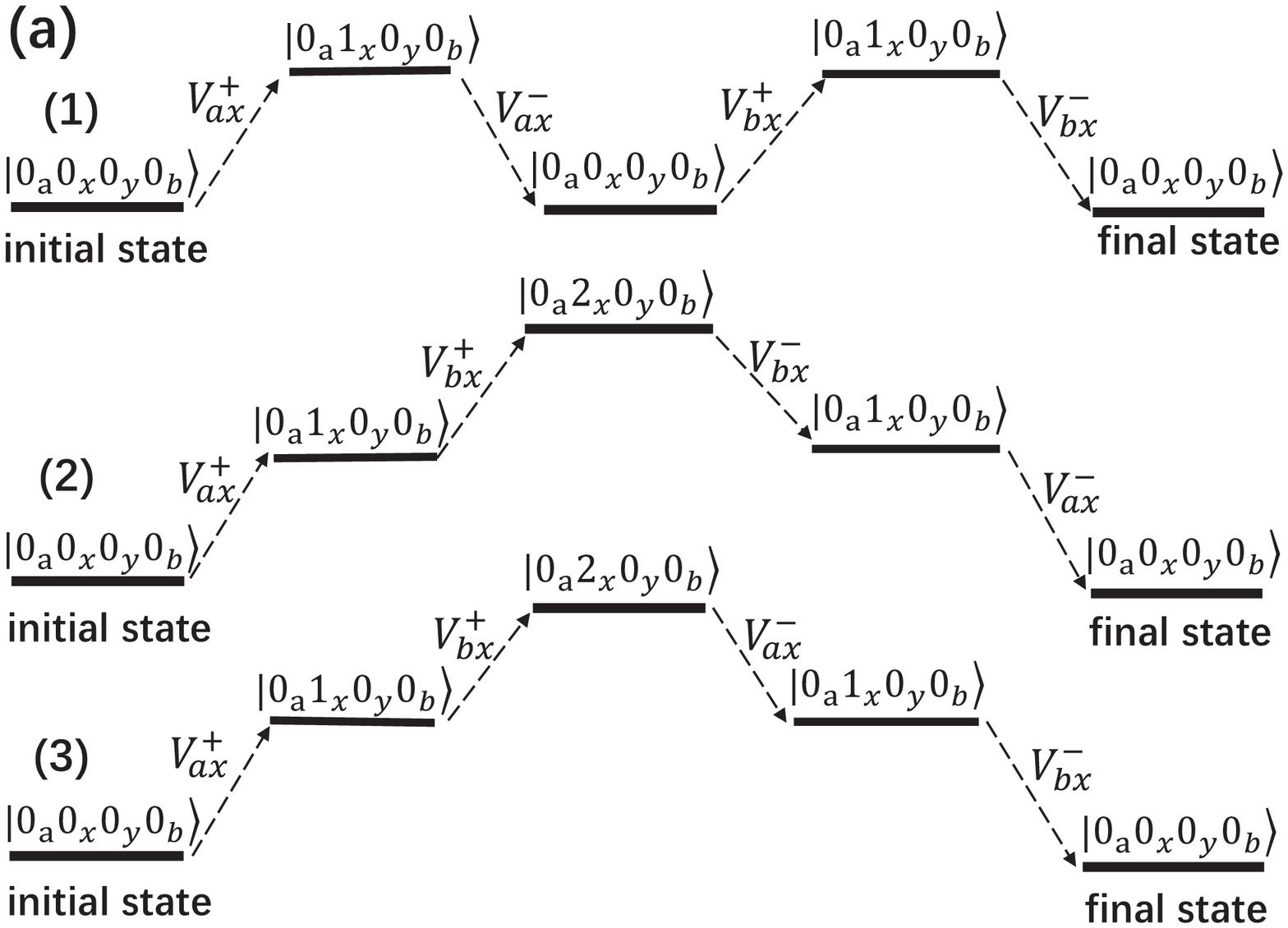}
\includegraphics[bb=0 200 575 485, width=6.8 cm, clip]{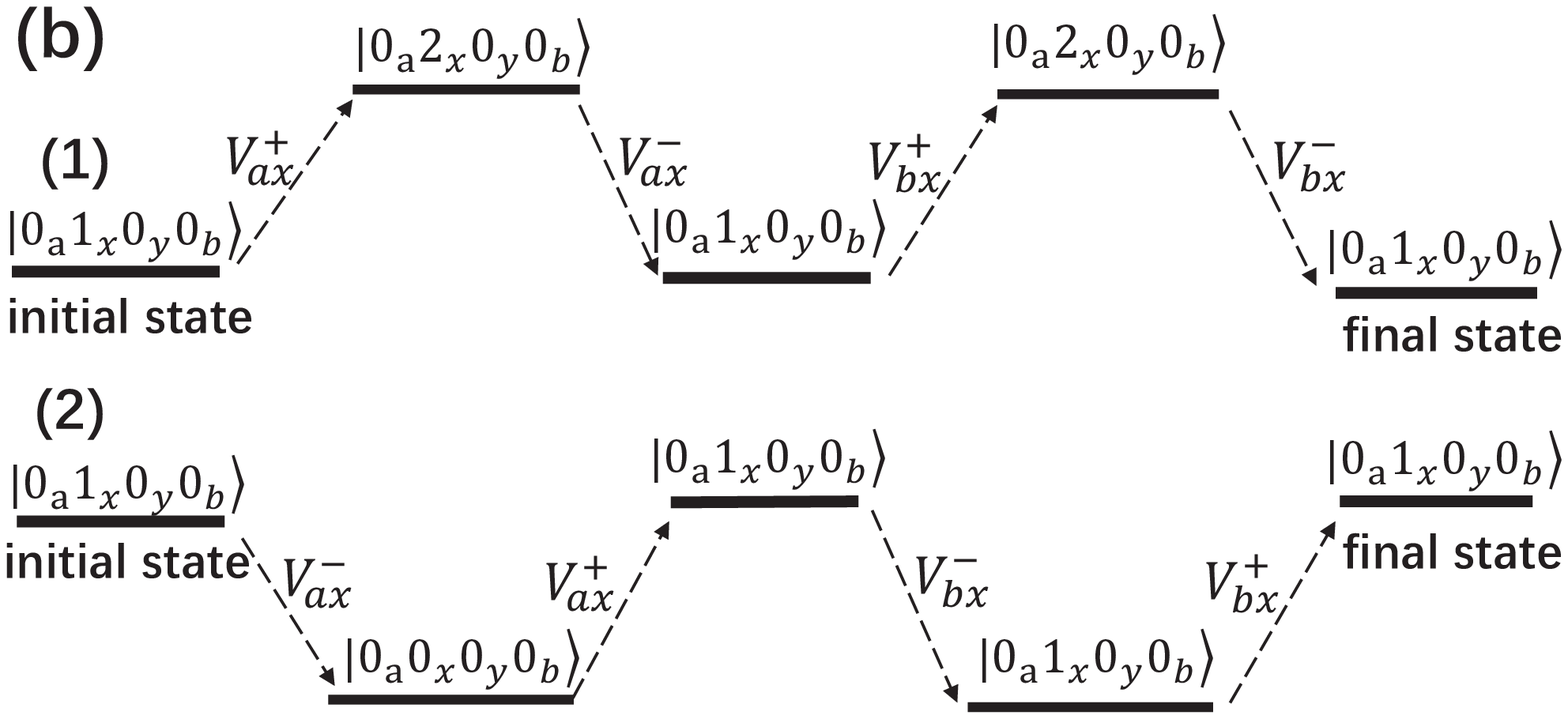}
\caption{(Color online) Energy-level diagrams of cross-kerr resonances.
The virtual photon exchange  among qubit \textbf{x}, resonator \textbf{a}, and resonator \textbf{b} with the qubit \textbf{x} initially in
(a) the ground state and (b) the first-excited state. The $V^{-}_{\lambda x}$ (or $V^{+}_{\lambda x}$)
describe the virtual photon annihilation (or creation) process through the interaction between qubit \textbf{x} and resonator $\lambda$,
 and  $V^{-}_{\lambda x}=(V^{+}_{\lambda x})^{\dagger}$.
   The energy-level diagram of cross-kerr resonances for qubit \textbf{y}  can be obtained by replacing the qubit \textbf{x} with qubit \textbf{y}.
}
\label{fig8}
  \end{figure}

If we incorporate the variations of  nonlinear term $(\alpha_{\beta}/2) a^{\dagger}_{\beta} a^{\dagger}_{\beta} a_{\beta} a_{\beta}$ during the decoupling processes of the qubit-resonator interactions, and then the cross-kerr resonances will contribute to the ZZ coupling\cite{Ferguson,Blais,Pople}. The resonator couplers are not pumped by  external fields,  the average cavity photon number is much smaller than one, so the single virtual photon exchanges will dominate the cross-kerr resonance processes.
 The cross-kerr resonance terms  $\nu_{ab,j_{\beta}} c^{\dagger}_{a}c_{a}c^{\dagger}_{b}c_{b}$ and
  $\nu_{ba,j_{\beta}} c^{\dagger}_{b}c_{b}c^{\dagger}_{a}c_{a}$ in Eq.(15) describe the physical processes of virtual photon exchange  between a qubit and two resonators, we plot the  energy-level diagrams of  single-virtual photon exchange process of cross-kerr resonance for qubit \textbf{x} in Fig.~\ref{fig8}. For simplicity, only  three lowest energy levels of superconducting artificial atoms are considered. The  virtual photon exchange processes of cross-kerr resonance for qubit \textbf{y}  can be obtained by replacing the \textbf{x}  with \textbf{y} in Fig.~\ref{fig8} .

\begin{figure}
\includegraphics[bb=-10 0 520 465, width=4.2 cm, clip]{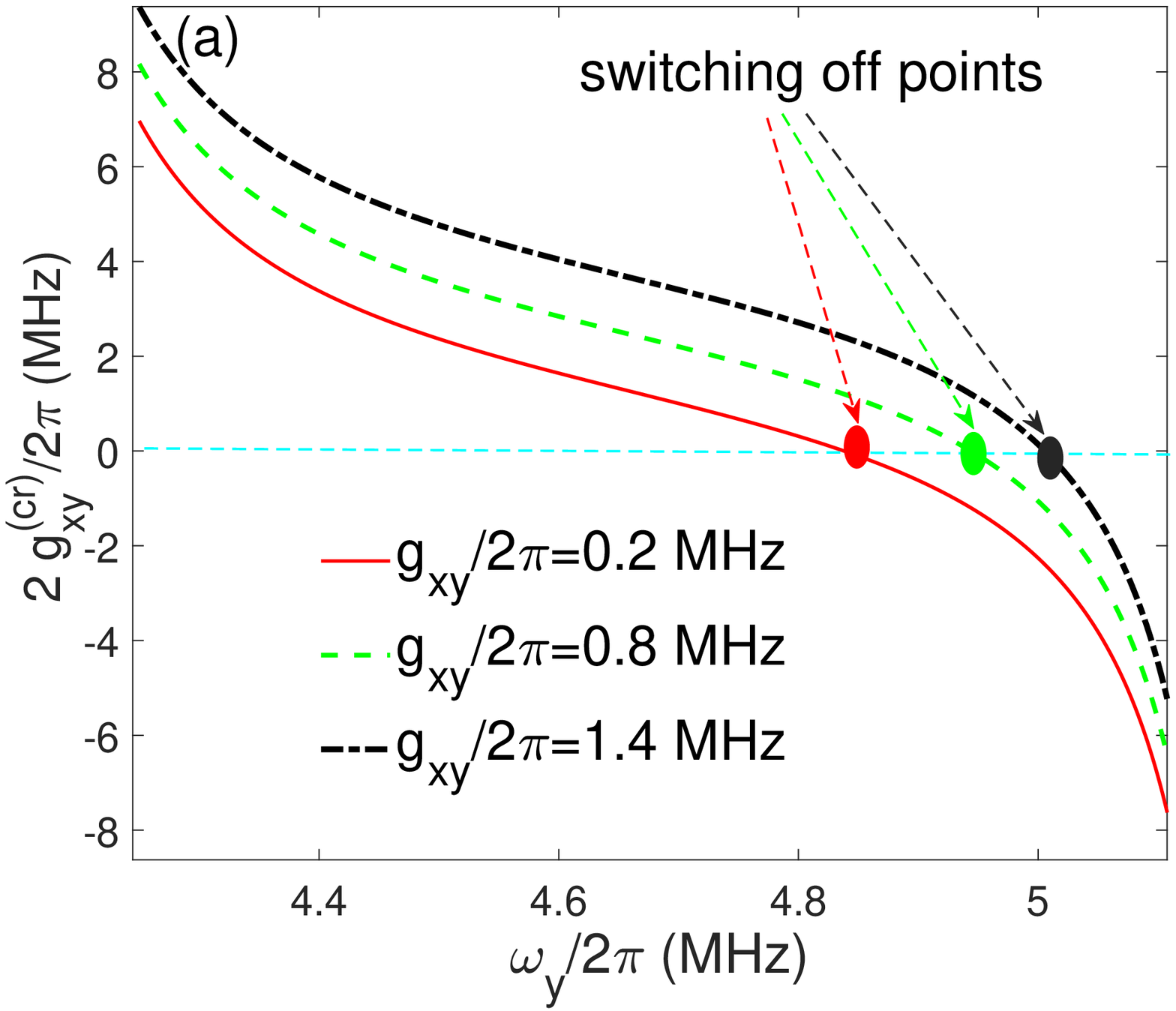}
\includegraphics[bb=0 0 551 440, width=4.3 cm, clip]{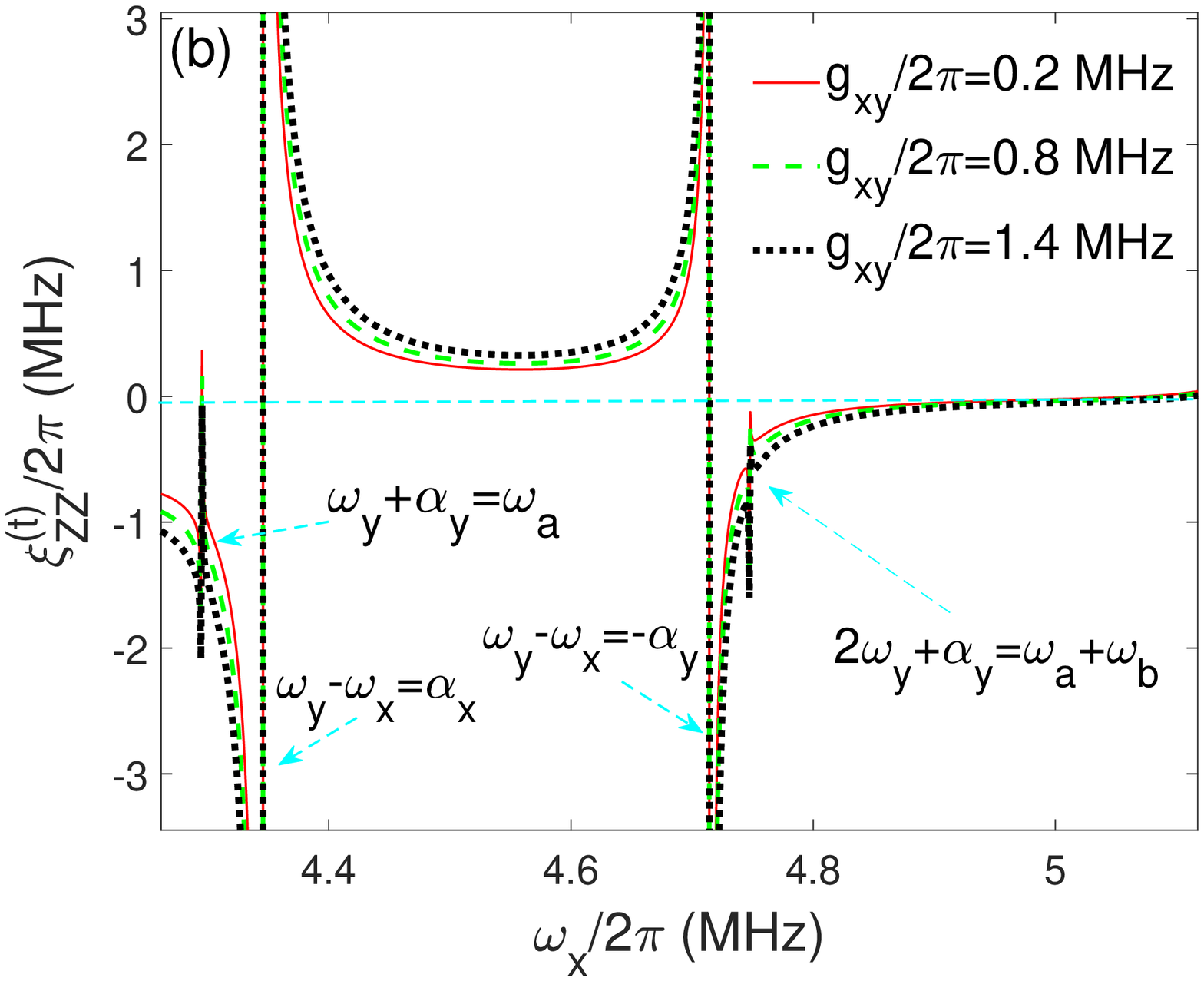}\\
\includegraphics[bb=0 0 470 415, width=4.3 cm, clip]{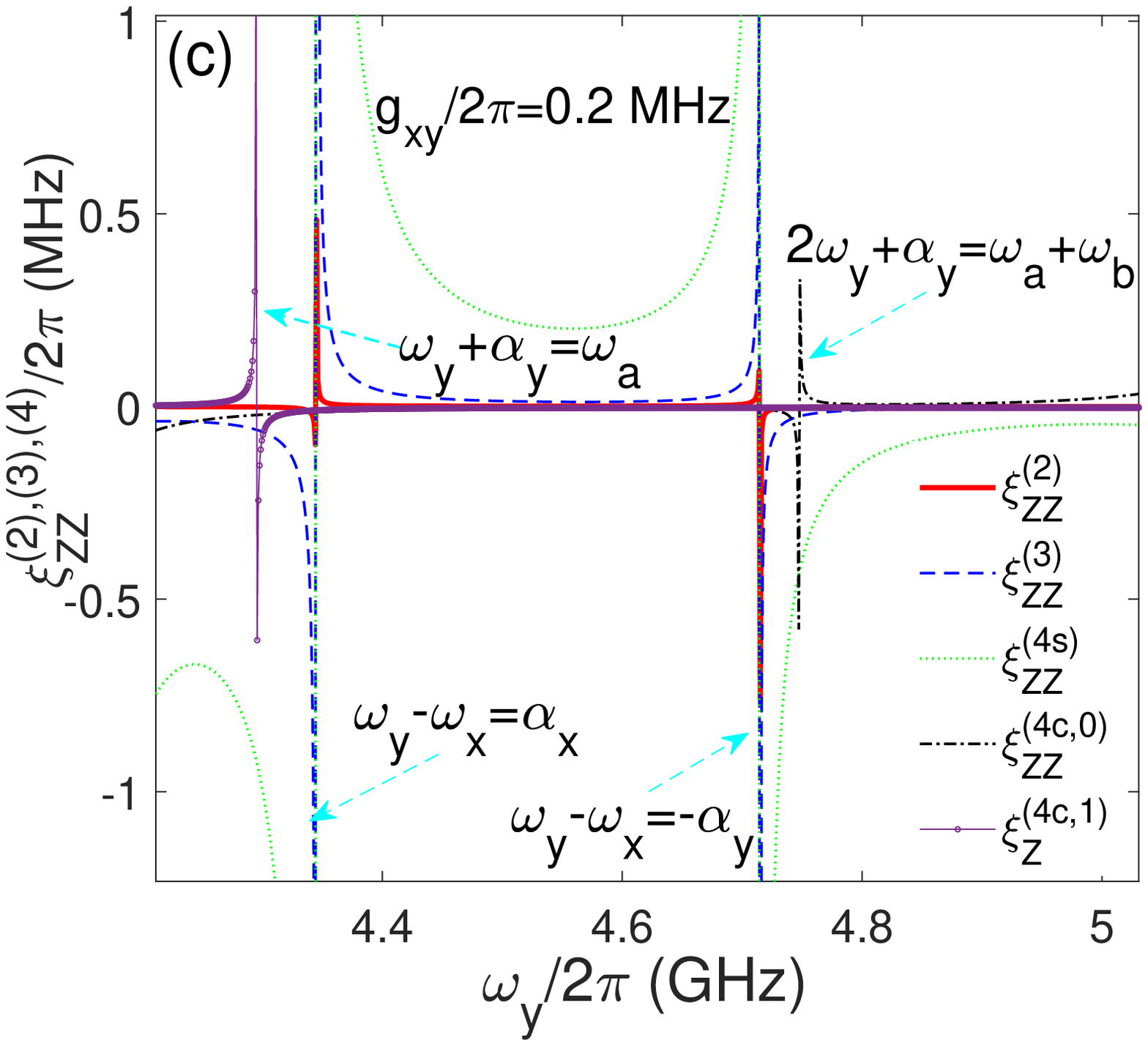}
\includegraphics[bb=0 0 455 410, width=4.2 cm, clip]{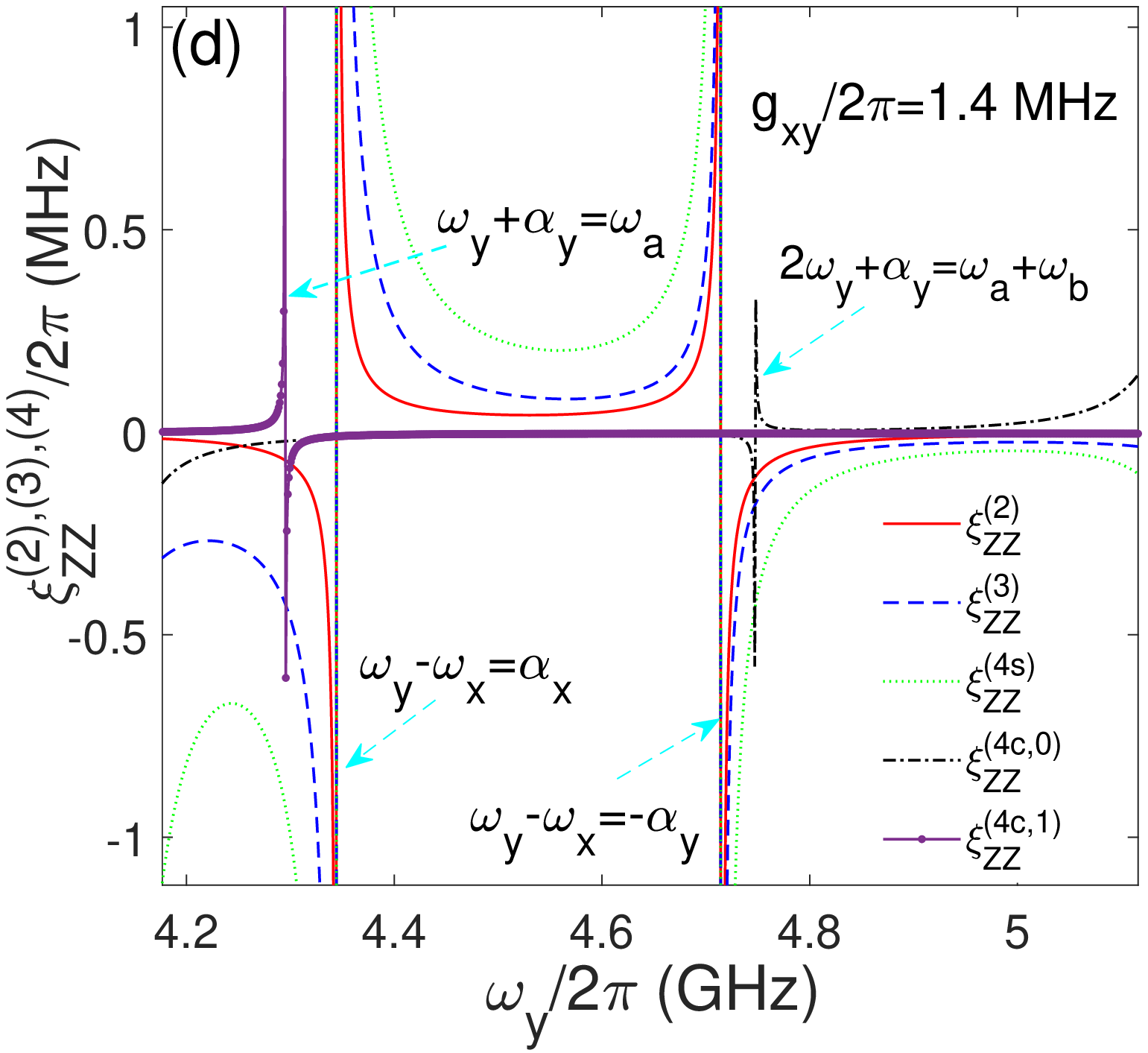}
\caption{(Color online) The  static ZZ coupling corrected by the cross-kerr resonances.
We plot (a) the effective qubit-qubit coupling and (b) the static ZZ coupling $\xi^{(t)}_{ZZ}=\xi^{(2)}_{ZZ}+\xi^{(3)}_{ZZ}+\xi^{(4s)}_{ZZ}+\xi^{(4c,0)}_{ZZ}-\xi^{(4c,1)}_{ZZ}$ at different direct qubit-qubit coupling strengths:
(1) $g_{xy}/2\pi=0.2$ MHz (red-solid curve); (2) $g_{xy}/2\pi=1.4$ MHz (green-dashed curve);
(3) $g_{xy}/2\pi=2$ MHz (black-dotted curve). The $\xi^{(2)}_{ZZ}$ (red-solid curve), $\xi^{(3)}_{ZZ}=\sum_{\lambda=a,b}\xi^{(3)}_{ZZ,\lambda}$ (blue-dashed curve), $\xi^{(4s)}_{ZZ}=\sum_{\lambda=a,b}\xi^{(4s)}_{ZZ,\lambda}$ (green-dotted curve),  $\xi^{(4c,0)}_{ZZ}=\sum_{\beta=x,y}\xi^{(4c,0)}_{ZZ,\beta}$ (black-dashed-dot curve), and $\xi^{(4c,1)}_{ZZ}=\sum_{\beta=x,y}\xi^{(4c,1)}_{ZZ,\beta}$ (marked purple-solid curve) are plotted with (c) $g_{xy}/2\pi=0.2$ MHz  and (d) $g_{xy}/2\pi=1.4$ MHz.
 The  parameters of four figures are the same as in Fig.2 except for $\omega_{x}/2\pi=4.52$ GHz.
}
\label{fig9}
\end{figure}

When  qubit \textbf{x} is initially  in the ground states, the Energy-level diagrams of cross-kerr resonance among qubit \textbf{x}, resonator \textbf{a}, and resonator \textbf{b} are shown in Fig.~\ref{fig8}(a).  The six  cross-kerr resonances  are attributed as  three types of virtual photon exchange processes among  qubit \textbf{x},  resonator \textbf{a}, and resonator \textbf{b}  \cite{Ferguson}.
The first type: the qubit \textbf{x}  absorbs a virtual photon from resonator \textbf{a} (or \textbf{b}) and transits  to the first-excited states  from the ground state, and it immediately returns the virtual  photon to  resonator \textbf{a} (or \textbf{b}) and decays to the ground state. Subsequently the qubit \textbf{x}  jumps to the first-excited state again by getting another virtual photon from  resonator \textbf{b} (or \textbf{a}), and  finally it emits the virtual photon to resonator \textbf{b} (or \textbf{a}) and decays to the ground state.
 The second type:  the qubit \textbf{x}  transits to the first-excited state from  the ground state by absorbing a virtual photon from resonator \textbf{a} (or \textbf{b}) and immediately jumps to the second-excited state by taking another virtual photon from  resonator \textbf{b} (or \textbf{a}).  Subsequently the qubit jumps to the first-excited state by emitting a virtual photon to the resonator \textbf{b} (or \textbf{a}), and finally decays to the ground state by emitting another virtual photon to resonator
 \textbf{a} (or \textbf{b}). The third type: the first two  transition processes are the same as the second type,   but the qubit firstly returns a virtual photon to  resonator \textbf{a} (or \textbf{b}) in the third transition process and transits to the first-excited state, and finally decays to the ground state by emitting another photon to resonator \textbf{b} (or \textbf{a}).  The  virtual photon exchange processes for cross-kerr resonances among qubit \textbf{y}, resonator \textbf{a}, and resonator \textbf{b} can be obtained by replacing the qubit \textbf{x} with qubit \textbf{y} in Fig.~\ref{fig8}(a). Adding together the contributions of the six type cross-kerr resonant processes, and we can obtain the energy level corrections  to the ground state  of qubit $\beta$ \cite{Ferguson},
\begin{eqnarray}\label{eq:11}
& &\xi^{(4c,0)}_{ZZ,\beta}|0_{\beta}\rangle\langle 0_{\beta}|
 =g^2_{a\beta}g^2_{b\beta}\bigg[\frac{2}{\Delta_{a\beta}\Delta_{b\beta}\omega_{\beta}}\\
 & &~~+\frac{1}{2\omega_{\beta}+\alpha_{\beta}-\omega_a-\omega_b}\bigg(\frac{2\omega_{\beta}-\omega_a-\omega_b}{\Delta_{a\beta}\Delta_{b\beta}}\bigg)^2\bigg]
 |0_{\beta}\rangle \langle 0_{\beta}|. \nonumber
\end{eqnarray}
For simplicity, we have neglected the small differences on the interactions of a resonator with different energy levels of superconducting artificial atom, that is  $g^{j_{\beta},j^{\prime}_{\beta}}_{\lambda}=g_{\lambda \beta}$.

When the qubit \textbf{x} is the initially in the first-excited state, the Energy-level diagrams of cross-kerr resonance among qubit \textbf{x}, resonator \textbf{a}, and resonator \textbf{b}  are shown in Fig.~\ref{fig8}(b). The four  cross-kerr resonances are attributed as  two types of virtual photon exchange processes.
The first type: the qubit \textbf{x}  in the first-excited state absorbs a virtual photon from resonator \textbf{a} (or \textbf{b}) and transits to the second-excited state,  and it immediately returns the photon to  resonator \textbf{a} (or \textbf{b}) and jumps to the first-excited state.  Subsequently the qubit jumps to the second-excited state again by getting another virtual photon from  resonator \textbf{b} (or \textbf{a}) and  finally it emits the photon to resonator \textbf{b} (or \textbf{a}) and jumps to the first-excited state. The second type: the qubit $x$ (in the first-excited state) emits a virtual photon to resonator \textbf{a} (or \textbf{b}) and decays to the ground state and immediately  transits to the first-excited state by absorbing a virtual photon from resonator \textbf{a} (or \textbf{b}). Subsequently the qubit  emits a virtual photon to resonator \textbf{b} (or \textbf{a}) and decays to the ground state and immediately absorbs another virtual photon from resonator \textbf{b} (or \textbf{a}) and finally jumps to the first-excited state. The  virtual photon exchange processes for cross-kerr resonances among qubit \textbf{y}, resonator \textbf{a}, and resonator \textbf{b} can be obtained by replacing the qubit \textbf{x} with qubit \textbf{y} in Fig.~\ref{fig8}(b). Adding together the contributions of the four type cross-kerr resonant processes, and we can obtain the energy level corrections  to  the first-excited state of qubit $\beta$ \cite{Ferguson},
 \begin{eqnarray}\label{eq:12}
\xi^{(4c,1)}_{ZZ,\beta}|1_{\beta}\rangle\langle 1_{\beta}|
&=&\frac{2g^2_{a\beta}g^2_{b\beta}}{\omega_{\beta}(\Delta_{a\beta}+\alpha_{\beta})(\Delta_{b\beta}+\alpha_{\beta})}|1_{\beta}\rangle\langle 1_{\beta}|\nonumber\\
&+&\frac{2g^2_{a\beta}g^2_{b\beta}}{\omega_{\beta}\Delta_{a\beta}\Delta_{b\beta}}|1_{\beta}\rangle\langle 1_{\beta}|.
\end{eqnarray}
Adding the corrections by the cross-kerr resonances, the total ZZ coupling  can be defined as
$\xi^{(t)}_{ZZ}=\xi^{(2)}_{ZZ}+\xi^{(3)}_{ZZ}+\xi^{(4s)}_{ZZ}+\xi^{(4c,0)}_{ZZ}-\xi^{(4c,1)}_{ZZ}$, with $\xi^{(3)}_{ZZ}=\sum_{\lambda=a,b}\xi^{(3)}_{ZZ,\lambda}$, $\xi^{(4s)}_{ZZ}=\sum_{\lambda=a,b}\xi^{(4s)}_{ZZ,\lambda}$, $\xi^{(4c,0)}_{ZZ}=\sum_{\beta=x,y}\xi^{(4c,0)}_{ZZ,\beta}$, and $\xi^{(4c,1)}_{ZZ}=\sum_{\beta=x,y}\xi^{(4c,1)}_{ZZ,\beta}$. We plot  the  second-order $\xi^{(2)}_{ZZ}$ (red-solid curves), third-order  $\xi^{(3)}_{ZZ}$ (blue-dashed curves), and fourth-order (self-kerr resonance) $\xi^{(4s)}_{ZZ}$ (green-dotted curves) static ZZ coupling in Fig.~\ref{fig9}(c) ($g_{xy}/(2\pi)=0.2$ MHz) and Fig.~\ref{fig9}(d) ($g_{xy}/(2\pi)=1.4$ MHz).  By reducing the direct qubit-qubit coupling strengths, the values of   static ZZ coupling curves  are apparently suppressed in Fig.~\ref{fig9}(c)  ($g_{xy}/(2\pi)=0.2$ MHz) compared with the result in Fig.~\ref{fig9}(d) ($g_{xy}/(2\pi)=1.4$ MHz).
The energy level corrections  to qubit's ground state ($\xi^{(4c,0)}_{ZZ}$) and first-excited state ($\xi^{(4c,1)}_{ZZ}$) by the cross-kerr resonances are respectively plotted in the black dashed-dot curves and  marked purple-solid curves in Figs.~\ref{fig9}(c) and ~\ref{fig9}(d).

The curves of effective qubit-qubit coupling  are plotted in Fig.~\ref{fig9}(a), and values of direct qubit-qubit coupling affect switching off positions.
 Similar with Fig.~\ref{fig6}(b), the two  poles locating at $\Delta_{xy}=\alpha_{x}$ and $\Delta_{xy}=-\alpha_{y}$  also  appear in each curve of static ZZ coupling $\xi^{(t)}_{ZZ}$ in Fig.~\ref{fig9}(b).  But there are two new poles in  each curve of static ZZ coupling $\xi^{(t)}_{ZZ}$ in  Fig.~\ref{fig9}(b) which should originate from the cross-kerr resonance\cite{Ferguson,Blais}.
  As indicated by Eqs.(19) and (20), the cross-kerr resonances through virtual photon exchanges induces additional poles for the static ZZ coupling  at  the point: $2\omega_{\beta}+\alpha_{\beta}=\omega_a+\omega_b$ (from Eq.(19)),  $\omega_\beta+\alpha_\beta=\omega_a$  (from Eq.(20)),  and $\omega_\beta+\alpha_\beta=\omega_b$  (from Eq.(20)).
So we can see two new poles  at $\omega_y+\alpha_y=\omega_a$ and $2\omega_y+\alpha_y=\omega_a+\omega_b$ in each curve  of Fig.~\ref{fig9}(b).  Another pole ($\omega_y+\alpha_y=\omega_b$) is out of the scope of the drawing. As shown  in Figs.~\ref{fig9}(c) and ~\ref{fig9}(d), the pole at $2\omega_{y}+\alpha_{y}=\omega_a+\omega_b$ only appears on the black dash-dotted curves which correspond to the level correction to qubits' ground states by the cross-kerr resonance $\xi^{(4c,0)}_{ZZ}$. While the pole at $\omega_{y}+\alpha_{y}=\omega_a$ only appears in the marked purple-solid curve which  describes the level correction to qubits' first-excited states by the cross-kerr resonance $\xi^{(4c,1)}_{ZZ}$.  In this article, we neglect the  level corrections of cross-kerr resonances to the double-excited state $|0 1_x 1_y 0\rangle$, which should correspond to more complex physical processes.

 \section{Conclusions}\label{conclusion}

In conclusion, we have studied the mechanism of the switching off  in the
superconducting  circuit consisting of two fixed-frequency  resonator couplers. The induced indirect qubit-qubit coupling by two resonators can be cancelled, so the switching off can be realized without the direct qubit-qubit coupling.
The frequencies of couplers can be much smaller than the single transmon-based coupler  circuit, and this  leaves wider available frequency spaces for couplers (or qubits), thus  the frequency crowding on the superconducting chip might be  relieved.

The weak direct qubit-qubit coupling can be used to suppress the static ZZ coupling in the double-coupler circuit, and  the  destructive interferences between double-path couplers can eliminate the static ZZ coupling, thus the quality of superconducting quantum chip might be enhanced.
Our proposed double-resonator couplers scheme can unfreeze some restrictions  on the superconducting quantum chip, mitigate the static ZZ coupling, and  also save the  dilution refrigerator lines, which might be a promising platform for superconducting quantum chip.

\section{ACKNOWLEDGMENTS}

H.W. is supported by the Natural Science Foundation of Shandong Province
under Grant No. ZR2023LZH002 and the Inspur artificial intelligence research institute.
Y.J.Z. is supported by Beijing Natural Science Foundation
under Grant No.  4222064 and NSFC under Grant No. 11904013.
X.-W.X. is supported by the National Natural Science Foundation of China
under Grant No.~12064010, and Natural Science Foundation of Hunan Province of China under Grant No.~2021JJ20036.

\section*{Appendix A: The numerical calculation of the energy levels}

\setcounter{equation}{0}
\renewcommand\theequation{A.\arabic{equation}}

\begin{figure}
\includegraphics[bb=0 0 375 295, width=2.7cm, clip]{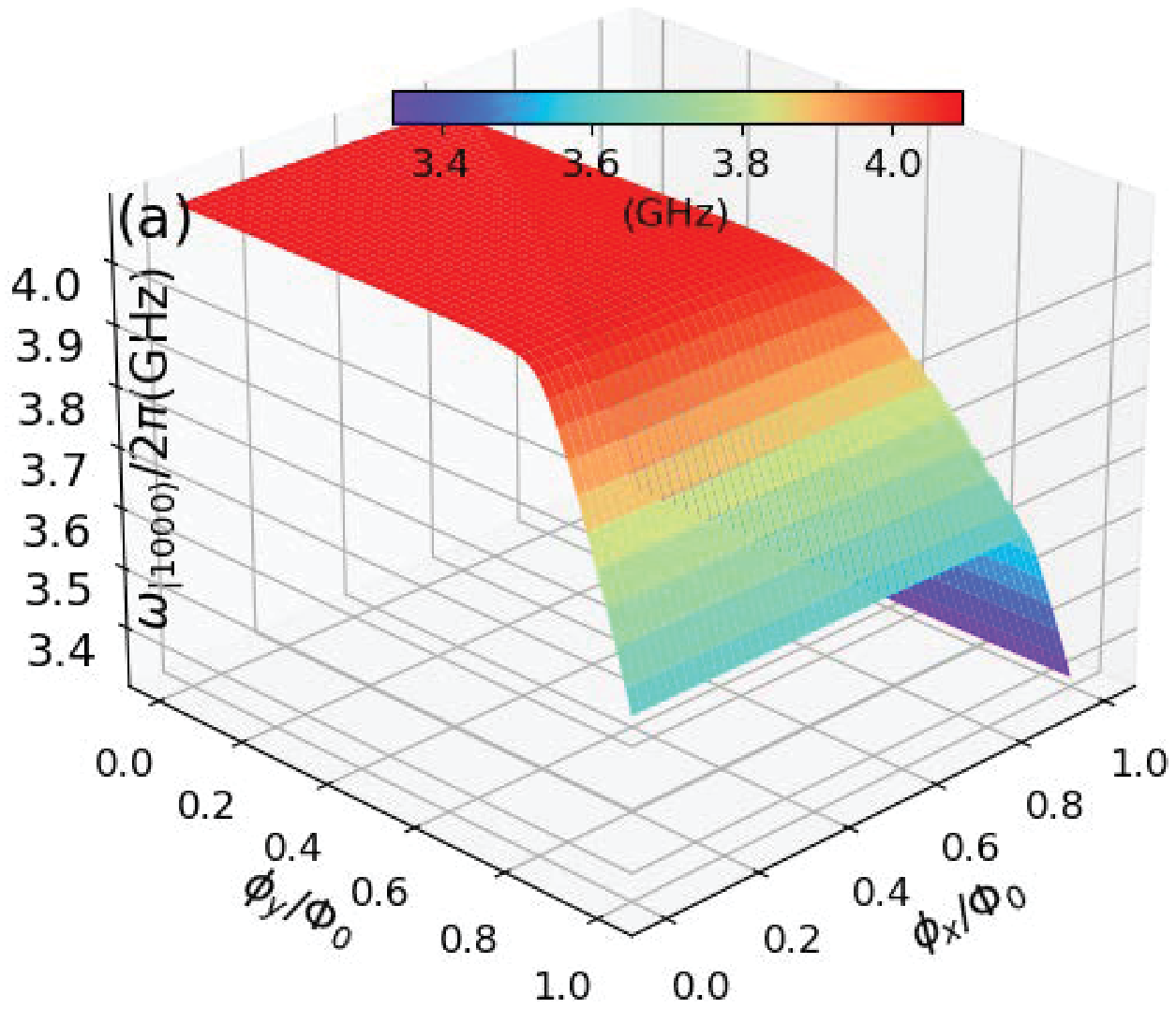}
\includegraphics[bb=0 0 375 300, width=2.7cm, clip]{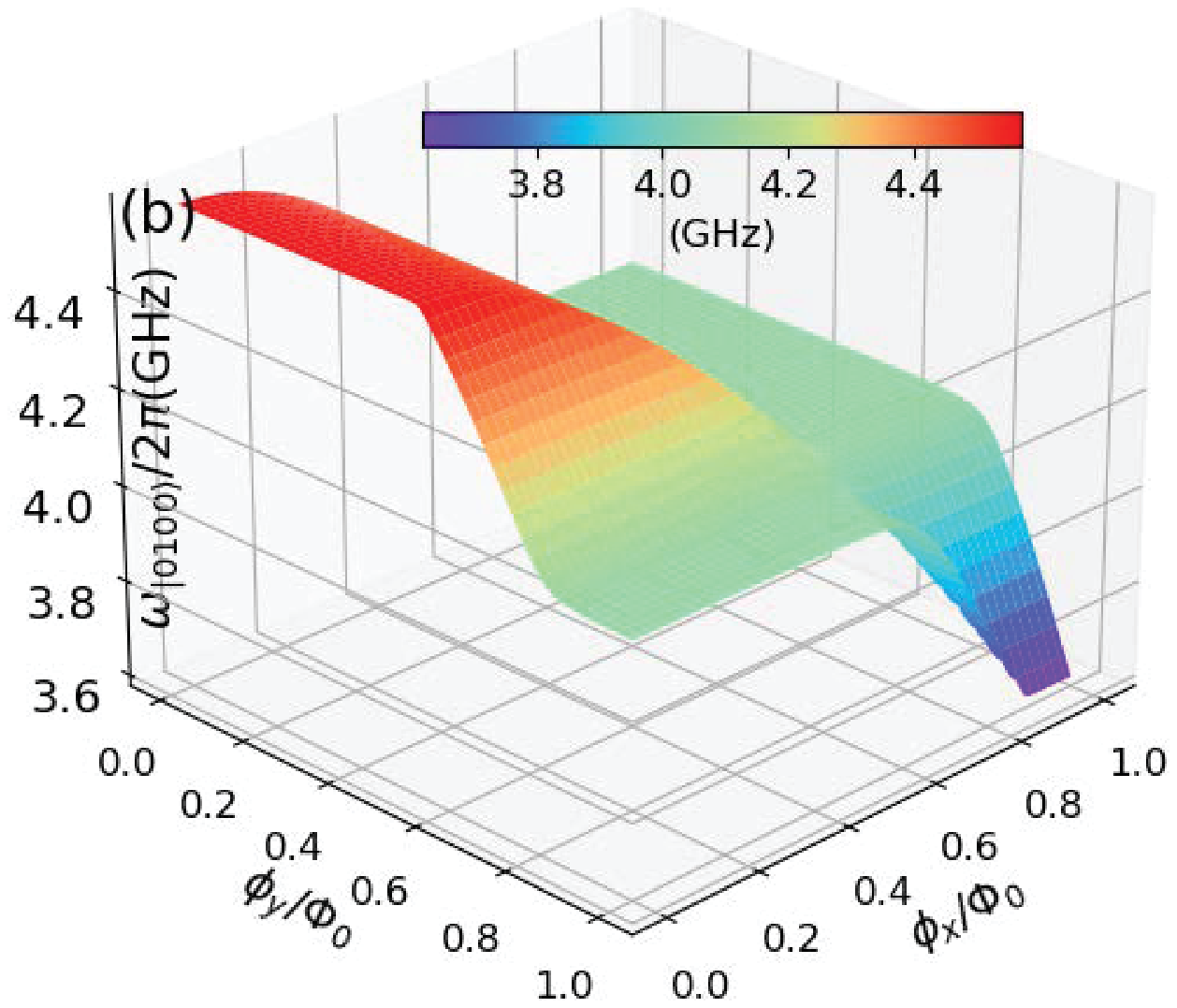}
\includegraphics[bb=0 0 375 300, width=2.7cm, clip]{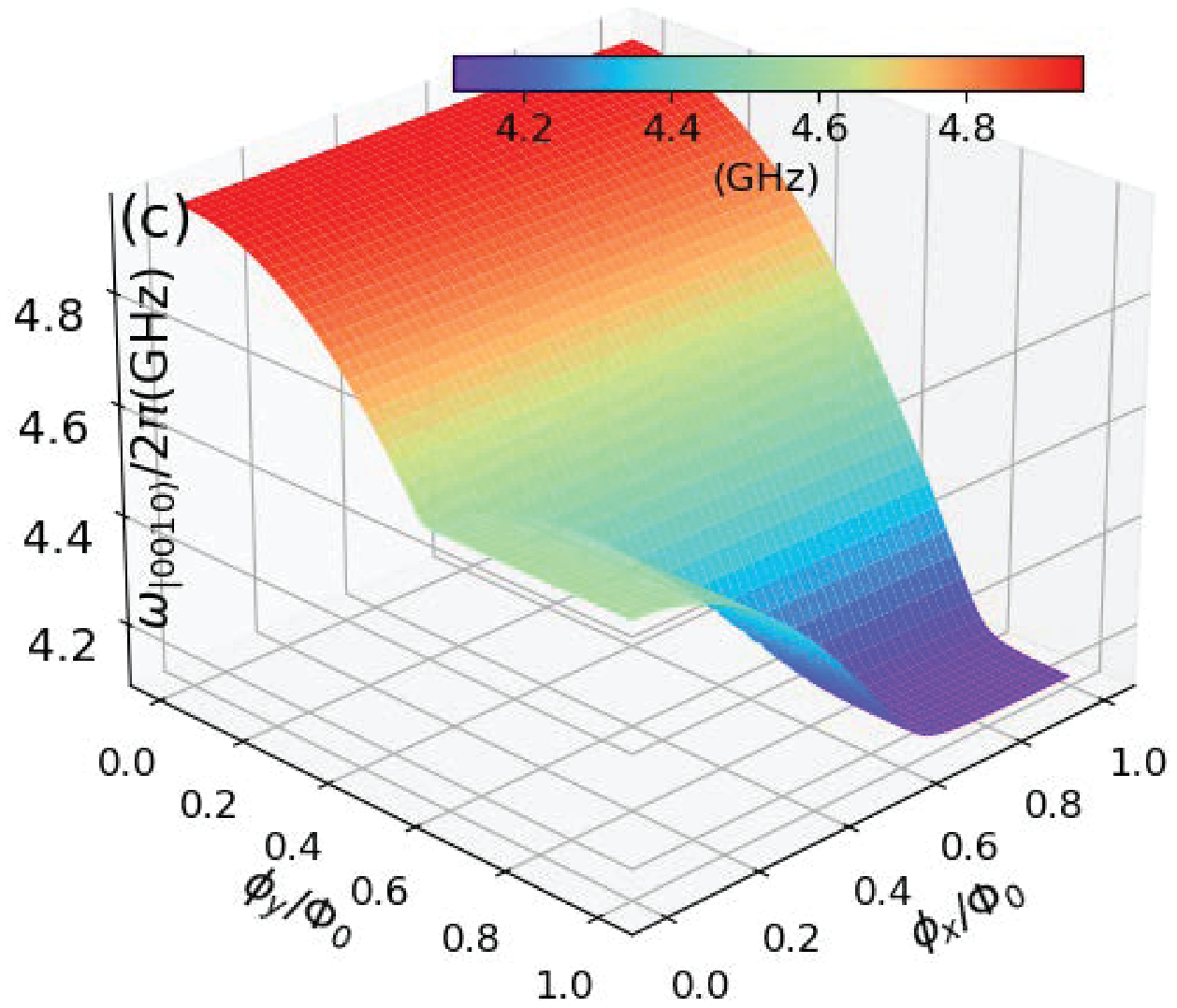}\\
\includegraphics[bb=0 0 375 300, width=2.7cm, clip]{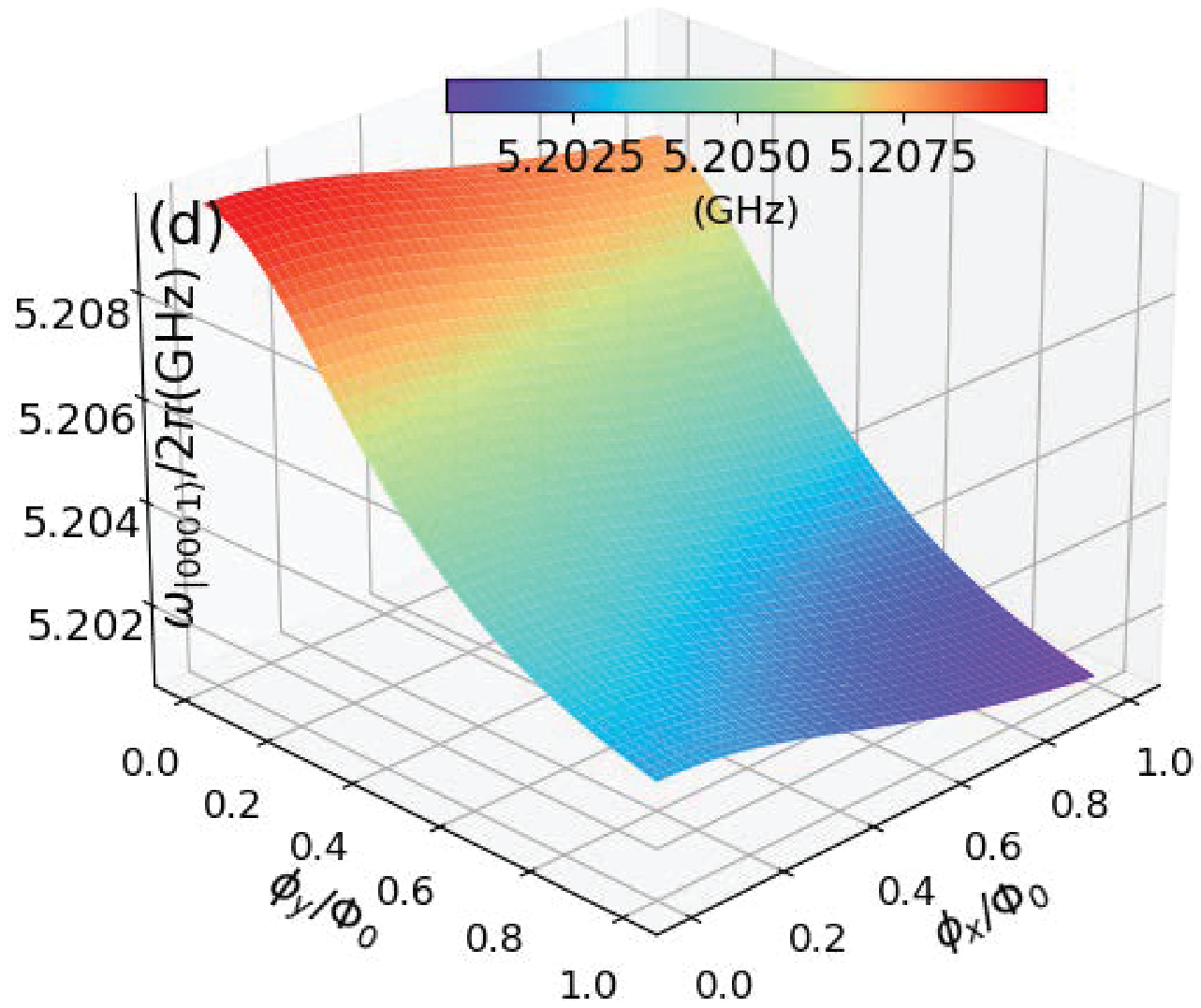}
\includegraphics[bb=0 0 375 300, width=2.7cm, clip]{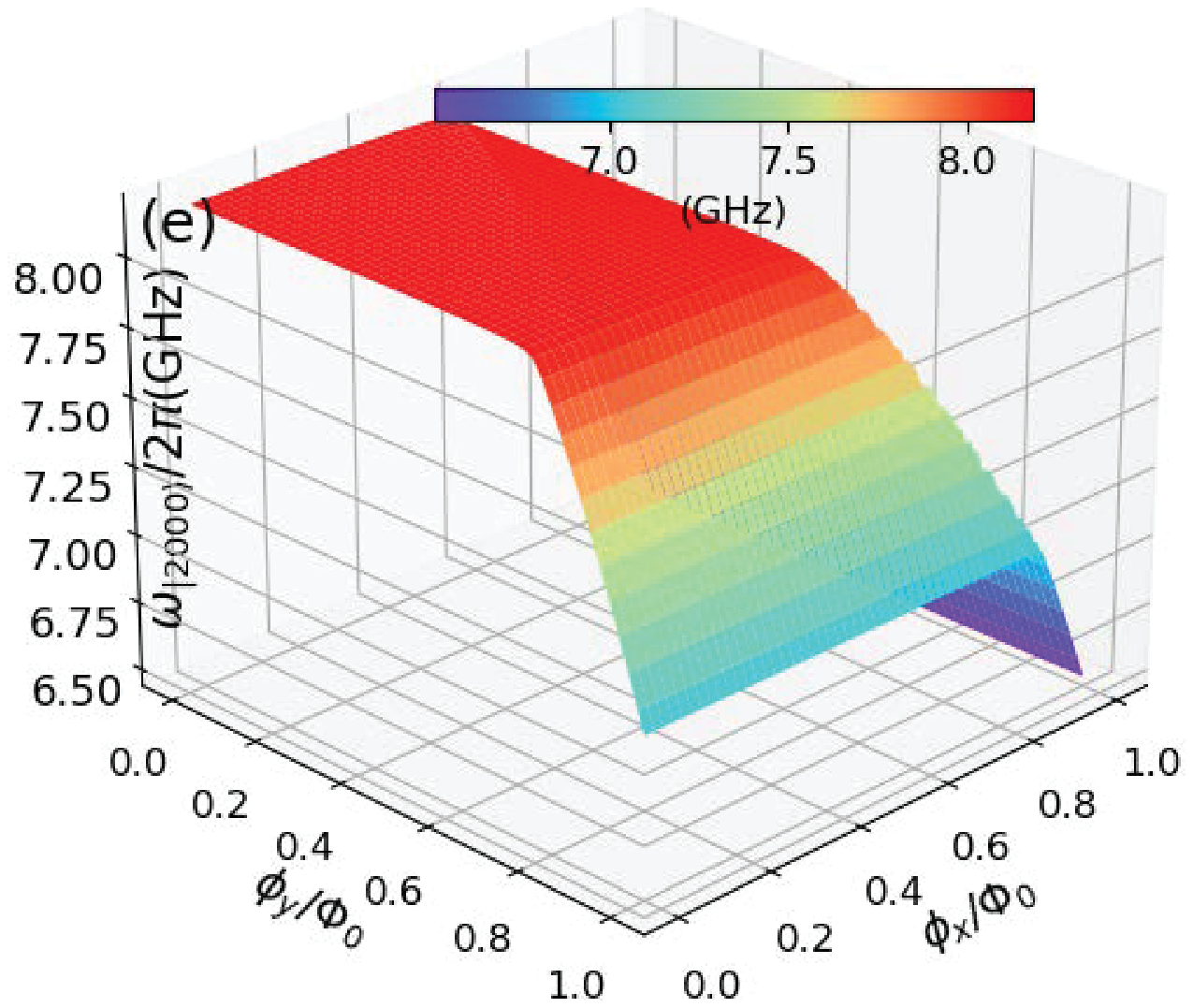}
\includegraphics[bb=0 0 375 300, width=2.7cm, clip]{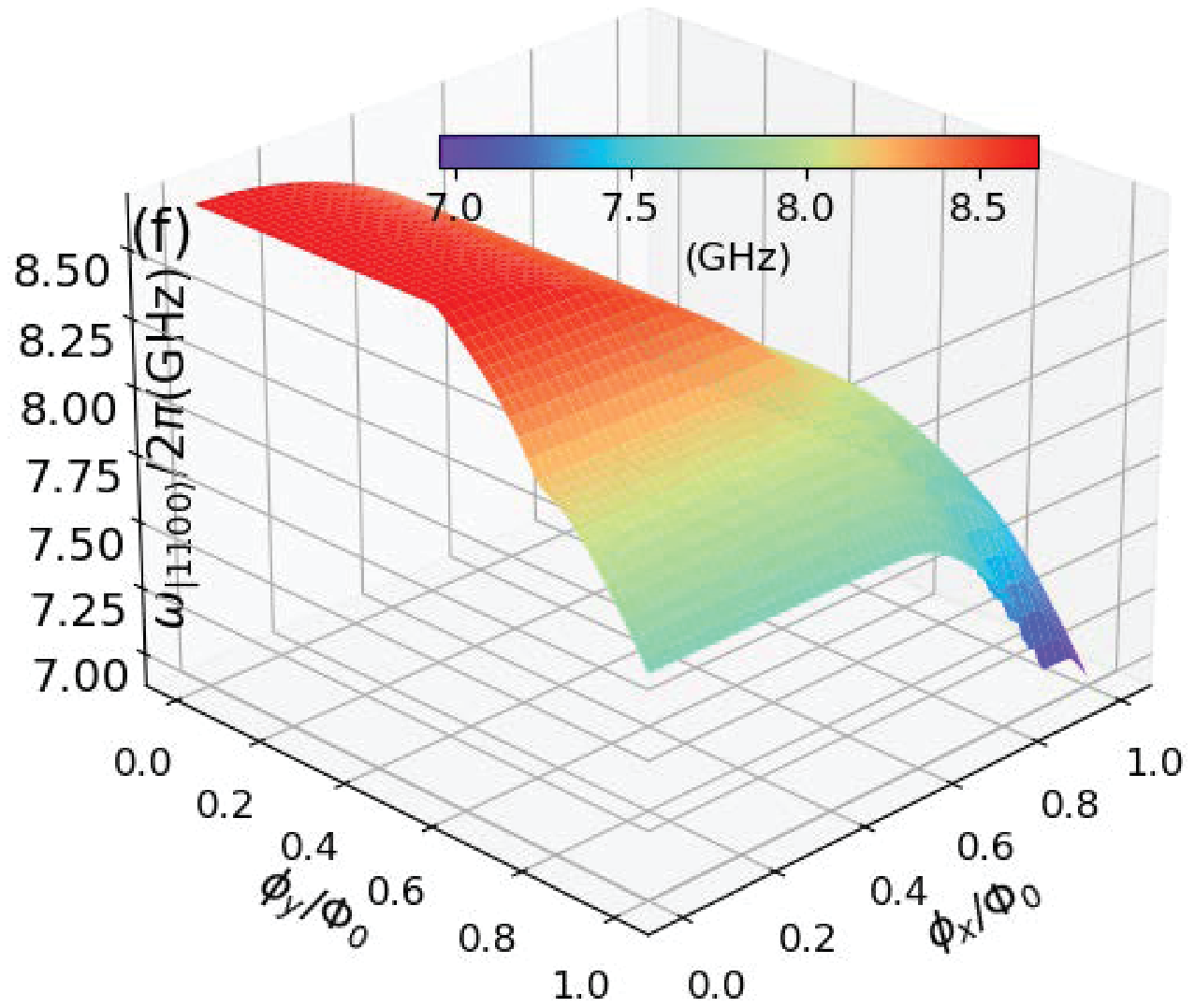}\\
\includegraphics[bb=0 0 375 300, width=2.7cm, clip]{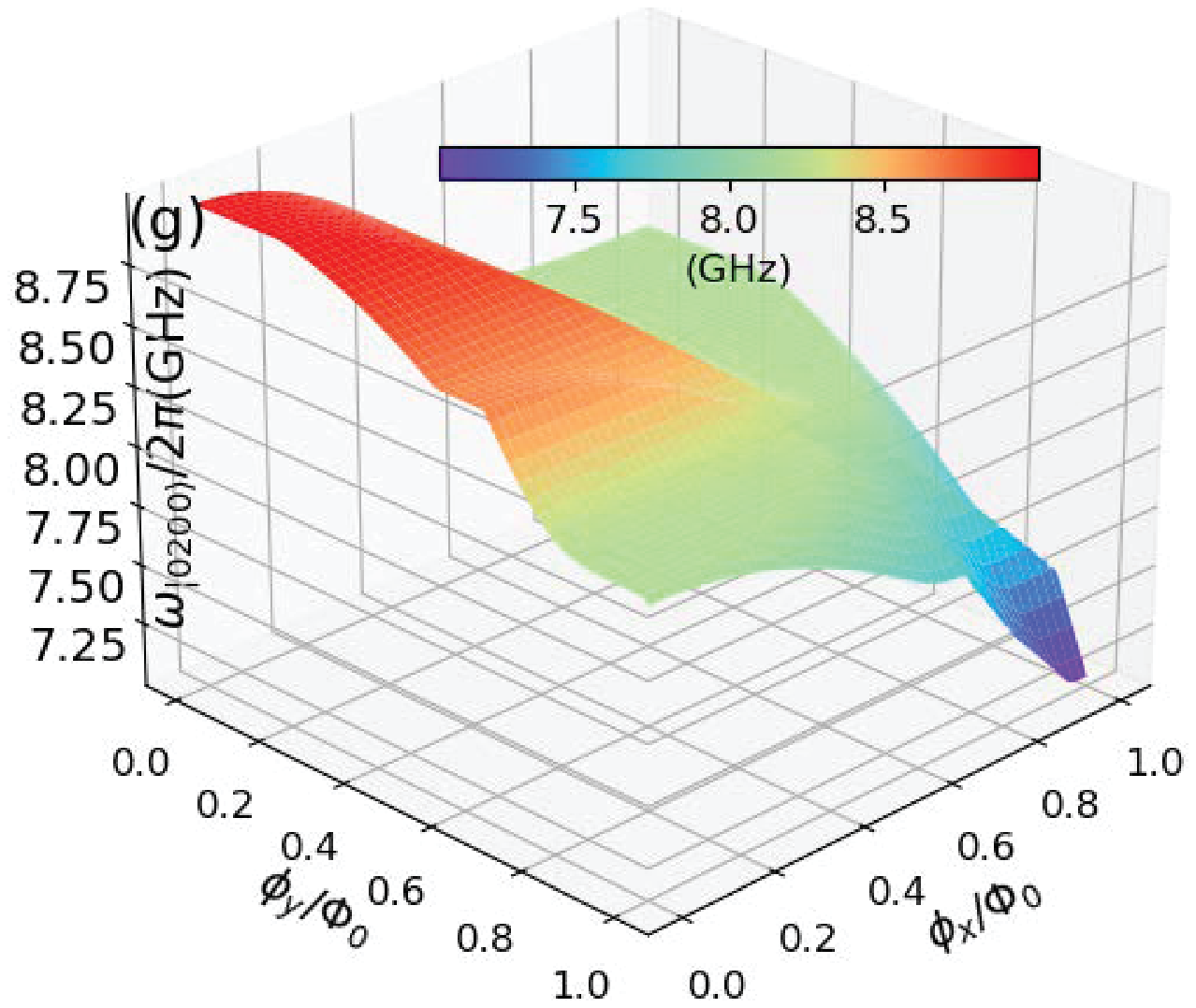}
\includegraphics[bb=0 0 375 300, width=2.7cm, clip]{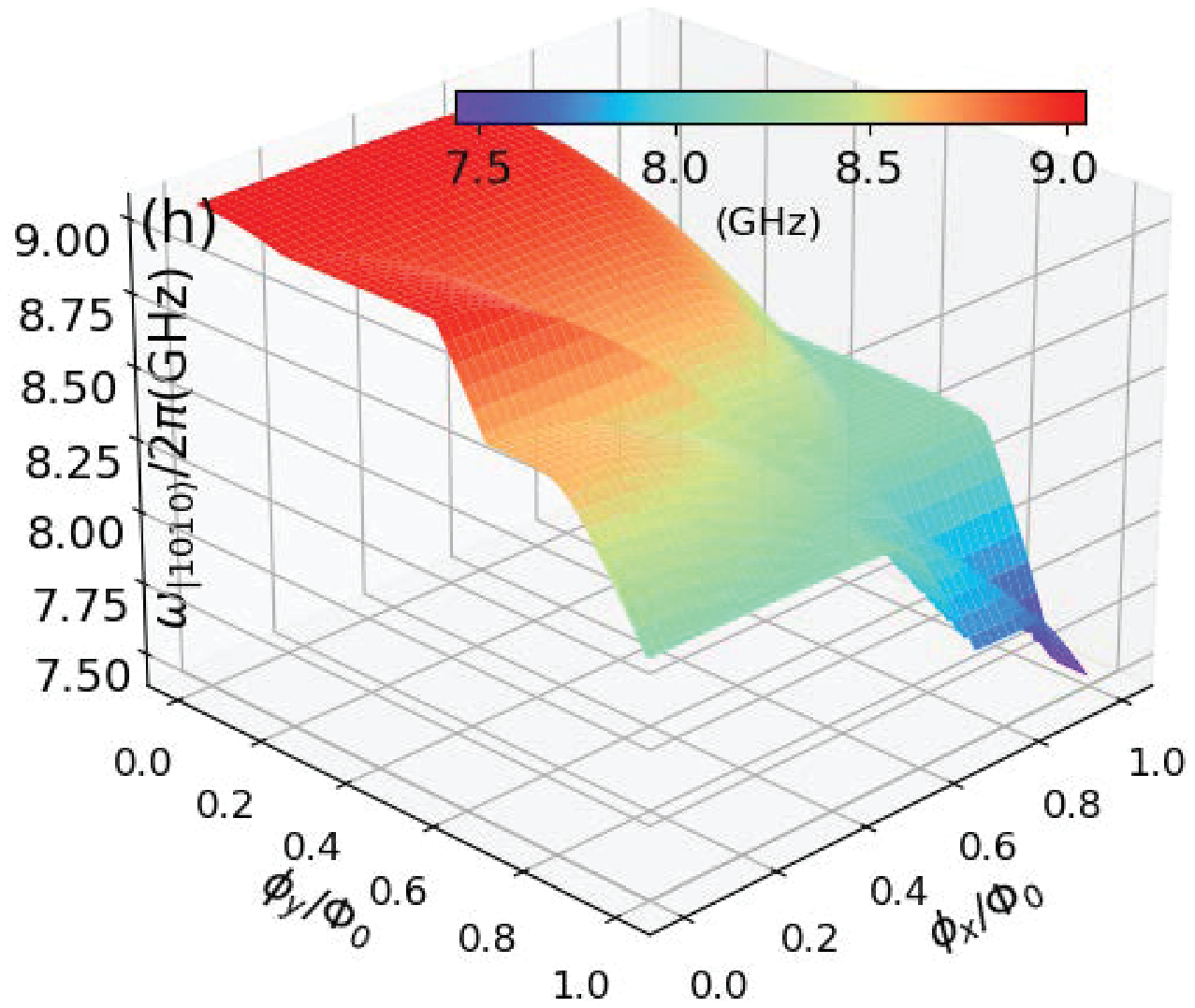}
\includegraphics[bb=0 0 375 300, width=2.7cm, clip]{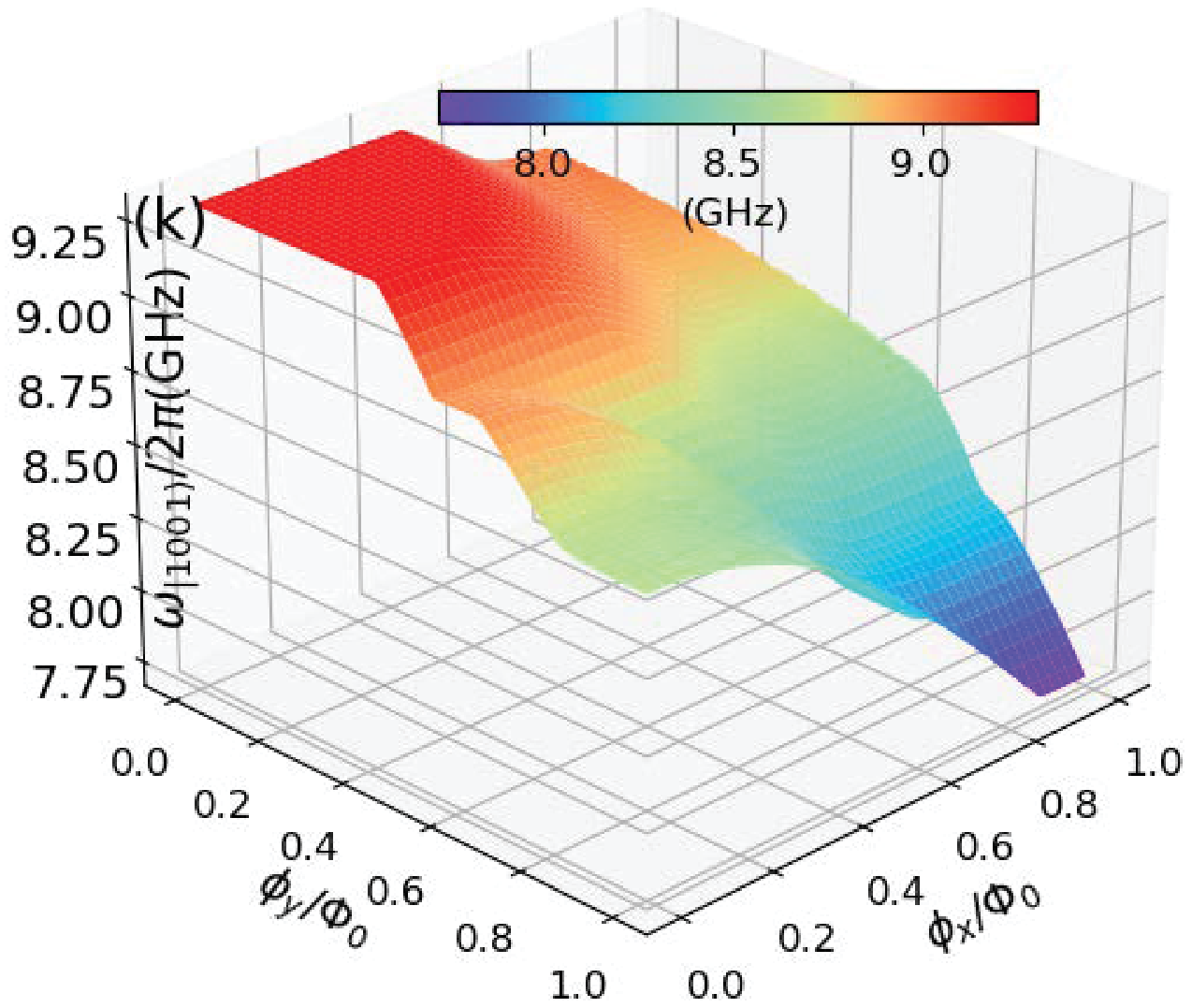}\\
\includegraphics[bb=0 0 375 300, width=2.7cm, clip]{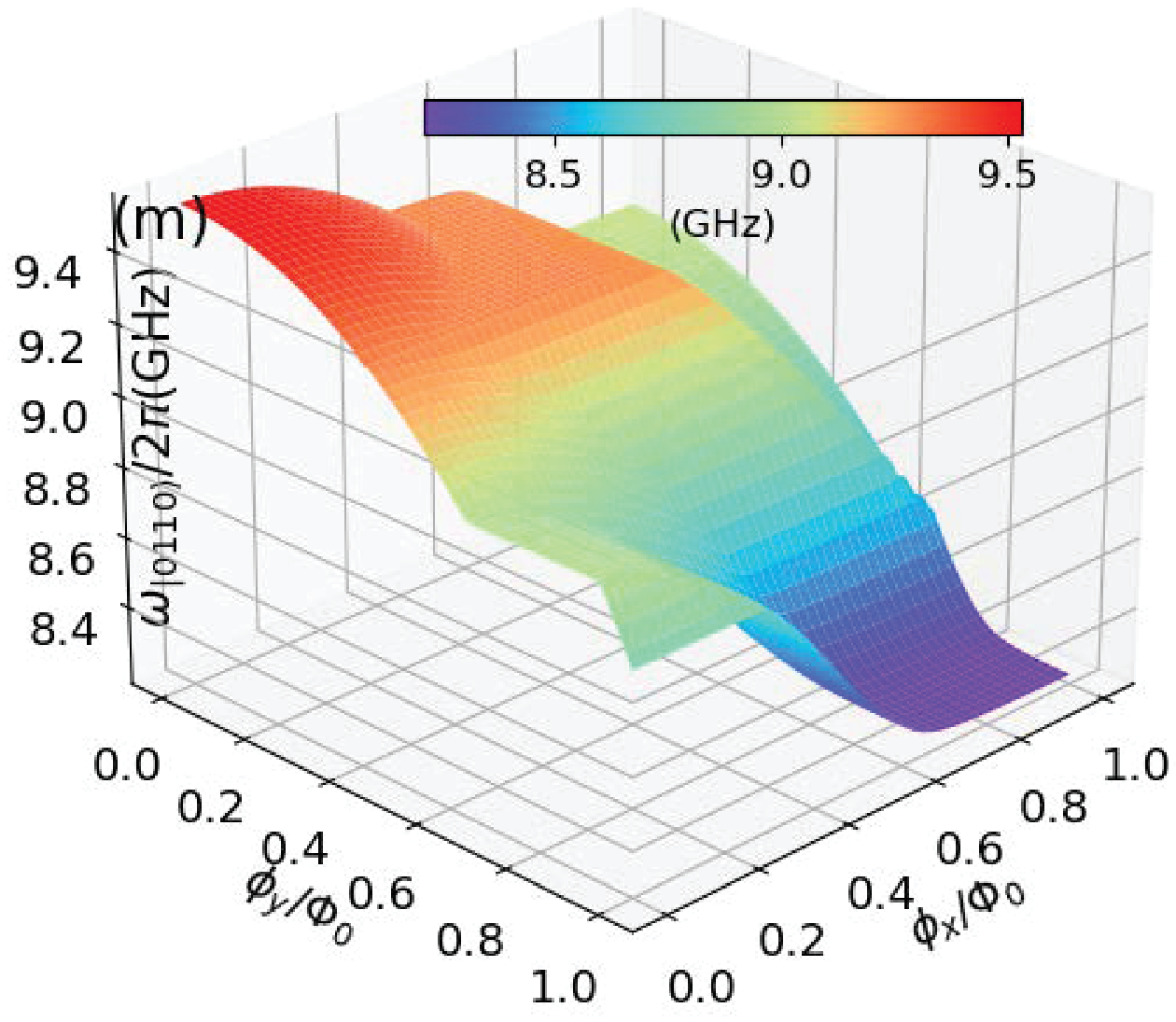}
\includegraphics[bb=0 0 375 300, width=2.7cm, clip]{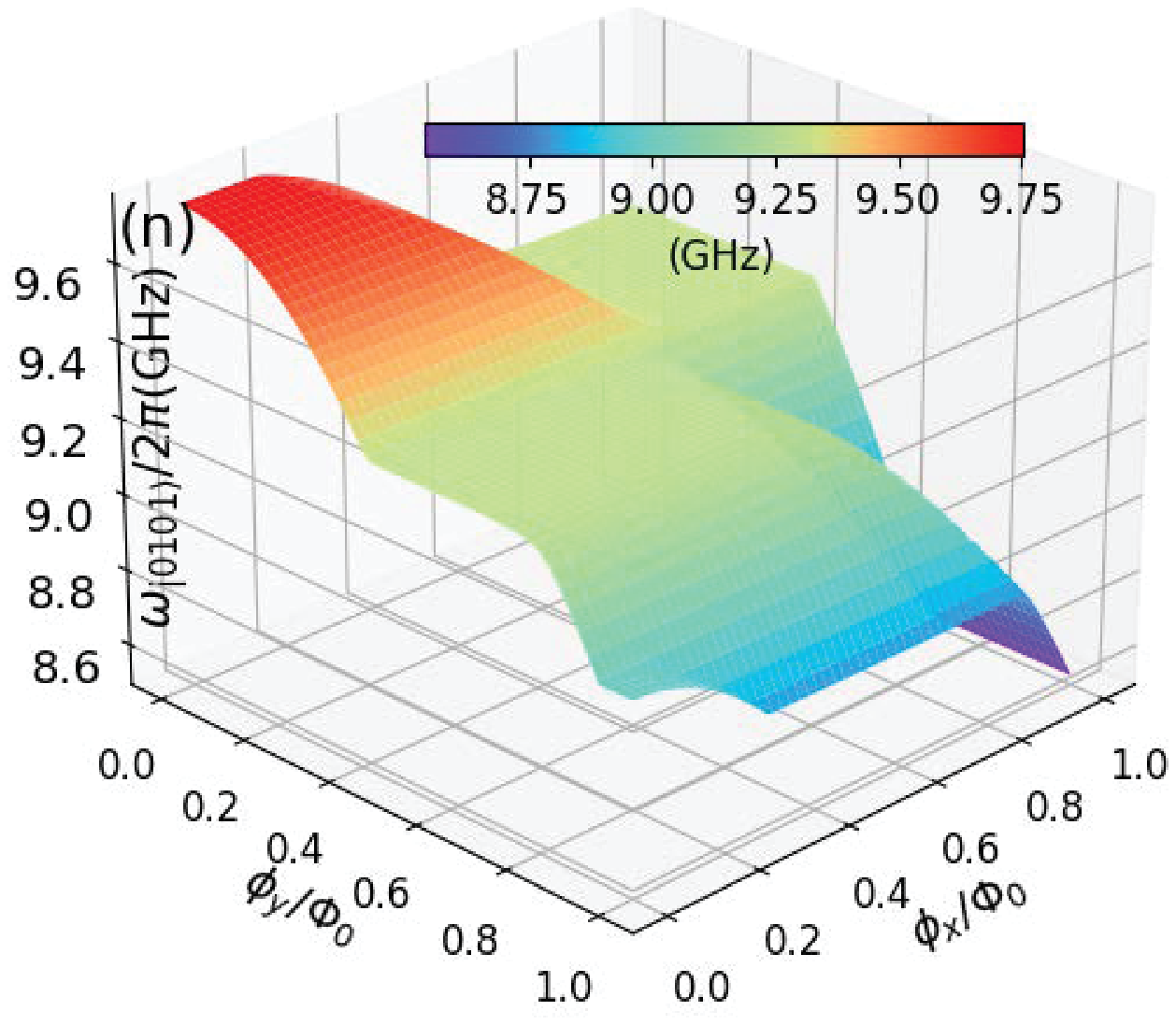}
\includegraphics[bb=0 0 375 300, width=2.7cm, clip]{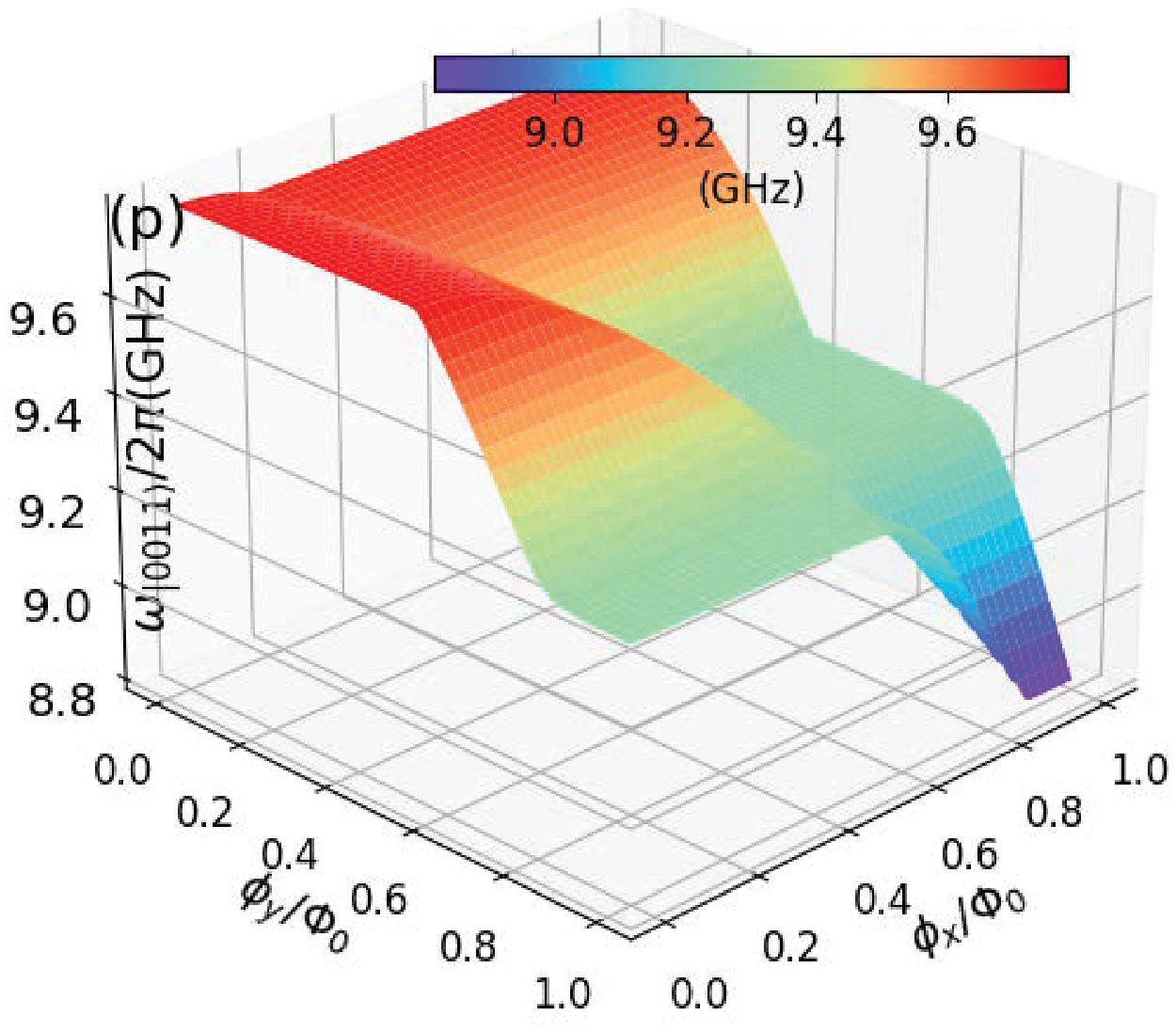}\\
\includegraphics[bb=0 0 375 300, width=2.9cm, clip]{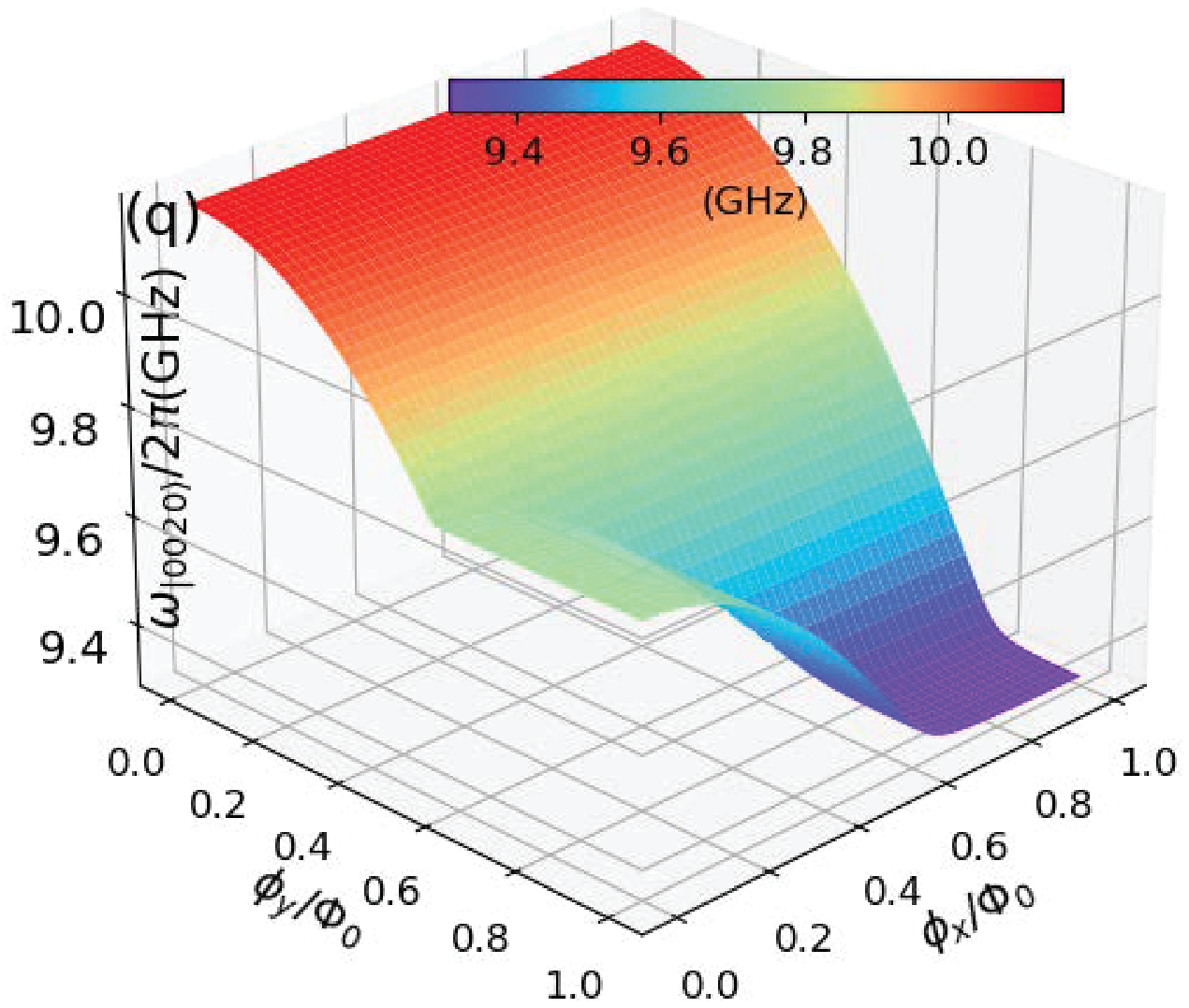}
\includegraphics[bb=0 0 375 300, width=2.9cm, clip]{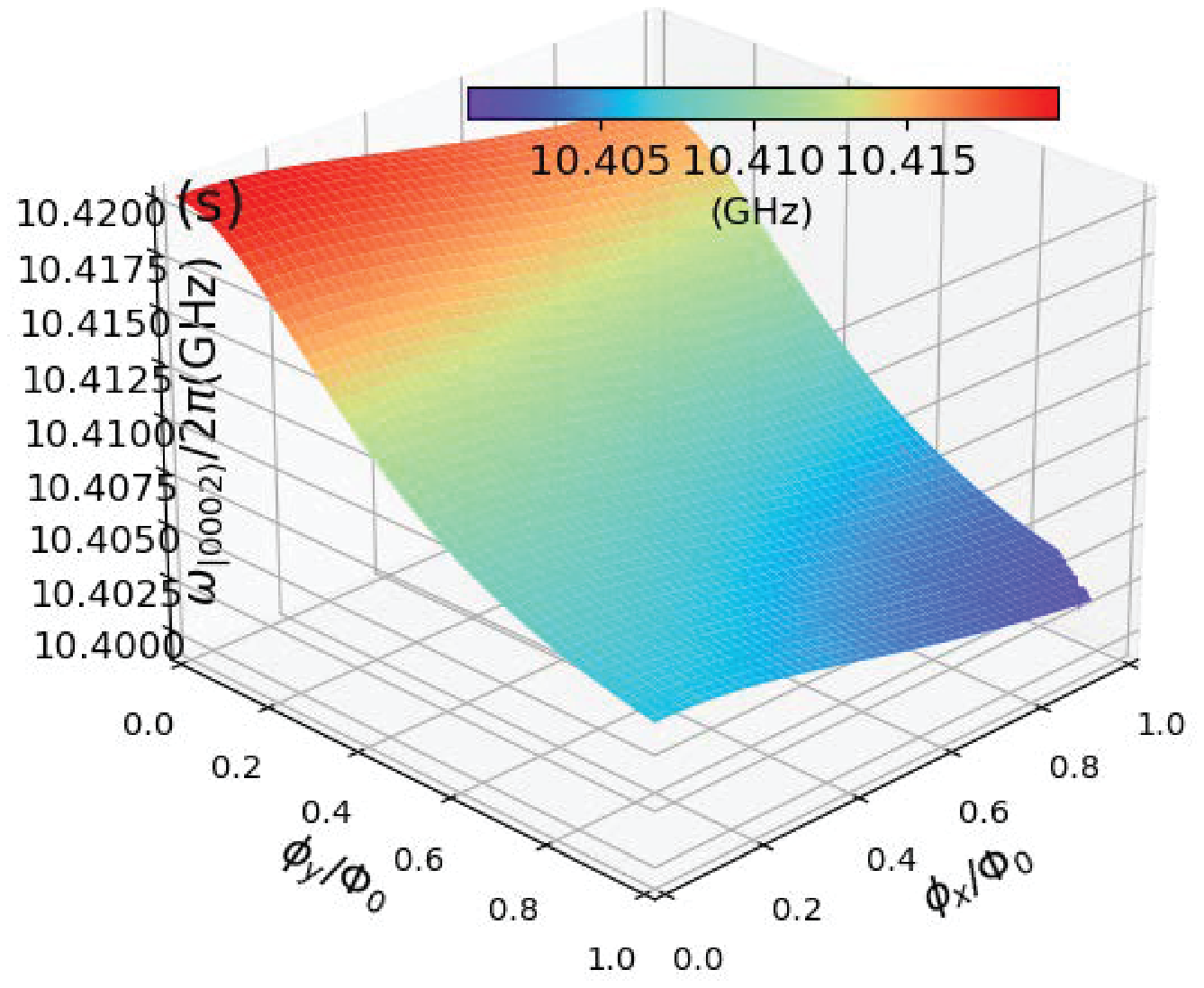}
\caption{(Color online) The two-dimensional surfaces of energy levels. The curved surfaces for single-excited state in the circuit are shown in (a)-(d),
and  (e)-(s) describe the energy level curved surfaces of double-excited states.
The maximal frequencies of two qubits are: $\omega^{(max)}_x/2\pi=4.56$ GHz, and $\omega^{(max)}_y/2\pi=5.12$ GHz.
The other parameters are :  $g_{xy}/2\pi=1$ MHz, $g_{ab}/2\pi=0.1$ MHz, $\omega_a/2\pi=4.10$ GHz,
$\omega_b/2\pi=5.20$ GHz, $\alpha_x/2\pi=-175$ MHz, $\alpha_y/2\pi=-195$ MHz,
$g_{ax}/2\pi=g_{ay}/2\pi=32$ MHz, and $g_{bx}/2\pi=g_{by}/2\pi=30 $ MHz.
}
\label{fig10}
\end{figure}

In this section, we use the numerical method to calculate the two-dimensional energy level curved surfaces of the double-resonator couplers circuit (Fig.~\ref{fig1}). Since the anharmonicities of Xmon qubit is very small,  we regard the superconducting artificial atom  as a multi-energy levels system in this section\cite{Ferguson,Sung}.
The Hamiltonian for the  circuit in Fig.~\ref{fig1} can be written as
\begin{eqnarray}\label{eq:13}
& &H_{m}/\hbar=\sum_{\lambda=a,b}\omega_{\lambda} c^{\dagger}_{\lambda}c_{\lambda} \nonumber\\
&+&\sum_{j_{\beta}}\omega_{j_{\beta}j^{\prime}_{\beta}} J^{z}_{j_{\beta}j^{\prime}_{\beta}}+g_{ab}\left(c^{\dagger}_a c_b+c^{\dagger}_b c_a\right)\nonumber\\
&+ &\sum_{\lambda=a,b ; \beta=x,y \atop  j_{\beta}, j^{\prime}_{\beta}=0,1,2,...} g^{j_{\beta},j^{\prime}_{\beta}}_{\lambda}\left(J^{+}_{j^{\prime}_{\beta}j_{\beta}} +J^{-}_{j_{\beta}j^{\prime}_{\beta}}\right)\left(c_{\lambda}+c^{\dagger}_{\lambda}\right)\\
&+ & \sum_{j_x,j^{\prime}_x,j^{\prime\prime}_y,j^{\prime\prime\prime}_y \atop=0,1,2,...} g^{j_x j^{\prime}_x,j^{\prime\prime}_y j^{\prime\prime\prime}_y}_{xy}\left(J^{+}_{j^{\prime}_x j_x}J^{-}_{j^{\prime\prime}_y j^{\prime\prime\prime}_y}+J^{+}_{j^{\prime\prime\prime}_y j^{\prime\prime}_y}J^{-}_{j_x j^{\prime}_x}\right),\nonumber
\end{eqnarray}
where $j_{\beta},j^{\prime}_{\beta}, j^{\prime\prime}_{\beta},j^{\prime\prime\prime}_{\beta}=0,1,2,3\cdot\cdot\cdot$, and they respectively label the $j_{\beta}$-th, $j^{\prime}_{\beta}$-th, $j^{\prime\prime}_{\beta}$-th, and $j^{\prime\prime\prime}_{\beta}$-th  quantum states of qubit $\beta$,
with $j^{\prime\prime\prime}_{\beta}>j^{\prime\prime}_{\beta}$ and $j^{\prime}_{\beta}>j_{\beta}$.
 Between the quantum states $|j_{\beta}\rangle$ and $|j^{\prime}_{\beta}\rangle$, we define the  angular momentum operators  as $\vec{J}_{j_{\beta}j^{\prime}_{\beta}}=[J^{x}_{j_{\beta}j^{\prime}_{\beta}}, J^{y}_{j_{\beta}j^{\prime}_{\beta}}, J^{z}_{j_{\beta}j^{\prime}_{\beta}}]$.
The ladder operators can be introduced by  $J^{+}_{j_{\beta}j^{\prime}_{\beta}}=|j^{\prime}_{\beta}\rangle\langle j_{\beta}|$
and $J^{-}_{j_{\beta}j^{\prime}_{\beta}}=|j_{\beta}\rangle \langle j^{\prime}_{\beta}|$,
thus we can get $J^{z}_{j_{\beta}j^{\prime}_{\beta}}=|j^{\prime}_{\beta}\rangle\langle j^{\prime}_{\beta}|-|j_{\beta}\rangle \langle j_{\beta}|$,  $J^{x}_{j_{\beta}j^{\prime}_{\beta}}=(J^{+}_{j_{\beta}j^{\prime}_{\beta}}+J^{-}_{j_{\beta}j^{\prime}_{\beta}})/2$,  and $J^{y}_{j_{\beta}j^{\prime}_{\beta}}=(J^{+}_{j_{\beta}j^{\prime}_{\beta}}-J^{-}_{j_{\beta}j^{\prime}_{\beta}})/(2i)$.
The corresponding transition frequency between  states $|j_{\beta}\rangle$ and $|j^{\prime}_{\beta}\rangle$ is defined as $\omega_{j_{\beta}j^{\prime}_{\beta}}$,
 the  $ g^{j_{\beta},j^{\prime}_{\beta}}_{\lambda}$ describe the corresponding coupling strengths with resonator coupler $\lambda$.
The $g^{j_x j^{\prime}_x,j^{\prime\prime}_y j^{\prime\prime\prime}_y}_{xy}$ describes the direct coupling strengths between the transition processes of $|j^{\prime}_{x}\rangle \leftrightarrow |j_{x}\rangle$  for qubits \textbf{x}   and   $|j^{\prime\prime\prime}_{y}\rangle \leftrightarrow |j^{\prime\prime\prime\prime}_{y}\rangle$ for qubit \textbf{y}.

With the QuTiP software, we calculate the two-dimensional curved surfaces  for the energy level of single-excited (Figs.~\ref{fig10}(a)-\ref{fig10}(d)) and double-excited  (Figs.~\ref{fig10}(e)-\ref{fig10}(s)) states. During the numerical calculations with the QuTiP software, we truncate to the second excited states  of  qubits and assume  $ g^{j_{\beta},j^{\prime}_{\beta}}_{\lambda}= g^{j^{\prime}_{\beta},j_{\beta}}_{\lambda}= g_{\lambda\beta}$ and $g^{j_x j^{\prime}_x,j^{\prime\prime}_y j^{\prime\prime\prime}_y}_{yx}=g^{j^{\prime}_y j_y,j^{\prime\prime\prime}_x j^{\prime\prime}_x}_{xy}=g_{xy}$.
Because of the anti-crossing effects, each curved surface in Fig.~\ref{fig10} can not  describe a total energy level of certain quantum state,
and we label the Z-axis of each figure by the corresponding state at zero magnetic flux.

\section*{Appendix B: Circuit Quantization}

\setcounter{equation}{0}
\renewcommand\theequation{B.\arabic{equation}}

In this section, we  conduct the quantization for  the superconducting circuit  in Fig.~\ref{fig1}.
 The kinetic energy of the  superconducting circuit  can be obtained as \cite{wang,Wu2}
\begin{eqnarray}\label{eq:14}
T&=&\frac{1}{2}\left(C_a \dot{\phi}^2_a+ C_b \dot{\phi}^2_b+ C_x \dot{\phi}^2_x+ C_y \dot{\phi}^2_y\right)\nonumber\\
&+&\frac{1}{2}C_{ab}\left( \dot{\phi}_a-\dot{\phi}_b\right)^2+\frac{1}{2}C_{xy}\left( \dot{\phi}_x-\dot{\phi}_y\right)^2 \nonumber\\
&+&\frac{1}{2}C_{ax}\left( \dot{\phi}_a-\dot{\phi}_x\right)^2+\frac{1}{2}C_{ay}\left( \dot{\phi}_a-\dot{\phi}_y\right)^2\nonumber\\
&+&\frac{1}{2}C_{bx}\left( \dot{\phi}_b-\dot{\phi}_x\right)^2+\frac{1}{2}C_{by}\left( \dot{\phi}_b-\dot{\phi}_y\right)^2.
\end{eqnarray}
As indicated by Fig.~\ref{fig1}(b),  the self-capacitances of the qubits and resonators is $C_\eta$ , and the relative capacitance between arbitrary two devices is defined as  $C_{\eta\eta^{\prime}}$ ( $C_{\eta\eta^{\prime}}=C_{\eta^{\prime}\eta}$), here $\eta,\eta^{\prime}=a,b,x,y$ with $\eta\neq\eta^{\prime}$.
Here  $\phi_a$ and $\phi_b$ are the respective magnetic fluxes of the circuit nodes of resonators \textbf{a} and \textbf{b}, while $\phi_x$ and $\phi_y$ are respective node fluxes of  qubits \textbf{x} and \textbf{y}.
If we define the vector $ \vec{\phi}=[\phi_a,\phi_b,\phi_x,\phi_y]$, thus the kinetic energy in Eq.(B.1)
can be written as $T=\frac{1}{2}\dot{\vec{\phi}}^{T} C \dot{\vec{\phi}}$, with
\begin{eqnarray}\label{eq:15}
C&=&\left(
 \begin{array}{cccc}
 C_{11}&  -C_{ab} & -C_{ax} & -C_{ay} \\
   -C_{ab}&  C_{22}& -C_{bx} & -C_{by} \\
    -C_{ax}&  -C_{bx}& C_{33} & -C_{xy} \\
     -C_{ay}&  -C_{by}& -C_{xy} & C_{44}. \\
           \end{array}
         \right),
 \end{eqnarray}
where we have defined  the coefficients:  $C_{11}=C_{a}+C_{ab}+C_{ax}+C_{ay}$,
$C_{22}= C_{ab}+C_{b}+C_{bx}+C_{by}$,
$C_{33}=C_{ax}+C_{bx}+C_{x}+C_{xy}$, and
$C_{44}= C_{ay}+C_{by}+C_{xy}+C_{y}$.

The  potential energy for the superconducting circuit  can be written as
\begin{eqnarray}\label{eq:16}
 U&=&\frac{\phi^2_a}{2L_a}+\frac{\phi^2_b}{2L_b}+E_{J_{x}}\left[1-\cos\left(\frac{2\pi}{\Phi_0}\phi_x\right)\right]\nonumber\\
 &+&E_{J_{y}}\left[1-\cos\left(\frac{2\pi}{\Phi_0}\phi_y\right)\right],
 \end{eqnarray}
 where $E_{J_{\beta}}=I_{c\beta}\Phi_0/2\pi$  is the Josephson energy of qubit $\beta$,   $I_{c\beta}$ is the corresponding  critical current, and $\Phi_0=h/2e$ is the flux quantum.

The Lagrangian of the superconducting circuit can be obtained by the definition $L = T-U$, thus the generalized momentum can be defined  as
$q_{\eta}=\partial L/ \partial \dot{\phi}_{\eta}$ ($\eta=a,b,x,y$), and it can be written  in the vector form as $\vec{q}=[q_a, q_b, q_x, q_y]$.
Thus the Hamiltonian of superconducting circuit   can be written as $H=\vec{q}\cdot \dot{\vec{\phi}}-L=\frac{1}{2}\vec{q}^{T}C^{-1}\vec{q}+U$,
 the  inverse matrix is defined as
\begin{eqnarray}\label{eq:17}
C^{-1}&=&\frac{A^{\ast}}{|C|}=\frac{1}{||C||}\left(
 \begin{array}{cccc}
 A_{11}&  A_{21} & A_{31} & A_{41} \\
   A_{12} &  A_{22}& A_{32} & A_{42}\\
   A_{13}&  A_{23}& A_{33} & A_{43}  \\
    A_{14}&  A_{24}& A_{34} & A_{44}.  \\
           \end{array}
         \right).
 \end{eqnarray}
where $A^{\ast}$ is the adjugate matrix of $A$.
With the conditions  $C_{ab}\ll  C_{xy}\ll  C_{ax}, C_{ay}, C_{bx}, C_{by}\ll C_{x}, C_{y} \ll C_{a},C_{b}$, thus $ \| C\|\approx C_{a} C_{b} C_{x} C_{y}$,  we get  approximate expressions for the elements in $A^{\ast}$ as
\begin{eqnarray}\label{eq:18}
A_{11}&=& C_{22}(C_{33}C_{44}-C^2_{xy})+C_{bx}(-C_{bx}C_{44}-C_{xy}C_{by})\nonumber\\
&-&C_{by}(C_{bx}C_{xy}+C_{33}C_{by})\approx C_{b}C_{x}C_{y},\nonumber\\
A_{12}&=&C_{ab}(C_{33}C_{44}-C^2_{xy})-C_{bx}(-C_{ax}C_{44} -C_{xy}C_{ay})\nonumber\\
&+&C_{by}(C_{ax}C_{xy}+C_{33}C_{ay})\nonumber\\
&\approx&  C_{ab}C_{x}C_{y}+C_{ax}C_{bx}C_y+C_{ay}C_{by}C_{x},\nonumber\\
A_{13}&=&C_{ab}(C_{bx}C_{44}+C_{by}C_{xy})+C_{22}(C_{ax}C_{44}+C_{xy}C_{ay})\nonumber\\
&-&C_{by}(C_{ax}C_{by}-C_{bx}C_{ay})\approx   C_{b}C_{y} C_{ax},\nonumber\\
A_{14}&=&C_{ab}(C_{bx}C_{xy}+C_{33}C_{by})+C_{22}(C_{ax}c_{xy}+C_{33}C_{ay})\nonumber\\
&+&C_{bx}(C_{ax}C_{by}-C_{bx}C_{ay})\approx  C_{b}C_{x}C_{ay},\\
A_{21}&=&C_{ab}(C_{33}C_{44}-C^2_{xy})+C_{ax}(C_{bx}C_{44}+C_{xy}C_{by})\nonumber\\
&+&C_{ay}(C_{bx}C_{xy}+C_{33}C_{by})\nonumber\\
&\approx&  C_{ab}C_{x}C_{y}+C_{ax}C_{bx}C_{y}+C_{ay}C_{by}C_{x},\nonumber\\
A_{22}&=&C_{11}(C_{33}C_{44}-C^2_{xy})-C_{ax}(C_{ax}C_{44}+C_{xy}C_{ay})\nonumber\\
&-&C_{ay}(C_{ax}C_{xy}+C_{33}C_{ay})\approx  C_{a}C_{x}C_{y},\nonumber\\
A_{23}&=&C_{11}(C_{bx}C_{44}+C_{xy}C_{by})+C_{ab}(C_{ax}C_{44}+C_{xy}C_{ay})\nonumber\\
&+& C_{ay}(C_{ax}C_{by}-C_{bx}C_{ay})\approx  C_{bx}C_{a} C_{y},\nonumber\\
A_{24} &=&C_{11}(C_{bx}C_{xy}+C_{33}C_{by})+C_{ab}(C_{ax}C_{xy}+C_{33}C_{ay})\nonumber\\
&-&C_{ax}(C_{ax}C_{by}-C_{bx}C_{ay})\approx  C_{a}C_{x}C_{by},\nonumber\\
A_{31} &=&-C_{ab}(-C_{bx}C_{44}-C_{by}C_{xy})+C_{ax}(C_{22}C_{44}-C^2_{by})\nonumber\\
&-&C_{ay}(-C_{22}C_{xy}-C_{bx}C_{by})\approx  C_{ax} C_{b}c_{y},\nonumber\\
A_{32}&=&C_{11}(C_{bx}C_{44}+C_{by}C_{xy})+C_{ax}(C_{ab}C_{44}+C_{ay}C_{by})\nonumber\\
&+&C_{ay}(C_{ab}C_{xy}-C_{bx}C_{by})\approx   C_{bx} C_{a}C_{y},\nonumber\\
A_{33}&=&C_{11}(C_{22}C_{44}-C^2_{by})+C_{ab}(-C_{ab}C_{44}-C_{ay}C_{by})\nonumber\\
&-&C_{ay}(C_{ab}C_{by}+C_{22}C_{ay})\approx  C_{a}C_{b}C_{y},\nonumber\\
A_{34}&=&C_{11}(C_{22}C_{xy}+C_{bx}C_{by})-C_{ab}(C_{ab}C_{xy}-C_{bx}C_{ay})\nonumber\\
&+&C_{ax}(C_{ab}C_{by}+C_{22}C_{ay})\nonumber\\
&\approx&  C_{xy}C_{a}C_{b}+C_{a}C_{bx}C_{by}+C_{b}C_{ax}C_{ay},\nonumber\\
A_{41}&=&C_{ab}(C_{bx}C_{xy}+C_{by}C_{33})+C_{ax}(C_{22}C_{xy}+C_{bx}C_{by})\nonumber\\
&+&C_{ay}(C_{22}C_{33}-C^2_{bx})\approx C_{ay} C_{b}C_{x},\nonumber\\
A_{42}&=&C_{11}(C_{bx}C_{xy}+C_{by}C_{33})+C_{ax}(C_{ab}C_{xy}-C_{ax}C_{by})\nonumber\\
&-&C_{ay}(-C_{ab}C_{33}-C_{ax}C_{bx})\approx  C_{a}C_{x}C_{by},\nonumber\\
A_{43}&=&C_{11}(C_{22}C_{xy}+C_{bx}C_{by})+C_{ab}(C_{ab}C_{xy}+C_{ax}C_{by})\nonumber\\
&+&C_{ay}(C_{ab}C_{bx}+C_{22}C_{ax})\nonumber\\
&\approx&  C_{xy}C_{a}C_{b}+C_{a}C_{bx}C_{by}+C_{b}C_{ax}C_{ay},\nonumber\\
A_{44}  &=&C_{11}(C_{22}C_{33}-C^2_{bx})+C_{ab}(-C_{ax}c_{33}-C_{ax}C_{bx})\nonumber\\
&-&C_{ax}(C_{ax}C_{bx}+C_{22}C_{ax})\approx  C_{a}C_{b}C_{x}.\nonumber
\end{eqnarray}

Thus the Hamiltonian of double-resonator couplers circuit  can be expressed as
 \begin{eqnarray}\label{eq:19}
 H&=&4 E_{C_a}(n_a)^2+4 E_{C_b}(n_b)^2+4 E_{C_x}(n_x)^2+4 E_{C_y}(n_y)^2\nonumber\\
 &+&\frac{\phi^2_a}{2L_a}+\frac{\phi^2_b}{2L_b}-E_{J_x}\cos\left(\frac{2\pi}{\Phi_0}\phi_x\right)
-E_{J_y}\cos\left(\frac{2\pi}{\Phi_0}\phi_y\right)\nonumber\\
 &+&8\frac{C_{ax}}{\sqrt{C_{a}C_{x}}}\sqrt{E_{C_a}E_{C_x}}(n_a n_x)\nonumber\\
 &+&8\frac{C_{ay}}{\sqrt{C_{a}C_{y}}}\sqrt{E_{C_a}E_{C_y}}(n_a n_y)\\
 &+&8\frac{C_{bx}}{\sqrt{C_{b}C_{x}}}\sqrt{E_{C_b}E_{C_x}}(n_b n_x)\nonumber\\
 &+&8\frac{C_{by}}{\sqrt{C_{b}C_{y}}}\sqrt{E_{C_b}E_{C_y}}(n_b n_y)\nonumber\\
 &+&8\left(1+\frac{C_{ax}C_{bx}}{C_{x}C_{ab}}+\frac{C_{ay}C_{by}}{C_{y}C_{ab}}\right)\frac{C_{ab}}{\sqrt{C_{a}C_{b}}}\sqrt{E_{C_a}E_{C_b}}(n_a n_b)\nonumber\\
 &+&8\left(1+\frac{C_{ax}C_{ay}}{C_{a}C_{xy}}+\frac{C_{bx}C_{by}}{C_{b}C_{xy}}\right)\frac{C_{xy}}{\sqrt{C_{x}C_{y}}}\sqrt{E_{C_x}E_{C_y}}(n_x n_y).\nonumber
 \end{eqnarray}
The two-body  coupling strengths can be defined as
\begin{eqnarray}\label{eq:20}
g_{\lambda\beta}&=&\frac{1}{2}\frac{C_{\lambda\beta}}{\sqrt{C_{\lambda}C_{\beta}}}\sqrt{\omega_{\lambda}\omega_{\beta}},\\
g_{ab}&=&\frac{1}{2}\left(1+\frac{C_{ax}C_{bx}}{C_{x}C_{ab}}+\frac{C_{ay}C_{by}}{C_{y}C_{ab}}\right)\frac{C_{ab}}{\sqrt{C_{a}C_{b}}}\sqrt{\omega_a \omega_b},\qquad\\
g_{xy}&=&\frac{1}{2}\left(1+\frac{C_{ax}C_{ay}}{C_{a}C_{xy}}+\frac{C_{bx}C_{by}}{C_{b}C_{xy}}\right)\frac{C_{xy}}{\sqrt{C_{x}C_{y}}}\sqrt{\omega_x \omega_y}.\qquad
\end{eqnarray}
The qubit-resonator interaction terms $g_{\lambda\beta}$ in Eq.(B.9) could induce indirect coupling between two qubits, which should be decoupled to obtained the effective qubit-qubit coupling.

\section*{Appendix C: Decoupling processes}

\setcounter{equation}{0}
\renewcommand\theequation{C.\arabic{equation}}

The Josephson  energy of  Xmon qubit is much larger than its capacitance energy,
 $E_{J_{\beta}}/E_{C_{\beta}}\gg 1$, and then we can approximately get $\cos(\phi_\beta)=1-\phi^2_\beta/2+\phi^4_\beta/24-...$.
 If we introduce the creation and annihilation operators by  the definitions: $\phi_\beta=\sqrt[4]{(2E_C/E_{J_{\beta}})}(a^{\dagger}_\beta+a_\beta)$ and $n_\beta=(i/2)\sqrt[4]{(2E_C/E_{J_{\beta}})}(a^{\dagger}_\beta-a_\beta)$,
then second-quantized    Hamiltonian can be obtained as $H_{tot}=\sum_{\lambda=a,b}{H_{\lambda}}+\sum_{\beta=x,y}{H_{\beta}}+\sum_{\lambda=a,b \atop \beta=x,y}H_{\lambda\beta}+H^{(r)}_{ab}+H^{(q)}_{xy}$, with
\begin{eqnarray}\label{eq:21}
H_{\lambda}&=&\frac{1}{2}\hbar\omega_{\lambda} c^{\dagger}_{\lambda} c_{\lambda},\\
H_{\beta}
&=&\frac{1}{2}\hbar\omega_{\beta} a^{\dagger}_{\beta} a_{\beta}+\frac{\alpha_{\beta}}{2} a^{\dagger}_{\beta} a^{\dagger}_{\beta} a_{\beta} a_{\beta},\\
H_{\lambda\beta}&=&\hbar g_{\lambda\beta}(c^{\dagger}_{\lambda}a_{\beta}+c_{\lambda}a^{\dagger}_{\beta}-c^{\dagger}_{\lambda}a^{\dagger}_{\beta}-c_{\lambda}a_{\beta}), \\
H^{(r)}_{ab}&=&\hbar g_{ab}(c^{\dagger}_{a}c_{b}+c_{a}c^{\dagger}_{b}-c^{\dagger}_{a}c^{\dagger}_{b}-c_{a}c_{b}), \\
H^{(q)}_{xy}&=&\hbar g_{xy}(a^{\dagger}_{x}a_{y}+a_{x}a^{\dagger}_{y}-a^{\dagger}_{x}a^{\dagger}_{y}-a_{x}a_{y}).
\end{eqnarray}
The transition frequencies of resonators and qubits are respectively defined as $\omega_{\lambda}=1/\sqrt{C_{\lambda} L_{\lambda}}$ and
$\omega_{\beta}=(\sqrt{8 E_{J_{\beta}}E_{C_{\beta}}}-E_{C_{\beta}})/\hbar$,
while the $\alpha_{\beta}=-E_{C_{\beta}}/\hbar$ describe the anharmonicity of qubit $\beta$.

In the  qubit-resonator dispersive coupling regimes, $g_{\lambda\beta}/|\Delta_{\lambda\beta}|\ll 1$, we can use the Schrieffer-Wolf transformation to decouple the variables of qubits and resonators. Since the resonator couplers are not pumped by the external fields, and the average cavity photon number should be much smaller than one, so the virtual photon exchanges will dominate the cross-kerr resonances.
We define  $S=\sum_{\lambda=a,b \atop \beta=x,y}[(g_{\lambda\beta}/\Delta_{\lambda\beta}) (c^{\dagger}_{\lambda} a_{\beta}-c_{\lambda} a^{\dagger}_{\beta})-(g_{\lambda\beta}/\Sigma_{\lambda\beta}) (c^{\dagger}_{\lambda} a^{\dagger}_{\beta}-c_{\lambda} a_{\beta})]$.
Under the Unitary transformation $H^{(d)}=\exp(-S)H\exp(S)$,  if we choose  $H_{\lambda\beta}+[H_{0}, S]=0$,
thus the decoupled Hamiltonian can be obtained as
\begin{eqnarray}\label{eq:22}
H^{(d)}&=&\hbar\omega^{(d)}_{a}c^{\dagger}_{a}c_{a}+\hbar\omega^{(d)}_{b}c^{\dagger}_{b}c_{b}+ \hbar \omega^{(d)}_{x}a^{\dagger}_{x}a_{x}+ \hbar \omega^{(d)}_{y}a^{\dagger}_{y}a_{y}\nonumber\\  &+&\frac{\hbar\tilde{\alpha}_x}{2}a^{\dagger}_{x}a^{\dagger}_{x}a_{x}a_{x}+\frac{\hbar\tilde{\alpha}_y}{2}a^{\dagger}_{y}a^{\dagger}_{y}a_{y}a_{y}\nonumber\\
&+ &\hbar g^{(d)}_{xy}(a^{\dagger}_{x}a_{y}+a^{\dagger}_{y}a_{x})+\hbar g^{(d)}_{ab}(c^{\dagger}_{a}c_{b}+c^{\dagger}_{b}c_{a}).
\end{eqnarray}
The rotating wave approximation has been used to derive the above formula, and the  constant terms were neglected.  We also neglected the effects of  small quantities ($H_{xy}$ and $H_{ab}$).
 The anharmonicities of qubits are considered as invariant during Unitary transformation ( $\tilde{\alpha}_{\beta} \approx \alpha_{\beta}$), so some high-order effects relating to high-excited states of superconducting artificial atoms are neglected.

 The transition frequencies of qubits, the resonant frequencies of resonators, the  qubit-qubit coupling strength, and the resonator-resonator coupling strength in decoupled coordinate are  obtained as
\begin{eqnarray}\label{eq:23}
\omega^{(d)}_{x}&=&\omega_{x}+\left(\frac{g^2_{a x}}{\Delta_{a x}}+\frac{g^2_{b x}}{\Delta_{b x}}-\frac{g^2_{a x}}{\Sigma_{a x}}-\frac{g^2_{b x}}{\Sigma_{b x}}\right),\\
\omega^{(d)}_{y}&=&\omega_{y}+\left(\frac{g^2_{a y}}{\Delta_{a y}}+\frac{g^2_{b y}}{\Delta_{b y}}-\frac{g^2_{a y}}{\Sigma_{a y}}-\frac{g^2_{b y}}{\Sigma_{b y}}\right),\\
\omega^{(d)}_{a}&=&\omega_{a}-\left(\frac{g^2_{a x}}{\Delta_{a x}}+\frac{g^2_{a y}}{\Delta_{a y}}-\frac{g^2_{a x}}{\Sigma_{a x}}-\frac{g^2_{a y}}{\Sigma_{a y}}\right),\\
\omega^{(d)}_{b}&=& \omega_{b}-\left(\frac{g^2_{b x}}{\Delta_{b x}}+\frac{g^2_{b y}}{\Delta_{b y}}-\frac{g^2_{b x}}{\Sigma_{b x}}-\frac{g^2_{b y}}{\Sigma_{b y}}\right),\\
g^{(d)}_{xy}&=&\frac{1}{2}\bigg(\frac{g_{a x}g_{a y}}{\Delta_{a y}}+\frac{g_{b x}g_{b y}}{\Delta_{b y}}+\frac{g_{a y}g_{a x}}{\Delta_{a x}}+\frac{g_{b y}g_{b x}}{\Delta_{b x}}\\
&-&\frac{g_{a x}g_{a y}}{\Sigma_{a y}}-\frac{g_{b x}g_{b y}}{\Sigma_{b y}}-\frac{g_{a y}g_{a x}}{\Sigma_{a x}}-\frac{g_{b y}g_{b x}}{\Sigma_{b x}}\bigg) +g_{xy},\nonumber\\
g^{(d)}_{ab}&=&\frac{1}{2}\bigg(\frac{g_{a x}g_{b x}}{\Delta_{b x}}+\frac{g_{a y}g_{b y}}{\Delta_{b y}}+\frac{g_{b x}g_{a x}}{\Delta_{a x}}+\frac{g_{b y}g_{a y}}{\Delta_{a y}}\\
&-&\frac{g_{a x}g_{b x}}{\Sigma_{b x}}-\frac{g_{a y}g_{b y}}{\Sigma_{b y}}-\frac{g_{b x}g_{a x}}{\Sigma_{a x}}-\frac{g_{b y}g_{a y}}{\Sigma_{a y}}\bigg)+g_{ab}.\nonumber
\end{eqnarray}
The $g^{(d)}_{xy}$ is the decoupled qubit-qubit coupling strength which can be used to analyze the switching off. The $g^{(d)}_{ab}$ is the  decoupled resonator-resonator and much smaller than the frequency detuning between two resonators ($g^{(d)}_{ab}\ll |\Delta_{xy}|$), thus its contributions are neglected in this article.

\section*{Appendix D:Calculations of High-energy level corrections}

\setcounter{equation}{0}
\renewcommand\theequation{D.\arabic{equation}}

In the current theoretical model, the kerr-nonlinear terms $H_{nl,\beta}=(\alpha_{\beta}/2) a^{\dagger}_{\beta} a^{\dagger}_{\beta} a_{\beta} a_{\beta}$ term are assumed  bo be  invariant ($\tilde{\alpha}_{\beta} \approx \alpha_{\beta}$) during the derivations of Eqs.(C.6).
This means that  some high-order effects and the contributions of high excited state of superconducting artificial atom are neglected, so the decoupled frequency and effective qubit-qubit coupling  in Eqs.(C.7)-(C.12) contain no information of qubits' anharmoncities.  But anharmoncity of  Xmon qubit is very small, thus the resonator  can couple to the high-excited states of  atoms, which should affect the transition frequencies of qubits and the effective qubit-qubit coupling strengths. As discussed by some theoretical work, the anharmonicity could induced fourth-order  self-kerr  and cross-kerr resonances, these effects could create corrections to the qubit's energy levels\cite{Blais,Ferguson}.

Bogoliubov transformation is used to analyze the higher-order effect during the  decoupling processes for the qubit-resonator interactions\cite{Blais}. To maintain consistency, in this article we still use the Schrieffer-Wolf transformation to analyze the contributions of the kerr-nonlinear terms $H_{nl,\beta}$  during the decoupling process.   In the qubit-resonator dispersive coupling regimes, $(g_{\lambda \beta}/\Delta_{\lambda \beta})\ll 1$ and $(g_{\lambda \beta}/\Sigma_{\lambda \beta}) \ll 1$,  we define
$S_{\lambda\beta}=(g_{\lambda\beta}/\Delta_{\lambda\beta}) (c^{\dagger}_{\lambda} a_{\beta}-c_{\lambda} a^{\dagger}_{\beta})-(g_{\lambda\beta}/\Sigma_{\lambda\beta}) (c^{\dagger}_{\lambda} a^{\dagger}_{\beta}-c_{\lambda} a_{\beta})$, here $S=\sum_{\lambda=a,b \atop \beta=x,y}S_{\lambda\beta}$.
Since $H_{nl,\beta}$ is a  small quantity, we can separately conduct the Unitary transform $H^{\prime}_{nl,\beta}=\exp(S)H_{nl,\beta}\exp(-S)$ \cite{Blais}, thus we get
\begin{eqnarray}\label{eq:24}
H^{\prime}_{nl,\beta}&=&H_{nl,\beta}+[S,H_{nl,\beta}]+\frac{1}{2!}[S,[S,H_{nl,\beta}]]\nonumber\\
&+&\frac{1}{3!}[S[S,[S,H_{nl,\beta}]]]+....
\end{eqnarray}

The  commutation relation for the first-order expansion term,
\begin{eqnarray}\label{eq:25}
& &[S_{\lambda\beta},H_{nl,\beta}]\nonumber\\
&= &\left[\frac{g_{\lambda \beta}}{\Delta_{\lambda \beta}} (c^{\dagger}_{\lambda} a_{\beta}-c_{\lambda} a^{\dagger}_{\beta})-\frac{g_{\lambda \beta}}{\Sigma_{\lambda \beta}} (c^{\dagger}_{\lambda} a^{\dagger}_{\beta}-c_{\lambda} a_{\beta}),\frac{\alpha_{\beta}}{2} a^{\dagger}_{\beta} a^{\dagger}_{\beta} a_{\beta} a_{\beta}\right]\nonumber\\
&=&\frac{g_{\lambda \beta}\alpha_{\beta}}{\Delta_{\lambda \beta}}\left(c^{\dagger}_{\lambda}a^{\dagger}_{\beta} a_{\beta} a_{\beta}+c_{\lambda} a^{\dagger}_{\beta} a^{\dagger}_{\beta} a_{\beta}\right)\nonumber\\
&+&\frac{g_{\lambda \beta}\alpha_{\beta}}{\Sigma_{\lambda \beta}}\left (c^{\dagger}_{\lambda}a^{\dagger}_{\beta} a^{\dagger}_{\beta} a_{\beta}+c_{\lambda}a^{\dagger}_{\beta}  a_{\beta} a_{\beta}\right ).
\end{eqnarray}
 Since $g_{xy}, g_{ab}\ll g_{\lambda\beta}$, we have neglected the  effects of the weak direct qubit-qubit and resonator-resonator interactions.

The  commutation relation for the second-order expansion term,
\begin{eqnarray}\label{eq:26}
& &[S_{\lambda\beta},[S_{\lambda\beta},H_{nl,\beta}]]\nonumber\\
&=&\left[\frac{4 g^2_{\lambda x}\alpha_{\beta}}{\Delta^2_{\lambda \beta}} c^{\dagger}_{\lambda}c_{\lambda}a^{\dagger}_{\beta}a_{\beta}
+\frac{4g^2_{\lambda \beta}\alpha_{\beta}}{\Sigma^2_{\lambda \beta}} c_{\lambda}c^{\dagger}_{\lambda} a^{\dagger}_{\beta}a_{\beta} \right] \nonumber\\
&+&\left(\frac{2g^2_{\lambda \beta}\alpha_{\beta}}{\Sigma^2_{\lambda \beta}}-\frac{2 g^2_{\lambda \beta}\alpha_{\beta}}{\Delta^2_{\lambda \beta}}\right) a^{\dagger}_{\beta}a^{\dagger}_{\beta}a_{\beta} a_{\beta}\nonumber\\
&+&\frac{ g^2_{\lambda \beta}\alpha_{\beta}}{\Delta^2_{\lambda \beta}} c^{\dagger}_{\lambda}c^{\dagger}_{\lambda} a_{\beta}a_{\beta}+\frac{g^2_{\lambda \beta}\alpha_{\beta}}{\Delta^2_{\lambda \beta}}c_{\lambda}c_{\lambda} a^{\dagger}_{\beta}a^{\dagger}_{\beta}\nonumber\\
&+&\frac{g^2_{\lambda \beta}\alpha_{\beta}}{\Sigma^2_{\lambda \beta}}c^{\dagger}_{\lambda}c^{\dagger}_{\lambda} a^{\dagger}_{\beta}a^{\dagger}_{\beta}+\frac{g^2_{\lambda \beta}\alpha_{\beta}}{\Sigma^2_{\lambda \beta}} c_{\lambda}c_{\lambda} a_{\beta}a_{\beta}\nonumber\\
&+&\frac{2g^2_{\lambda \beta}\alpha_{\beta}}{\Delta_{\lambda \beta}\Sigma_{\lambda \beta}}c^{\dagger}_{\lambda} c^{\dagger}_{\lambda} a^{\dagger}_{\beta}a_{\beta}
+\frac{4 g^2_{\lambda \beta}\alpha_{\beta}}{\Delta_{\lambda \beta}\Sigma_{\lambda \beta}} c_{\lambda}c_{\lambda}a^{\dagger}_{\beta}a_{\beta}\nonumber\\
&+ &\frac{g^2_{\lambda \beta}\alpha_{\beta}}{\Delta_{\lambda \beta}\Sigma_{\lambda \beta}} (2 c^{\dagger}_{\lambda}c_{\lambda}+1) a_{\beta}a_{\beta}\nonumber\\
&+&\frac{g^2_{\lambda \beta}\alpha_{\beta}}{\Delta_{\lambda \beta}\Sigma_{\lambda \beta}} (2 c^{\dagger}_{\lambda}c_{\lambda}+1) a^{\dagger}_{\beta}a^{\dagger}_{\beta}
\end{eqnarray}

Up to the second-order expanding terms, keeping the energy and particle number conservation  terms, we can get the nonlinear term for qubit $\beta$ in the decoupled coordinate as
\begin{eqnarray}\label{eq:27}
H^{\prime}_{nl,\beta}&=&\sum_{\lambda=a,b}\left(\frac{g^2_{\lambda \beta}\alpha_{\beta}}{\Sigma^2_{\lambda \beta}}-\frac{g^2_{\lambda \beta}\alpha_{\beta}}{\Delta^2_{\lambda \beta}}\right) a^{\dagger}_{\beta}a^{\dagger}_{\beta}a_{\beta} a_{\beta}\nonumber\\
&+&\sum_{\lambda=a,b}\left[\frac{2 g^2_{\lambda x}\alpha_{\beta}}{\Delta^2_{\lambda \beta}} c^{\dagger}_{\lambda}c_{\lambda}a^{\dagger}_{\beta}a_{\beta}+\frac{2 g^2_{\lambda \beta}\alpha_{\beta}}{\Sigma^2_{\lambda \beta}} c_{\lambda}c^{\dagger}_{\lambda} a^{\dagger}_{\beta}a_{\beta} \right] \nonumber\\
& &+\sum_{\lambda=a,b}\left[\frac{ g^2_{\lambda \beta}\alpha_{\beta}}{2\Delta^2_{\lambda \beta}} c^{\dagger}_{\lambda}c^{\dagger}_{\lambda} a_{\beta}a_{\beta}+\frac{g^2_{\lambda \beta}\alpha_{\beta}}{2\Delta^2_{\lambda \beta}}c_{\lambda}c_{\lambda} a^{\dagger}_{\beta}a^{\dagger}_{\beta}\right]\nonumber\\
& &+\sum_{\lambda=a,b}\frac{g_{\lambda \beta}\alpha_{\beta}}{\Delta_{\lambda \beta}}\left(c^{\dagger}_{\lambda}a^{\dagger}_{\beta} a_{\beta} a_{\beta}+c_{\lambda} a^{\dagger}_{\beta} a^{\dagger}_{\beta} a_{\beta}\right).
\end{eqnarray}
In the right side of Eq.(D.4), the  first line  describes the self-kerr resonance, while the second line labels to the cross-kerr resonance.
The third line describe the double virtual processes between one qubit  and one resonator. The fourth line describes combined physical  processes that the qubit and the resonator exchange a virtual photon, and  the qubit simulatively participates the self-excitation and subsequent self-annihilation processes.

\end{document}